\documentclass[%
 aip,
 amsmath,amssymb,
 reprint,
]{revtex4-2}

\usepackage[dvipsnames,monochrome]{xcolor}
\usepackage{tabularx}
\usepackage{graphicx}
\usepackage{dcolumn}
\usepackage{bm}
\usepackage[utf8]{inputenc}
\usepackage[T1]{fontenc}
\usepackage{mathptmx}
\usepackage{etoolbox}
\usepackage{array}

\usepackage{placeins}
\usepackage[caption=false]{subfig} 
\usepackage{epsf}
\usepackage{epsfig}
\usepackage{psfrag}
\usepackage{wrapfig}

\usepackage{tikz}
\usetikzlibrary{calc, arrows.meta, quotes, angles, positioning}

\usepackage{mathrsfs}
\usepackage{amsthm}
\usepackage{stmaryrd}
\usepackage{bbding}
\usepackage{amsfonts}
\usepackage{esint}

\usepackage{multirow}
\usepackage{diagbox}
\usepackage{verbatim}
\usepackage{makecell}
\usepackage{float}
\usepackage{makeidx}
\usepackage{setspace}

\usepackage[ruled,linesnumbered]{algorithm2e}

\usepackage{hyperref}

\newtheorem{remark}{Remark}[section]

\makeatletter
\def\@email#1#2{%
 \endgroup
 \patchcmd{\titleblock@produce}
  {\frontmatter@RRAPformat}
  {\frontmatter@RRAPformat{\produce@RRAP{*#1\href{mailto:#2}{#2}}}\frontmatter@RRAPformat}
  {}{}
}%
\makeatother

\begin{document}
	
\preprint{AIP/123-QED}

\title{A Three-Dimensional Two-Temperature Gas-Kinetic Scheme with Generalized Kinetic Boundary Condition for Hypersonic SBLI}
	
	% \author[XJTU]{Xingjian Gao}
	% \ead{gxjacjm88@stu.xjtu.edu.cn}
		
	% \author[XJTU]{Xing Ji\corref{cor1}}
	% \ead{jixing@xjtu.edu.cn}
	
	% \author[CJLU]{Hualin Liu}
	% \ead{hualinliu@cjlu.edu.cn}
	
    % \author[XJTU]{Gang Chen}
	% \ead{aachengang@xjtu.edu.cn}

	% \address[XJTU]{Shaanxi Key Laboratory of Environment and Control for Flight Vehicle, 
	% State Key Laboratory for Strength and Vibration of Mechanical Structures, School of Aerospace Engineering, 
	% Xi'an Jiaotong University, Xi'an 710049, China}
	% \address[CJLU]{ College of Science, China Jiliang University, Hangzhou, China}
	% \cortext[cor1]{Corresponding author}

\author{Xingjian Gao}
% \author{Xingjian Gao(高行健)}
% \email{gxjacjm88@stu.xjtu.edu.cn}
\affiliation{
	State Key Laboratory for Strength and Vibration of Mechanical Structures, 
	Shaanxi Key Laboratory of Environment and Control for Flight Vehicle, 
	School of Aerospace Engineering, 
	Xi'an Jiaotong University, Xi'an 710049, China
}

\author{Hualin Liu}
% \author{Hualin Liu(刘华林)}
% \email{hualinliu@cjlu.edu.cn}
\affiliation{
	College of Sciences, China Jiliang University, Hangzhou, China
}

% \author{Xian Wang}
% % \author{Xian Wang(王娴)}
% % \email{wangxian@mail.xjtu.edu.cn}
% \affiliation{
% 	State Key Laboratory for Strength and Vibration of Mechanical Structures, 
% 	Shaanxi Key Laboratory of Environment and Control for Flight Vehicle, 
% 	School of Aerospace Engineering, 
% 	Xi'an Jiaotong University, Xi'an 710049, China
% }

\author{Fengxiang Zhao}
% \author{Fengxiang Zhao(赵丰祥)}
\affiliation{
Department of Mathematics, Hong Kong University of Science and Technology, Clear Water Bay, Kowloon, Hong Kong, China
}

\author{Xing Ji}
% \author{Xing Ji(姬兴)}
\email{jixing@xjtu.edu.cn}
\affiliation{
	State Key Laboratory for Strength and Vibration of Mechanical Structures, 
	Shaanxi Key Laboratory of Environment and Control for Flight Vehicle, 
	School of Aerospace Engineering, 
	Xi'an Jiaotong University, Xi'an 710049, China
}

% \author{Gang Chen}
% % \author{Gang Chen(陈刚)}
% % \email{aachengang@xjtu.edu.cn}
% \affiliation{
% 	State Key Laboratory for Strength and Vibration of Mechanical Structures, 
% 	Shaanxi Key Laboratory of Environment and Control for Flight Vehicle, 
% 	School of Aerospace Engineering, 
% 	Xi'an Jiaotong University, Xi'an 710049, China
% }

\date{\today}

\begin{abstract}
Accurate prediction of aerothermal loads in hypersonic flows is critical yet challenging due to the coupling of Shock-Wave/Boundary-Layer Interactions (SBLI) and 
thermal non-equilibrium. This work presents the development of a three-dimensional two-temperature Gas-Kinetic Scheme (3D 2T-GKS) on unstructured meshes. The scheme 
resolves translational-rotational and vibrational energy modes within a unified kinetic framework. A key innovation is the integration of a Generalized Kinetic Boundary 
Condition (GKBC), which physically decouples the thermal accommodation of vibrational energy from the translational-rotational mode, thereby offering a more accurate model 
for gas-surface interactions. Additionally, a Discontinuity Feedback Factor (DFF) is employed to capture strong shock waves with reduced numerical dissipation compared to 
classical limiters. The method is rigorously validated against standard experimental benchmarks, including the sharp double-cone and hollow cylinder-flare configurations. 
Numerical results demonstrate that the proposed solver, augmented by the GKBC, accurately captures complex wave structures, separation topologies, and surface heat flux 
distributions. These findings confirm the robustness and fidelity of the 3D 2T-GKS for simulating complex hypersonic non-equilibrium flows.
\end{abstract}

\maketitle

% \begin{keyword}
% 	gas-kinetic scheme; two-temperature kinetic model; thermodynamic non-equilibrium; 
% 	vibration mode; hypersonic
% \end{keyword}

\maketitle

\section{Introduction}

The development of advanced hypersonic vehicles requires addressing critical challenges in aerothermodynamics~\cite{Anderson1989}. During atmospheric reentry, flight velocities 
can reach several kilometers per second~\cite{Bose2013, Olynick1999} , generating strong bow shocks that elevate post-shock gas temperatures to thousands of Kelvin. 
Under these conditions, the flow field exhibits complex high-temperature phenomena, including the excitation of internal energy modes such as translational, rotational,
and vibrational modes~\cite{Park1989-2}. The relaxation of these energy modes and the flow 
evolution typically occur on comparable time scales, leading to a regime known as thermal non-equilibrium~\cite{Bird1994}.

Beyond thermodynamic effects, a ubiquitous and critical phenomenon in hypersonic aerodynamics is Shock-Wave/Boundary-Layer Interaction (SBLI)~\cite{Babinsky2011}. 
Frequently occurring on control surfaces and body flaps, SBLI induces complex flow features such as flow separation, reattachment, 
and shock-shock interference, often resulting in severe localized aerodynamic heating~\cite{Edney1968}. 
In realistic flight scenarios, this gas dynamic phenomenon is often intricately coupled with significant thermal non-equilibrium effects. The mutual interaction 
between the viscous separation zone and the relaxation of internal energy modes adds layers of complexity to the flow physics~\cite{Candler2019}.

To study these coupled phenomena, extensive experimental investigations have been conducted in the LENS shock tunnels at the Calspan University of Buffalo Research Center 
(CUBRC) using canonical configurations, such as the double cone and hollow cylinder-flare~\cite{Holden2003, Holden2004}. 
These experiments serve as rigorous benchmarks for Computational Fluid Dynamics (CFD) 
validation~\cite{Candler2001, Candler2002, Candler2003, Druguet2003, Harvey2001, Holden2006, MacLean2004, Knight2018, Hao2022}. 
While these methods have achieved a reasonable degree of accuracy in predicting general flow features, there remains room for improvement, 
particularly in the precise prediction of the separation region size and the accurate modeling of flows in rarefied regions~\cite{Candler2002, Hao2022}.
Building upon these foundational studies, the research landscape has progressively expanded to address more complex flow physics, including Shock-Wave/Turbulent Boundary-Layer 
Interactions (SWTBLI)~\cite{Holden2015, Dai2022} and strong chemical non-equilibrium in diverse atmospheric compositions~\cite{Holden2013, Nompelis2010, Hao2017}. 
On the methodological side, high-fidelity approaches have been developed to improve predictive capabilities~\cite{Harvey2003, Grover2025}. 
Meanwhile, investigations of flow three-dimensionality, global instabilities, and the effects of freestream disturbances have provided insights into the underlying mechanisms 
governing separation and unsteadiness~\cite{Hao2022, Li2025, Kieweg2019, Hong2022}.

Despite these high-fidelity advancements, conventional CFD methods based on the Navier-Stokes (NS) equations remain the workhorses for engineering simulations due to their 
computational efficiency. However, while augmented with multi-temperature models, they face intrinsic 
limitations. Fundamentally, the NS equations rely on linear constitutive relations derived under the assumption of small Knudsen numbers (near-equilibrium). In regions with 
strong gradients, such as shock layers and the Knudsen layer near the wall, the gas distribution function deviates from the equilibrium state, rendering these linear 
assumptions insufficient. Furthermore, in standard finite volume solvers, surface quantities such as heat flux are typically evaluated via discrete finite-difference 
approximations based on the state of the first near-wall cell and the boundary condition. This approach implicitly assumes a linear macroscopic profile within the first 
grid cell, making the accuracy highly sensitive to the near-wall grid resolution (first cell height) and lacking the physical fidelity to capture the complex non-equilibrium 
gas distribution function at the gas-solid interface~\cite{Lofthouse2008}.

As a promising alternative, the Gas-Kinetic Scheme (GKS), based on the mesoscopic Bhatnagar-Gross-Krook (BGK) model, provides a unified framework that naturally incorporates 
both inviscid and viscous fluxes through the evolution of a microscopic distribution function~\cite{Xu2001}. Crucially, GKS allows for the natural coupling with 
multi-temperature thermodynamic models~\cite{Xu2004-2T, Cao2022-3T, Liu2021-3T}. 
Our previous work established a robust and accurate two-temperature GKS~\cite{Gao2025} by coupling the kinetic method with the widely used 
Park model~\cite{Park1989-1}. This model distinguishes between translational-rotational ($T_{tr}$) and vibrational ($T_{v}$) energy modes. The physical rationale lies 
in the nature of thermal non-equilibrium in air and nitrogen: the rotational temperature tends to rapidly equilibrate with the translational one, whereas vibrational 
relaxation is significantly slower. However, despite these successes in two dimensions, the extension of multi-temperature gas-kinetic schemes to complex three-dimensional 
SBLI problems on unstructured meshes remains relatively unexplored.

A critical aspect in hypersonic aerothermodynamics is the Gas-Surface Interaction (GSI). In the near-continuum transition regime, velocity slip and temperature jump become 
significant~\cite{Lofthouse2008}. More importantly, experiments indicate that vibrational energy accommodates to the wall at a much slower rate compared to translational and 
rotational modes~\cite{Black1974}.
However, the standard kinetic Maxwell boundary condition typically couples the momentum and thermal accommodation coefficients~\cite{Li2005}. This formulation implicitly assumes 
that all internal energy modes accommodate to the wall state at the same rate as momentum exchange. Consequently, it forces the often-frozen vibrational energy to equilibrate 
instantaneously at the wall, leading to a significant overprediction of surface heat flux~\cite{Xu2005}. In contrast, the kinetic nature of GKS allows for the derivation of 
Generalized Kinetic Boundary Conditions (GKBC) based on particle scattering kernels. This approach effectively decouples the accommodation processes, allowing for distinct 
accommodation coefficients for different energy modes.

The primary objective of this study is to validate the robustness and accuracy of the 3D 2T-GKS integrated with the proposed GKBC. The paper is organized as follows. 
Section 2 introduces the numerical algorithm, including the 3D kinetic model and the generalized boundary condition. Section 3 presents the results and discussion, 
beginning with the validation against two standard benchmarks (sharp double cone and hollow cylinder-flare), followed by a parametric analysis of Reynolds number effects 
via density variation, and concluding with a three-dimensional double-cone configuration at a $2^\circ$ angle of attack to further assess the capability of the proposed method 
under asymmetric flow conditions. Conclusions are drawn in Section 4.

\section{Gas-kinetic models and macroscopic governing equations}

In this section, the extended kinetic model and its derived macroscopic equations in 
three dimensions for diatomic gases are presented. 

\subsection{Non-equilibrium translational-rotational and vibrational temperature model}
The Boltzmann equation describes the behavior of a many-particle kinetic system
through the evolution of a single-particle gas distribution function.
The right-hand side represents binary molecular collisions, which are valid over a wide range of pressures.
The Bhatnagar--Gross--Krook (BGK) model is usually applied for 
the simplification of the collision term in Boltzmann equations~\cite{Bhatnagar1954}. 
In equilibrium flows, all energy modes (translational, rotational, and vibrational) are assumed to share a 
common temperature. However, this assumption becomes inaccurate for non-equilibrium flows 
because of the different temperatures for the translational, rotational and vibrational energy modes. 
In this subsection, we propose a BGK model in which translational and rotational energies are assumed to 
be equilibrated, while the vibrational energy remains in non-equilibrium. Although BGK models for 
non-equilibrium vibrational energy have been introduced in earlier studies, this is the first formulation 
that distinguishes between a translational-rotational equilibrium and vibrational non-equilibrium within a 
BGK framework, and couples it with the Park two-temperature model in the context of GKS.
For the non-equilibrium two-temperature diatomic gas flow, 
the above-mentioned BGK model can be extended in the following form:
\begin{equation}\label{bgk_nonequ}
\frac{\partial f}{\partial t} + \mathbf{u}\cdot\nabla f
= \frac{{f}^{eq} - f}{\tau } + \frac{g - {f}^{eq}}{Z_v\tau } 
= \frac{{f}^{eq} - f}{\tau } + {Q}_{s},
\end{equation}
where $f$ is the distribution function, defined as the number density of molecules at the position $(x, y, z)$ 
and particle velocity $(u, v, w)$ at time $t$, and $g$ denotes the local equilibrium state, represented as a 
Maxwellian distribution constructed from local macroscopic quantities.
To model thermal non-equilibrium, an intermediate equilibrium distribution $f^{eq}$ is introduced, 
characterized by two distinct temperatures: a translational-rotational temperature and a 
vibrational temperature.
$\tau=\mu/p$ is the characteristic relaxation time, 
$Q_s$ is inelastic collision operator, accounts for the energy exchange between 
translational-rotational and vibrational modes, serving as a source in the macroscopic three-dimensional 
flow evolution equations. The coefficient \(Z_v\) is termed the vibrational collision number. 
It is a dimensionless multiplier relating the vibrational relaxation time \(\tau_v\) 
to the characteristic relaxation time \(\tau\) of the total energy, 
i.e., \(\tau_v = Z_v\,\tau\); thus \(Z_v\) scales the overall relaxation timescale 
to the vibrational mode.
The left-hand side of the equation represents the free transport of molecules in 
physical space, while the right-hand side models the relaxation process due to particle collisions.
The intermediate equilibrium state $f^{eq}$ is expressed as follows:
\begin{equation}\label{f^eq}
	\begin{aligned}
{f}^{eq} &= \rho {\left( \frac{{\lambda }_{tr}}{\pi }\right) }^{(K_r+3)/2} 
{e}^{-{\lambda }_{tr}\left\lbrack  {{\left( u - U\right) }^{2} + {\left( v - V\right) }^{2}
+ {\left( w - W\right) }^{2} +{\xi }_{r}^{2} }\right\rbrack}  \\
&\quad \times {\left( \frac{{\lambda }_{v}}{\pi }\right) }^{K_v/2}
{e}^{-{\lambda }_{v}{\xi }_{v}^{2}}.
	\end{aligned}
\end{equation}
Here, $\rho$ is the density, and $(U, V, W)$ are the macroscopic fluid velocities in the $x$-, $y$-, 
and $z$- directions, respectively. 
The parameter $\lambda_{tr} = m/2kT_{tr}$ is related to the translational-rotational temperature $T_{tr}$,  
while $\lambda_v = m/2kT_v$ accounts for the vibrational temperature $T_v$. 
For three-dimensional non-equilibrium diatomic gas, the internal variable $\xi$ represents the 
contributions from rotational and vibrational modes, 
with no internal translational degree of freedom included.
since the internal translational degree of freedom along the $z$-direction is not included. 
Its expression is given by $\xi^2 = \xi_r^2 + \xi_v^2$, 
where $\xi_r$ and $\xi_v$ correspond to rotational and vibrational energies, with $K_r$ and $K_v$ 
degrees of freedom, respectively. 
The vibrational degrees of freedom $K_v$ are determined from the vibrational-energy 
equation~\cite{Bird1994}.
\begin{equation}\label{K_v}
	K_{v}=\frac{2\theta_{v}/T_{v}}{e^{\theta_{v}/T_{v}}-1},
\end{equation}
where $\theta_v$ is the vibrational characteristic temperature. For nitrogen, $\theta_v=3393~\mathrm{K}$ 
is used in this paper. 
The RHS collision term in Eq.~\eqref{bgk_nonequ} consists of two terms corresponding to elastic and inelastic 
collisions, respectively, where the relaxation process becomes $f \to f^{eq} \to g$. 
In the elastic collision stage, internal energy exchange between translational-rotational and vibrational 
modes is prohibited. Over a time scale $\tau$, the gas relaxes from its initial non-equilibrium state to 
an intermediate equilibrium state $f^{eq}$, where translational and rotational energies follow a Maxwellian 
distribution characterized by a translational-rotational temperature $T_{tr}$, and the vibrational energy 
follows a Maxwellian distribution at a vibrational temperature $T_v$.

Subsequently, during the inelastic collision stage, energy exchange between translational-rotational and 
vibrational modes occurs over a time scale $Z_v \tau$, and the distribution function further relaxes to 
the final equilibrium state $g$, where all energy modes share a common equilibrium temperature and follow 
a full Maxwellian distribution. The coefficient $Z_v$ is termed the vibrational collision number. 

The relation between mass $\rho$, momentum $(\rho U, \rho V, \rho W)$, total energy 
$\rho E$, and vibrational energy $\rho E_v$ with the distribution function $f$ is given by
\begin{equation}\label{macro&micro}
\mathbf{W} = \left( \begin{matrix} \rho \\  {\rho U} \\  {\rho V}\\  {\rho W} \\  {\rho E} \\ 
	 \rho {E}_{v} \end{matrix}\right)  = \int {\psi }_{\alpha }{f\mathrm{d}u\mathrm{d}
	 v\mathrm{d}w \mathrm{d}{\xi}_{r}\mathrm{d}{\xi}_{v}},\alpha  = 1,2,3,4,5,6.
\end{equation}
\textcolor{blue}{
$\mathbf{W}$ represents the matrix composed of all the aforementioned conserved quantities together with the 
vibrational energy.}
The vibrational energy $\rho E_v$ can be calculated using the relation $\rho E_v=\frac{K_v}{2}\rho RT_v$. 
The integration is performed over the entire velocity space and internal degrees of freedom space, 
with limits from $-\infty$ to $+\infty$.
The detailed formulas for the moment calculation are given in \cite{Gao2025}.
$\psi_\alpha$ is the component of the vector for moments as follows:
\begin{equation}
	\begin{aligned}
\psi_\alpha  &= {\left( {\psi }_{1},{\psi }_{2},{\psi }_{3},{\psi }_{4},{\psi }_{5}
,{\psi }_{6}\right) }^{T} \\
&= {\left( 1,u,v,w,\frac{1}{2}\left( {u}^{2} + {v}^{2} + {w}^{2} + {\xi }_{r}^{2} 
+ {\xi }_{v}^{2}\right) 
,\frac{1}{2}{\xi }_{v}^{2}\right) }^{T}.
	\end{aligned}
\end{equation}
Using the formulas in \cite{Gao2025}, 
the flux expressions can be further obtained as follows for subsequent use.
\begin{equation}\label{Flux}
	\mathbf{F} = \int u{\psi }_{\alpha }{f\mathrm{d}u\mathrm{d}v\mathrm{d}w
	\mathrm{d}{\xi}_{r}\mathrm{d}{\xi}_{v}},\alpha  = 1,2,3,4,5,6.
\end{equation}
As a separate vibrational temperature $T_v$ is introduced, the constraint of
vibrational energy relaxation has to be imposed on the above
extended kinetic model to self-consistently determine all
unknowns. However, since only mass, momentum, and total energy are conserved during particle 
collisions, while vibrational energy undergoes exchange with translational-rotational modes, 
the original compatibility condition for the collision term is no longer strictly satisfied. 
Instead, a modified compatibility condition is imposed, where the vibrational energy relaxation 
appears as a non-conservative source term in the macroscopic equations.
\begin{equation}\label{cc}
\int(\frac{f^{eq}-f}{\tau}+Q_{s})\psi_{\alpha}\mathrm{d}u\mathrm{d}v\mathrm{d}w
\mathrm{d}{\xi}_{r}\mathrm{d}{\xi}_{v}=\mathbf{S}=(0,0,0,0,0,s)^{T}, 
\end{equation}
where $\alpha=1,2,3,4,5,6$.
The source term for the vibrational energy $s$ is from the energy
exchange between translational-rotational and vibrational energies during
inelastic collision. Which is modeled through the Landau--Teller--Jeans-type relaxation model as follows~\cite{Benettin1997},
\begin{equation}\label{source}
s=\frac{(\rho E_v)^{eq}-\rho E_v}{Z_v\tau}.
\end{equation}
The equilibrium energy $(\rho E_{v})^{eq}$ is determined by the
assumption $T_{tr} = T_v = T^{eq}$ such that
\begin{equation}\label{source_cal}
	\begin{aligned}
	\rho E_{v}^{eq}&=\frac{K_{v}}{2}\rho RT^{eq},\\
	T^{eq}&=\frac{(3+K_{r}) T_{tr}+K_{v} T_{v}}{3+K_{r}+K_{v}}.
    \end{aligned}
\end{equation}
The product of the particle collision time $\tau$ and the vibrational collision number $Z_v$ represents the 
relaxation time for the vibrational energy to equilibrate with the translational-rotational energy. 
The value of $Z_v$ is calculated using the classical Millikan-White expression for continuum flows
\cite{Millikan1963,Chapman1970}, with additional corrections applied to account for non-equilibrium 
effects in the vibrational mode\cite{Lumpkin1991}:
\begin{equation}
Z_v=\frac{3+K_{r}}{3+K_{r}+K_{v}}  \hspace{3pt} \frac{c_1}{T_{tr}^\omega} 
\exp \left({\frac{c_2}{T_{tr}^{1/3}}}\right),
\end{equation}
where $c_1 = 9.1$, $c_2 = 220$, and $\omega = 0.74$.
In order to simulate the flow with any realistic Prandtl number, a modification of the heat flux in the 
energy transport is used in GKS, which is also implemented in the present study.

\subsection{Macroscopic equation corresponding to the current method}
Based on the intermediate state given by Eq.~\eqref{f^eq}, with vibrational energy exchange frozen,  
the first-order Chapman--Enskog expansion of the non-equilibrium distribution function $f$ yields the 
following expression~\cite{Chapman1970}:
\begin{equation}
f=f^{eq}+\epsilon f_{1}=f^{eq}-\tau(\frac{\partial f^{eq}}{\partial t}+u\frac{\partial f^{eq}}
{\partial x}+v\frac{\partial f^{eq}}{\partial y}+w\frac{\partial f^{eq}}{\partial z}).
\end{equation}
where $\epsilon$ is a small dimensionless quantity.
The corresponding macroscopic non-equilibrium translational-rotational and vibrational 
two-temperature macroscopic equations in two dimensions can be
derived as follows, and the detailed derivation is provided in \cite{Gao2025}. 
\begin{equation}\label{NS}
    \frac{\partial W}{\partial t}+\frac{\partial F}{\partial x}+\frac{\partial G}{\partial y}
	+\frac{\partial H}{\partial z}
    =\frac{\partial F_{v}}{\partial x}+\frac{\partial G_{v}}{\partial y}
	+\frac{\partial H_{v}}{\partial z}+S,
\end{equation}
with
\begin{widetext}
\begin{equation}
	\begin{gathered}
    W=\begin{pmatrix}\rho\\ \rho U\\ \rho V\\ \rho W \\ \rho E\\\rho E_v \end{pmatrix}, \quad
    F=\begin{pmatrix}\rho U\\ \rho U^2+p\\ \rho U V\\ \rho U W \\ (\rho E+p)U\\ \rho E_v U\end{pmatrix}, \quad
    G=\begin{pmatrix}\rho V\\ \rho V U\\ \rho V^2+p\\ \rho V W \\ (\rho E+p)V\\ \rho E_v V\end{pmatrix}, \quad
    H=\begin{pmatrix}\rho W\\ \rho W U\\ \rho W V\\ \rho W^2+p \\ (\rho E+p)W\\ \rho E_v W\end{pmatrix}, \\
    F_v=\begin{pmatrix}0\\ \tau_{xx}\\ \tau_{xy}\\ \tau_{xz} \\ U\tau_{xx}+V\tau_{xy}+W\tau_{xz}+q_x\\U\tau_{tr-v}+q_{vx}
	\end{pmatrix}, \quad
    G_v=\begin{pmatrix}0\\ \tau_{yx}\\ \tau_{yy}\\ \tau_{yz} \\ U\tau_{yx}+V\tau_{yy}+W\tau_{yz}+q_y\\V\tau_{tr-v}+q_{vy}
	\end{pmatrix}, \quad
    H_v=\begin{pmatrix}0\\ \tau_{zx}\\ \tau_{zy}\\ \tau_{zz} \\ U\tau_{zx}+V\tau_{zy}+W\tau_{zz}+q_z\\W\tau_{tr-v}+q_{vz}
	\end{pmatrix},
    \end{gathered}
\end{equation}
\end{widetext}
where $\rho E = \frac{1}{2} \rho(U^2 + V^2 + W^2 + (3+K_r)RT_{tr}+K_v RT_v)$ is the total energy and
$\rho E_{v} = \frac{K_v}{2}\rho RT_v$ is the vibrational energy. The pressure $p$ is related to 
the translational-rotational temperature as $p = \rho RT_{tr}$. In particular, 
the viscous normal stress terms are
\begin{equation}
    \begin{aligned}
    \tau_{xx} =&\tau p\left[2\frac{\partial U}{\partial x}
    -\frac{2}{3+K_r}\left(\frac{\partial U}{\partial x}+\frac{\partial V}{\partial y}+\frac{\partial W}{\partial z}\right)\right] \\
    &-\frac{\rho K_v}{2(K_r+K_v+3)Z_v}\left(\frac{1}{\lambda_{tr}}-\frac{1}{\lambda_{v}}\right), \\
    \tau_{yy} =&\tau p\left[2\frac{\partial V}{\partial y}
    -\frac{2}{3+K_r}\left(\frac{\partial U}{\partial x}+\frac{\partial V}{\partial y}+\frac{\partial W}{\partial z}\right)\right] \\
    &-\frac{\rho K_v}{2(K_r+K_v+3)Z_v}\left(\frac{1}{\lambda_{tr}}-\frac{1}{\lambda_{v}}\right), \\
    \tau_{zz} =&\tau p\left[2\frac{\partial W}{\partial z}
    -\frac{2}{3+K_r}\left(\frac{\partial U}{\partial x}+\frac{\partial V}{\partial y}+\frac{\partial W}{\partial z}\right)\right] \\
    &-\frac{\rho K_v}{2(K_r+K_v+3)Z_v}\left(\frac{1}{\lambda_{tr}}-\frac{1}{\lambda_{v}}\right),
    \end{aligned}
\end{equation}
with the viscous shear stress component given by
\begin{equation}
    \begin{gathered}
	\tau_{xy}=\tau_{yx}=\tau p\left(\frac{\partial U}{\partial y}+\frac{\partial V}{\partial x}\right), \\
    \tau_{xz}=\tau_{zx}=\tau p\left(\frac{\partial U}{\partial z}+\frac{\partial W}{\partial x}\right), \\
    \tau_{yz}=\tau_{zy}=\tau p\left(\frac{\partial V}{\partial z}+\frac{\partial W}{\partial y}\right),
    \end{gathered}
\end{equation}
and the heat conduction components are expressed as
\begin{equation}\label{q}
    \begin{gathered}
    q_{x}=\tau p \left[\frac{5+K_r}{4}\frac{\partial}{\partial x}\left(\frac{1}{\lambda_{tr}}\right)
	+\frac{K_v}{4}\frac{\partial}{\partial x}\left(\frac{1}{\lambda_v}\right)\right], \\
    q_{y}=\tau p\left[\frac{5+K_r}{4}\frac{\partial}{\partial y}\left(\frac{1}{\lambda_{tr}}\right)
	+\frac{K_v}{4}\frac{\partial}{\partial y}\left(\frac{1}{\lambda_{v}}\right)\right], \\
    q_{z}=\tau p\left[\frac{5+K_r}{4}\frac{\partial}{\partial z}\left(\frac{1}{\lambda_{tr}}\right)
	+\frac{K_v}{4}\frac{\partial}{\partial z}\left(\frac{1}{\lambda_{v}}\right)\right].
    \end{gathered}
\end{equation}
The following terms contribute to the governing equation of vibrational energy $\rho E_v$:
\begin{equation}
	\begin{gathered}
    \tau_{tr-v}=\frac{(K_r+3)\rho K_v}{4(K_r+K_v+3)Z_v}\left(\frac{1}{\lambda_{tr}}-\frac{1}{\lambda_v}\right), \\
    q_{vx}=\tau p \frac{K_v}{4}\frac{\partial}{\partial x}\left(\frac{1}{\lambda_v}\right), \\
    q_{vy}=\tau p \frac{K_v}{4}\frac{\partial}{\partial y}\left(\frac{1}{\lambda_v}\right), \\
    q_{vz}=\tau p \frac{K_v}{4}\frac{\partial}{\partial z}\left(\frac{1}{\lambda_v}\right).
    \end{gathered}
\end{equation}
The source term in Eq.~\eqref{NS} is defined in Eq.~\eqref{source}.

\section{Numerical Method}
\subsection{Gas-kinetic scheme on the framework of finite volume method}
First, take the moments of the BGK model Eq.~\eqref{bgk_nonequ} in the velocity and 
internal state spaces. Then integrate it over a control volume $\Omega_i$, we obtain
\begin{equation}\label{integrate_bgk}
\begin{split}
{\int }_{{\Omega }_{i}}\int \left( {{f}_{t} + \mathbf{u} \cdot  \nabla f}\right) 
\psi \mathrm{d}\Xi \mathrm{d}V = \\ {\int }_{{\Omega }_{i}}\int 
(\frac{{f}^{eq} - f}{\tau } + \frac{g - {f}^{eq}}{Z_v\tau })  \psi \mathrm{d}\Xi \mathrm{d}V ,
\end{split}
\end{equation}
where $\mathrm{d}\Xi$ denotes 
$\mathrm{d}u \mathrm{d}v \mathrm{d}\xi_t \mathrm{d}\xi_r \mathrm{d}\xi_v$ 
and $\mathrm{d}V$ is the integration of control volume. 
It should be noted that $\nabla f$ represents the divergence of $f$ in physical space, 
which is independent of $(\boldsymbol{u}, \boldsymbol{\xi})$. Therefore, we obtain:
\begin{equation}
\int \left( {\mathbf{u} \cdot  \nabla f}\right) \psi \mathrm{d}\Xi  = \int \nabla  
\cdot  \left( {\mathbf{u}f}\right) \psi \mathrm{d}\Xi  
= \nabla  \cdot  \int \mathbf{u}{f\psi }\mathrm{d}\Xi.
\end{equation}
Based on Eq.~\eqref{macro&micro}, Eq.~\eqref{Flux}, and the modified compatibility condition in 
Eq.~\eqref{cc}, the integral form is obtained from Eq.~\eqref{integrate_bgk}:
\begin{equation}\label{integral_fvm}
{\int }_{{\Omega }_{i}}{\mathbf{W}}_{t}\mathrm{\;d}V + {\int }_{{\Omega }_{i}}\nabla  
\cdot  \mathbf{F}\mathrm{d}V = {\int }_{{\Omega }_{i}} \mathbf{S} \mathrm{\;d}V.
\end{equation}
The integral form in Eq.~\eqref{integral_fvm} is discretized using the finite volume method (FVM),
\begin{equation}
\frac{\mathrm{d}{\overline{\mathbf{W}}}_{i}}{\mathrm{\;d}t} =  - \frac{1}{\left| {\Omega }_{i}
\right| }{\int }_{{\Omega }_{i}}\nabla  \cdot  \mathbf{F}\mathrm{d}V + \mathbf{S},
\end{equation}
where $|{\Omega }_{i}|$ is the volume of the control volume 
and $\overline{\mathbf{W}_i}$ represents the cell-averaged conserved variables in cell $i$.
From Gauss's theorem, the semi-discrete form of FVM is written as
\begin{equation}\label{semi-discrete fvm}
\begin{aligned}
\frac{\mathrm{d}{\overline{\mathbf{W}}}_{i}}{\mathrm{\;d}t} &= \mathcal{L}\left( 
	{\mathbf{W}}_{i}\right)  =  - \frac{1}{\left| {\Omega }_{i}\right| }
	{\oint}_{\partial {\Omega }_{i}}\mathbf{F} \cdot  \mathbf{n}\mathrm{d}s + \mathbf{S}  \\
	&= - \frac{1}{\left| {\Omega }_{i}\right| }\mathop{\sum }
	\limits_{{p = 1}}^{{N}_{f}}{\int }_{{\Gamma }_{ip}}\mathbf{F} \cdot  
	{\mathbf{n}}_{p}\mathrm{\;d}s + \mathbf{S},	
\end{aligned}
\end{equation}
where $\mathcal{L}\left( {\mathbf{W}}_{i}\right)$ represents the cell residual, 
$\partial {\Omega }_{i}$ denotes the boundary of the control volume, 
$\mathrm{d}s$ is the area of the corresponding boundary surface, 
and $\mathbf{n}_p$ is the unit outward normal vector of the interface.
$\partial {\Omega }_{i}$ is expressed as the union of all its boundary faces, as given below.
\begin{equation}
	\partial {\Omega }_{i} = \mathop{\bigcup }\limits_{{p = 1}}^{{N}_{f}}{\Gamma }_{ip},
\end{equation}
where ${\Gamma }_{ip}$ is the neighboring interface of the cell ${\Omega }_{i}$, 
${N}_{f}$ is the number of cell interfaces.
Numerical method is used to evaluate the surface integral of fluxes,
\begin{equation}
	{\int }_{{\Gamma }_{ip}}\mathbf{F} \cdot  {\mathbf{n}}_{p}\mathrm{\;d}s = \left| 
		{\Gamma }_{ip}\right| \mathbf{F}\left( {{\mathbf{x}}_{p},t}\right)  \cdot  {\mathbf{n}}_{p},
\end{equation}
where $\left|{\Gamma }_{ip}\right|$ is the area of the mesh face.

\subsection{Gas-kinetic solver}
In the two-temperature GKS, the equilibrium state $g$ 
has been superseded by the intermediate equilibrium state $f^{eq}$ in 
computational implementation. To maintain notation consistency with one-temperature GKS literature, 
the symbol $g$ is hereby formally redefined as $g \equiv f^{eq}$ for all subsequent derivations.

A finite volume method is used to solve the BGK-type model. The general integral solution of $f$ 
in Eq.~\eqref{bgk_nonequ} at a cell interface $\mathbf{x}=(0,0,0)^T$ at time $t$ is expressed as
\begin{equation}
	\begin{aligned}
f\left( \mathbf{x},t,\mathbf{u},{\xi }_{r},{\xi }_{v}\right)  
&= \frac{1}{\tau } \int_{0}^{t} g\left( \mathbf{x}^{\prime},
{t}^{\prime },\mathbf{u},{\xi }_{r},{\xi }_{v}\right) {e}^{-\left( {t - {t}^{\prime }}\right) 
/\tau }d{t}^{\prime } \\
&\quad + {e}^{-t/\tau }{f}_{0}\left( \mathbf{x} - \mathbf{u}t,
\mathbf{u},{\xi }_{r},{\xi }_{v}\right),
    \end{aligned}
\end{equation}
where $\mathbf{x}' =\mathbf{x}-\mathbf{u}(t-t')$ is the trajectory of particle
motion, and $f_0$ is the initial gas distribution function at the beginning of each time step.
For viscous flow, the physical collision time $\tau$ is defined as 
\begin{equation}
\tau= \frac{\mu}{p},
\end{equation}
where $\mu$ is the dynamic viscosity. To properly capture discontinuities with additional
numerical dissipation, the numerical collision time is modified as
\begin{equation}\label{tau}
\tau =\frac{\mu }{p} + C\frac{\left| {p}_{L} - {p}_{R}\right| }{\left| {p}_{L} + {p}_{R}\right|}\Delta t,
\end{equation}
where $C$ is set to 1.0 in the computation. $p_L$ and $p_R$ denote the 
pressures on the left- and right-hand sides at the cell interface, 
which reduces to $\tau = \mu/p$ in smooth flow regions. $\Delta t$ is the 
time step determined according to the Courant--Friedrichs--Lewy (CFL) condition.
The dynamic viscosity coefficient $\mu$ is computed as a function of temperature.
Sutherland's law~\cite{Sutherland1893} is applied for $T \le 1000\,\mathrm{K}$, while a
high-temperature correlation is adopted for $T > 1000\,\mathrm{K}$\cite{Gupta1990}:
\begin{equation}
\mu(T)=
\begin{cases}
\mu_{\mathrm{ref}}
\left( \dfrac{T}{T_{\mathrm{ref}}} \right)^{3/2}
\dfrac{T_{\mathrm{ref}} + S}{T + S},
& T \le 1000\,\mathrm{K}, \\[6pt]
0.1 \exp\left[
C_{\mu} + \left(A_{\mu} \ln T + B_{\mu}\right)\ln T
\right],
& T > 1000\,\mathrm{K},
\end{cases}
\end{equation}
where $\mu_{\mathrm{ref}} = 1.663\times 10^{-5}\,\mathrm{Pa}\cdot\mathrm{s}$,
$T_{\mathrm{ref}} = 273\,\mathrm{K}$, $S = 107\,\mathrm{K}$, 
$A_{\mu} = 0.0203$, $B_{\mu} = 0.4329$, and $C_{\mu} = -11.8153$.
The temperature $T$ denotes the translational-rotational temperature $T_{tr}$.
The initial gas distribution function $f_0$ can be constructed as
\begin{equation}\label{f}
f_0=\begin{cases}g^l(1-\tau({a}_{{x}_{i}}^l u_i+A^l)-t({a}_{{x}_{i}}^l u_i)),\quad x_1\leq0\\
	g^r(1-\tau({a}_{{x}_{i}}^r u_i+A^r)-t({a}_{{x}_{i}}^r u_i)),\quad x_1>0\end{cases},
\end{equation}
where $g^l$ and $g^r$ are related to the macroscopic values reconstructed at the two sides of a cell
interface. The microscopic slope vectors ${a}_{{x}_{i}}^{l,r}$ and scalars $A^{l,r}$ can be calculated using the 
macroscopic slopes. The specific calculations of microscopic slopes are shown in \cite{Gao2025}. The above equation can be simplified as
\begin{equation}
f_0=\begin{cases}f_0^l,\quad x_1\leq0\\
	f_0^r,\quad x_1>0\end{cases},
\end{equation}
the above equation can be further simplified as follows:
\begin{equation}
f_0=f_0^l\mathbb{H}(x_1)+f_0^r(1-\mathbb{H}(x_1)),
\end{equation}
where $\mathbb{H}(x_1)$ is the Heaviside function and $x_1$ represents the local coordinate in the normal direction of the interface. After determining the kinetic part $f_0$, 
the equilibrium state $g$ part can be expressed as
\begin{equation}\label{g}
\begin{aligned}
\frac{1}{\tau } &\int_{0}^{t} g\left( \mathbf{x}^{\prime },
{t}^{\prime },u_i,{\xi }_{r},{\xi }_{v}\right) {e}^{-\left( {t - {t}^{\prime }}\right) 
/\tau }d{t}^{\prime } \\
& =C_1g^c+C_2 ({a}_{{x}_{i}}^c u_i) g^c+C_3 A^c g^c,
\end{aligned}
\end{equation}
where $g^c$ is the Maxwellian equilibrium state located at an interface, which can
be determined through the compatibility condition 
\begin{equation}
 \int \mathbf{\psi }{g}^{c}\mathrm{\;d}\Xi  
 = {\mathbf{W}}^{c} = {\int }_{{u}_{1} > 0}
 \mathbf{\psi }{g}^{l}\mathrm{\;d}\Xi  + {\int }_{{u}_{1} < 0}\mathbf{\psi }{g}^{r}\mathrm{\;d}\Xi .
\end{equation}
The coefficients ${a}_{{x}_{i}}^c, A^c$ in Eq.~\eqref{g} are defined from the expansion of the
$g^c$, the coefficients $C_{1,2,3}$ in Eq.~\eqref{g} are given by
\begin{equation}
C_{1}=1-e^{-t/\tau},C_{2}=(t+\tau)e^{-t/\tau}-\tau,C_{3}=t-\tau+\tau e^{-t/\tau}.
\end{equation}
The details of the calculation of each microscopic term's coefficients ${a}_{{x}_{i}}^{l,r,c}$ and $A^{l,r,c}$ 
in Eq.~\eqref{f} and Eq.~\eqref{g} from macroscopic quantities are given in \cite{Gao2025}.
Then the second-order time dependent gas distribution function at a cell interface is
\begin{equation}\label{f_all}
	\begin{aligned}
f&\left( \mathbf{x},t,\mathbf{u},{\xi }_{r},{\xi }_{v}\right) 
= \left( {1 - {e}^{-t/\tau }}\right) {g}^{c} \\
&+ \left( {\left( {t + \tau }\right) {e}^{-t/\tau } 
- \tau }\right) ({a}_{{x}_{i}}^c u_i){g}^{c} + \left( {t - \tau  + \tau {e}^{-t/\tau }}\right) 
{A}^{c}{g}^{c} \\
&+{e}^{-t/\tau }{g}^{l}\left\lbrack  {1 - \left( {\tau  + t}\right) 
({a}_{{x}_{i}}^l u_i) - \tau {A}^{l}}\right\rbrack  \mathbb{H}\left( {u}_{1}\right) \\ 
&+{e}^{-t/\tau }{g}^{r}\left\lbrack  {1 - \left( {\tau  + t}\right) ({a}_{{x}_{i}}^r u_i)
- \tau {A}^{r}} \right\rbrack  \left( {1 - \mathbb{H}\left( {u}_{1}\right) }\right).
\end{aligned}
\end{equation}
The gas distribution function \(f\) is substituted into Eq.~\eqref{Flux} to obtain the flux \(\mathbf{F}\)
in the semi-discrete finite volume formulation of Eq.~\eqref{semi-discrete fvm}.
\begin{remark}
The Prandtl number of real gases depends on temperature. 
For instance, nitrogen has $Pr \approx 0.72$ at room temperature, 
and the value increases at higher temperatures. 
However, the BGK model intrinsically gives a fixed Prandtl number of unity.
From the BGK-based macroscopic heat flux Eq.~\eqref{q},
it follows that the translational-rotational thermal conductivity derived from the BGK flux
corresponds to a Prandtl number of 1. 
To correct this discrepancy and match realistic Prandtl numbers, 
the translational-rotational heat flux in the GKS energy flux is modified as~\cite{Xu2001},
\begin{equation}\label{flux fix}
	{F}_{\rho E}^{new} = {F}_{\rho E} + \left( {\frac{1}{Pr} - 1}\right) q,
\end{equation}
where $F_{\rho E}$ refers to the total-energy component of the flux in
Eq.~\eqref{Flux}, and $F_{\rho E}^{new}$ represents its corrected form.
In smooth flow regions, the time-dependent translational-rotational heat flux can be approximated by
\begin{equation}
\begin{aligned}
q_s &= - \tau \int g^c\, u_1 \Big[
    \left( a^c_{x_i} u_i + A^c \right) (\psi_5 - \psi_6)
    - U \left( a^c_{x_i} u_i u_1 + A^c u_1 \right) \\
&\quad - V \left( a^c_{x_i} u_i v_1 + A^c v_1 \right)
    - W \left( a^c_{x_i} u_i w_1 + A^c w_1 \right)
\Big] \, \mathrm{d}\Xi.
\end{aligned}
\end{equation}
This approximation neglects higher-order non-equilibrium terms and is sufficient
for implementing the Prandtl number correction in near-equilibrium flows.

It is important to note that in the current two-temperature model, the vibrational energy 
is solved via a separate conservation equation with its own independent heat flux terms. 
Therefore, the heat flux $q$ appearing in the total energy correction Eq.~\eqref{flux fix} 
physically represents the heat transport due to translational and rotational modes only. 
Consequently, the Prandtl number $Pr$ used in the correction must be the 
translational-rotational Prandtl number, determined exclusively by the translational 
and rotational properties of the gas.

The specific Prandtl number is defined as
\begin{equation}
    Pr = \frac{c_{p,tr} \mu}{\kappa_{t} + \kappa_r},
\end{equation}
where $\mu$ is the dynamic viscosity, and $\kappa_{t}$ and $\kappa_r$ are the 
translational and rotational thermal conductivities, respectively. 
$c_{p,tr}$ is the translational-rotational specific heat ratio at constant pressure, given by
\begin{equation}
	\begin{aligned}
    c_{p,tr} &= c_{v,t} + c_{v,r} + R, \\
	&=\frac{3}{2}R+\frac{K_r}{2}R+R
	\end{aligned}
\end{equation}
where $R$ is the specific gas constant. The thermal conductivities are related to the 
viscosity and specific heats through the dimensionless Eucken factors $f$~\cite{Eucken1913}:
\begin{equation}
    f_{t} = \frac{\kappa_{t}}{c_{v,t}\mu }, \quad
    f_{r} = \frac{\kappa_{r}}{c_{v,r}\mu }.
\end{equation}
Mason and Monchick~\cite{Mason1962} proposed accurate expressions for the translational and internal 
Eucken factors. For nitrogen, the curve-fitting formulations for the Eucken factor are employed as follows:
\begin{widetext}
\begin{equation}
\begin{aligned}
f_t &=
\begin{cases} 
1.80, & T_{t} \leq 100\,\mathrm{K}, \\  
\displaystyle \frac{2.435 T_{t}^{3} + 47.1 T_{t}^{2} + 3.42 T_{t} + 0.8337}
{T_{t}^{3} + 61.13 T_{t}^{2} - 4.624 T_{t} + 0.8556},
& 100\,\mathrm{K} < T_{t} \leq 3000\,\mathrm{K}, \\  
2.42, & T_{t} > 3000\,\mathrm{K}, 	
\end{cases} \\[1.0ex]
f_{\text{int}} &=
\begin{cases} 
1.88, & T_{\text{int}} \leq 100\,\mathrm{K}, \\  
\displaystyle \frac{1.373 T_{\text{int}}^{3} + 125.7 T_{\text{int}}^{2} - 3.87 T_{\text{int}} + 0.2182}
{T_{\text{int}}^{3} + 39.89 T_{\text{int}}^{2} + 8.668 T_{\text{int}} + 0.9401},
& 100\,\mathrm{K} < T_{\text{int}} \leq 3000\,\mathrm{K}, \\  
1.40, & T_{\text{int}} > 3000\,\mathrm{K}, 
\end{cases} \\[1.0ex]
f_v &= 1.37,
\end{aligned}
\end{equation}
\end{widetext}
where the effective internal temperature $T_{\text{int}}$ is a weighted average of the 
rotational and vibrational temperatures, defined as
\begin{equation}
    T_{\text{int}} = \frac{K_{r}T_{r} + K_{v}T_{v}}{K_{r} + K_{v}},
\end{equation}
with $K_r$ and $K_v$ representing the degrees of freedom for rotation and vibration, respectively. 
The rotational Eucken factor $f_{r}$ is derived from the coupling relation:
\begin{equation}
    f_{r}c_{v,r} + f_{v}c_{v,v} = f_{\text{int}}\left( c_{v,r} + c_{v,v}\right).
\end{equation}
Finally, the correct Prandtl number for Eq.~\eqref{flux fix} is obtained by substituting the derived 
conductivities:
\begin{equation}
    Pr = \frac{c_{v,t} + c_{v,r} + R}{c_{v,t} f_t + c_{v,r} f_r}.
\end{equation}
This ensures that the heat flux correction properly reflects the transport properties of the 
translational and rotational modes, which are the primary contributors to the heat flux $q$ in 
the kinetic scheme.
It is worth noting that a similar discrepancy exists between the numerical vibrational thermal 
conductivity inherent to the BGK scheme and the physical value corresponding to the vibrational 
Eucken factor $f_v$. Theoretically, a flux correction analogous to Eq.~\eqref{flux fix} could be 
applied to the vibrational energy equation. However, the magnitude of the vibrational heat flux is 
typically negligible compared to the translational-rotational component, having a minimal impact on 
the overall flow field resolution. More importantly, the vibrational energy evolution is strongly 
coupled with stiff relaxation source terms. Introducing an additional flux correction term 
can significantly deteriorate the numerical robustness, 
particularly in regions with strong shock waves. Consequently, to maintain numerical stability 
without compromising physical accuracy, no Prandtl number correction is applied to the vibrational 
heat flux in the present study.
\end{remark}

\subsection{Spatial reconstruction via weighted least-squares}
In this study, the computational domain is discretized using unstructured grids to handle complex geometries. 
To achieve second-order spatial accuracy, a linear reconstruction of flow variables is performed within each control volume. 
For a given cell \( i \), the value of a conservative variable \( \phi \) at the face center \( \mathbf{x}_f \) is reconstructed as
\begin{equation}
	\phi_f = \phi_i + \nabla \phi_i \cdot (\mathbf{x}_f - \mathbf{x}_i),
\end{equation}
where \( \phi_i \) is the cell-averaged value, \( \nabla \phi_i \) is the reconstructed gradient vector 
at the cell centroid, 
\( \mathbf{x}_i \) denotes the centroid of cell \( i \), and \( \mathbf{x}_f \) is the position vector of 
the face center.

The accuracy of this reconstruction relies critically on the precise computation of the cell-centered 
gradient \( \nabla \phi_i \). 
To ensure robustness on arbitrary unstructured meshes, particularly those characterized by high aspect 
ratios, significant non-orthogonality, and grid skewness, a weighted least-squares (WLSQ) method is 
employed. 
Unlike methods that depend on specific element shapes, the least-squares construction relies simply on 
a stencil of neighboring points. 
For the cell-centered finite volume scheme used in this work, this stencil is naturally defined by the 
immediate geometric environment: it consists of the centroid of the cell \( i \) itself and the centroids 
of all its face-neighboring cells (denoted as the set \( k \in \mathcal{N}(i) \)).

The gradient is then determined by minimizing the weighted sum of the squared differences between the linear approximation and the actual values at these neighboring centroids. The objective function to be minimized is given as:
\begin{equation}
\sum_{k=1}^{N} w_{ik}^2 E_{ik}^2.
\end{equation}
In three dimensions, the error term $E_{ik}$ is given as
\begin{equation}
E_{ik}^2 = \left( \Delta \phi_{ik} - \left( (\phi_x)_i \Delta x_{ik} + (\phi_y)_i \Delta y_{ik} + (\phi_z)_i \Delta z_{ik} \right) \right)^2,
\end{equation}
where $\Delta \phi_{ik}$ represents the difference $\phi_k - \phi_i$, and $\Delta x_{ik}, \Delta y_{ik}, \Delta z_{ik}$ 
represent the geometric differences $x_k - x_i$, $y_k - y_i$, and $z_k - z_i$, respectively. 
$w_{ik}$ is a weighting factor. Dropping the $i$ subscripts on the gradients for clarity, a system of 
three equations for the three gradient components $\phi_x, \phi_y, \phi_z$ is obtained by solving the 
minimization problem, i.e., setting the partial derivatives with respect to each component to zero:
\begin{equation}
\frac{\partial \sum_{k=1}^{N} w_{ik}^2 E_{ik}^2}{\partial \phi_x} = 0, \quad
\frac{\partial \sum_{k=1}^{N} w_{ik}^2 E_{ik}^2}{\partial \phi_y} = 0, \quad
\frac{\partial \sum_{k=1}^{N} w_{ik}^2 E_{ik}^2}{\partial \phi_z} = 0.
\end{equation}
Using straightforward algebra, the following system of linear equations is obtained:
\begin{equation}
\begin{pmatrix} 
L_{xx} & L_{xy} & L_{xz} \\ 
L_{xy} & L_{yy} & L_{yz} \\ 
L_{xz} & L_{yz} & L_{zz} 
\end{pmatrix}
\begin{pmatrix} \phi_x \\ \phi_y \\ \phi_z \end{pmatrix}
=
\begin{pmatrix} R_x \\ R_y \\ R_z \end{pmatrix},
\end{equation}
where the coefficients of the matrix (symmetric moments of the stencil) are defined as:
\begin{equation}
\begin{aligned}
L_{xx} &= \sum_{k=1}^{N} w_{ik}^2 \Delta x_{ik}^2, & L_{xy} &= \sum_{k=1}^{N} w_{ik}^2 \Delta x_{ik} \Delta y_{ik}, \\
L_{xz} &= \sum_{k=1}^{N} w_{ik}^2 \Delta x_{ik} \Delta z_{ik}, & L_{yy} &= \sum_{k=1}^{N} w_{ik}^2 \Delta y_{ik}^2, \\
L_{yz} &= \sum_{k=1}^{N} w_{ik}^2 \Delta y_{ik} \Delta z_{ik}, & L_{zz} &= \sum_{k=1}^{N} w_{ik}^2 \Delta z_{ik}^2,
\end{aligned}
\end{equation}
and the right-hand side vector components are:
\begin{equation}
\begin{aligned}
R_x &= \sum_{k=1}^{N} w_{ik}^2 \Delta \phi_{ik} \Delta x_{ik}, \\
R_y &= \sum_{k=1}^{N} w_{ik}^2 \Delta \phi_{ik} \Delta y_{ik}, \\
R_z &= \sum_{k=1}^{N} w_{ik}^2 \Delta \phi_{ik} \Delta z_{ik}.
\end{aligned}
\end{equation}
In practice, all the matrix terms ($L_{xx}, \dots, L_{zz}$) depend only on the grid geometry and metrics, 
so they can be precomputed and stored to improve efficiency. 
To address accuracy issues on highly stretched meshes, an inverse distance weighting is used in this work:
\begin{equation}
w_{ik} = \frac{1}{\sqrt{\Delta x_{ik}^2 + \Delta y_{ik}^2 + \Delta z_{ik}^2}}.
\end{equation}
With this weighting, the system is much better conditioned compared to the unweighted case ($w_{ik}=1$), 
scaling as $\mathcal{O}(1)$ rather than exhibiting ill-conditioning due to higher-order scaling terms. 
The system is solved locally for each cell to obtain the reconstructed gradient vector $\nabla \phi_i$.

Finally, to suppress non-physical oscillations and ensure numerical stability near strong discontinuities, 
the computed gradient is modified by applying the discontinuity feedback factor (DFF):
\begin{equation}
	\widetilde{\nabla \phi_i} = \alpha_i \nabla \phi_i,
\end{equation}
where \( \alpha_i \) is the limitation coefficient determined by the DFF sensor.

\subsection{Discontinuity feedback factor}
To suppress non-physical oscillations near strong discontinuity, the reconstructed gradients 
obtained in the previous section are limited using the Discontinuity Feedback Factor (DFF)~\cite{Ji2021}. 
The DFF acts as a scalar limiter $\alpha_i \in [0, 1]$, modifying the slope as:
\begin{equation}
	\widetilde{\nabla \phi}_i = \alpha_i \nabla \phi_i.
\end{equation}
When the flow is smooth, $\alpha_i \to 1$, preserving the high-order accuracy; when discontinuities are 
detected, 
$\alpha_i \to 0$, reducing the reconstruction to first-order accuracy to ensure monotonicity.

The calculation of the final DFF involves three steps: interface discontinuity detection, cell-based 
accumulation, and neighborhood averaging.

First, a discontinuity strength parameter $A_f$ is calculated at each face center $f$ of the 
cell $\Omega_i$. 
For three-dimensional flows, this parameter accounts for jumps in pressure and Mach number components:
\begin{equation}
	\begin{aligned}
	A_f &= \frac{\left| {p}_{L} - {p}_{R}\right| }{{p}_{L}} 
	+ \frac{\left| {p}_{L} - {p}_{R}\right| }{{p}_{R}} 
    + \left( \mathrm{Ma}_{n}^{L} - \mathrm{Ma}_{n}^{R}\right)^{2} \\
    &\quad + \left( \mathrm{Ma}_{t1}^{L} - \mathrm{Ma}_{t1}^{R}\right)^{2} 
    + \left( \mathrm{Ma}_{t2}^{L} - \mathrm{Ma}_{t2}^{R}\right)^{2},
	\end{aligned}
\end{equation}
where $p_L, p_R$ denote the pressure at the left and right sides of the interface, and $\mathrm{Ma}_{n}$, 
$\mathrm{Ma}_{t1}$, $\mathrm{Ma}_{t2}$ represent the Mach number components in the normal and two 
tangential directions, respectively. 
To filter out weak fluctuations, a thresholding strategy is applied to define the interface discontinuity 
indicator $D_f$:
\begin{equation}
    D_f = \begin{cases} 
    \epsilon A_f^2, & \text{if } A_f^2 \geq 0.5, \\
    0, & \text{otherwise},
    \end{cases}
\end{equation}
where $\epsilon$ is a tunable scaling parameter controlling the sensitivity of the limiter. 
For challenging simulations where numerical robustness is critical, this parameter can be increased to 
amplify the limiting effect, thereby effectively suppressing non-physical 
oscillations. In the present study, a fixed value of $\epsilon = 1.0$ is adopted for all computations.

Second, a preliminary local limiting factor $\phi_i$ is computed for cell $\Omega_i$ by accumulating 
the indicators from all its $N_f$ faces:
\begin{equation}
    \phi_i = \frac{1}{1 + \sum_{f=1}^{N_f} D_f}.
\end{equation}

Finally, to enhance the robustness of the reconstruction stencil, the final DFF $\alpha_i$ is determined by taking the 
harmonic mean of the preliminary factors $\phi$ of all neighboring cells sharing a face with cell $\Omega_i$:
\begin{equation}
    \alpha_i = \frac{N_f}{\sum_{k \in \mathcal{N}(i)} \frac{1}{\phi_k}},
\end{equation}
where $\mathcal{N}(i)$ represents the set of face-neighboring cells. 

Special attention is required for grid elements near solid boundaries to prevent the limiter from deteriorating the accuracy of the viscous sublayer and the gas-surface 
interaction. Specifically, the discontinuity strength parameter $A_f$ is forcibly set to zero in two scenarios.

First, for the lateral interfaces shared by two wall-adjacent cells (i.e., faces where both the left and right neighboring cells are located in the first layer of the mesh), 
$A_f = 0$ is imposed. This treatment ensures that the limiter is not triggered by flow gradients in the tangential direction, thereby preserving the high-order reconstruction 
along the wall.

Second, for the Generalized Kinetic Boundary Condition (GKBC) introduced in the following section, which inherently permits physical discontinuities such as velocity slip and 
temperature jump across the gas-solid interface, $A_f$ is also set to zero at wall boundary faces to avoid artificial damping of these physically meaningful jumps.

\subsection{Time stepping strategy}
The global time step $\Delta t$ is determined by the minimum local time step allowed by the 
Courant-Friedrichs-Lewy (CFL) condition over the entire computational domain:
\begin{equation}
    \Delta t = \min_{i} \left( \frac{\text{CFL}}{\Lambda_{i}^c + \Lambda_{i}^v} \right).
\end{equation}
Here, $\Lambda_{i}^c$ and $\Lambda_{i}^v$ represent the convective and viscous spectral radius for cell $i$, 
respectively, defined as:
\begin{equation}
    \Lambda_{i}^c = \frac{|\mathbf{U}| + c}{\Delta l}, \quad
    \Lambda_{i}^v = \frac{\mathcal{C}}{\rho} \left( \frac{\mu}{Pr} \right) \frac{1}{\Delta l^2},
\end{equation}
where $\Delta l$ is the characteristic length scale, $c$ is the speed of sound, and the coefficient 
$\mathcal{C} = \max(4/3, \gamma)$ accounts for the viscous stability limit.

However, this explicit time step is often overly restrictive for high-speed viscous flows. To accelerate 
the convergence towards steady state and effectively handle the stiffness arising from the vibrational 
relaxation source terms, an implicit time integration scheme based on the Lower-Upper Symmetric 
Gauss-Seidel (LU-SGS) method is employed.
The first-order backward Euler discretization of the governing equations can be written as:
\begin{equation}
    \left[ \frac{\Omega_i}{\Delta t} \mathbf{I} + \frac{\partial \mathbf{R}(\mathbf{W})}{\partial \mathbf{W}} 
	\right] \Delta \mathbf{W}_i = \mathbf{R}_i^n,
\end{equation}

where $\Delta \mathbf{W}_i = \mathbf{W}_i^{n+1} - \mathbf{W}_i^n$ is the solution update, $\Omega_i$ is the cell 
volume, and $\mathbf{R}_i^n$ is the residual vector at time step $n$ including both net flux and vibrational source 
terms.

Directly inverting the large sparse matrix on the left-hand side is computationally expensive. To avoid the explicit 
storage of the Jacobian matrix $\partial \mathbf{R}/\partial \mathbf{W}$, the Jacobian of the numerical flux is 
approximated using the Lax-Friedrichs (L-F) splitting method. 
It is important to note that in the GKS framework, the numerical flux naturally couples inviscid and viscous 
contributions. In the implicit construction, this total flux Jacobian across a face $f$ is split into positive 
and negative contributions:
\begin{equation}
    \mathbf{A}^{\pm} = \frac{1}{2} \left( \mathbf{A} \pm \lambda_{f} \mathbf{I} \right),
\end{equation}
where $\lambda_{f}$ is the spectral radius at the interface.
Consequently, the system is solved via forward and backward sweeps. The update equation for cell $i$ is formulated as:
\begin{equation}
    \mathbf{D}_i \Delta \mathbf{W}_i = \mathbf{R}_i^n - \sum_{k \in \mathcal{N}(i)} 
	\delta \mathbf{F}^{\sigma}_{ik} S_{ik},
\end{equation}
where $S_{ik}$ is the face area, and $\mathcal{N}(i)$ denotes the set of neighboring cells. 
The term $\delta \mathbf{F}^{\sigma}_{ik}$ on the right-hand side represents the implicit variation of the numerical 
flux from the neighbor $k$.
In the present matrix-free implementation, the term $\mathbf{A}^{\pm} \Delta \mathbf{W}_k$ is evaluated directly 
as the increment of the L-F split fluxes:
\begin{equation}
    \delta \mathbf{F}^{\pm} \approx \mathbf{F}^{\pm}(\mathbf{W}_k + \Delta \mathbf{W}_k) 
	- \mathbf{F}^{\pm}(\mathbf{W}_k),
\end{equation}
with the split fluxes defined as $\mathbf{F}^{\pm}(\mathbf{W}) = \frac{1}{2} (\mathbf{F}(\mathbf{W}) 
\cdot \mathbf{n} \pm \lambda \mathbf{W})$. This formulation incorporates the contribution of the total flux 
Jacobian and the spectral radius term without explicit Jacobian construction.

The spectral radius $\lambda_{f}$ used in the splitting is augmented with a viscous correction to enhance robustness:
\begin{equation}
    \lambda_{f} = \left| \mathbf{U} \cdot \mathbf{n} \right| + c + \frac{10 \mu}{\min(\rho_L, \rho_R) \Delta l},
\end{equation}
where $c$ is the speed of sound, $\Delta l$ is the characteristic length scale, and the factor 10 is empirically 
chosen to ensure stability.

The diagonal matrix $\mathbf{D}_i$ on the left-hand side lumps the time term, the spectral radius of all faces, 
and the implicit treatment of the stiff source terms:
\begin{equation}
    \mathbf{D}_i = \left( \frac{\Omega_i}{\Delta t} + \frac{1}{2} \sum_{k \in \mathcal{N}(i)} \lambda_{ik} 
	S_{ik} \right) \mathbf{I} - \Omega_i \frac{\partial \mathbf{S}}{\partial \mathbf{W}},
\end{equation}
where $\mathbf{S}$ is the vector of source terms. In the current two-temperature model, the Jacobian 
$\partial \mathbf{S} / \partial \mathbf{W}$ is non-zero only for the vibrational energy equation, ensuring 
the stability of the stiff relaxation processes.
The implicit time step is determined by scaling the explicit limit according to the CFL ratio: 
\begin{equation}
\Delta t = \frac{\text{CFL}_{\text{imp}}}{\text{CFL}_{\text{expl}}} \Delta t_{\text{expl}}.
\end{equation}
Although theoretically unconditionally stable, the implicit CFL number ($\text{CFL}_{\text{imp}}$) is practically restricted to $10 \sim 20$ in 
the present study. This limitation is necessary to address the strong non-linearity of the Landau-Teller relaxation and the reduced coupling accuracy inherent in the 
matrix-free diagonal approximation. A moderate CFL ensures numerical stability by preventing unphysical temperature excursions during strong shock interactions.

To enable parallel execution of the Gauss-Seidel sweeps on shared-memory architectures, a graph coloring strategy 
is adopted. Cells are grouped into independent color sets such that no two cells of the same color share a common 
interface, allowing concurrent updates within each color group without data race.

\subsection{Generalized Kinetic Wall Boundary Condition (GKBC)}

In the continuum regime, the no-slip boundary condition is traditionally employed for viscous flows. However, in the near-continuum and slip flow regimes, particularly for 
high-speed non-equilibrium flows, the gas-surface interaction becomes significant. Unlike macroscopic solvers that rely on gradient-based slip corrections, the GKS operates 
directly on the microscopic particle distribution function, providing a natural and superior framework to capture velocity slip and temperature jump phenomena from a bottom-up 
perspective.

Despite this advantage, the standard Maxwell boundary condition widely adopted in GKS simulations typically assumes that the diffusely reflected particles are fully accommodated 
to the wall temperature. This inherently couples the momentum and thermal accommodation coefficients (i.e., $\sigma = \alpha$), which contradicts experimental observations where 
momentum and energy accommodations often exhibit distinct behaviors. 
Although some kinetic boundary models (e.g. Ref.~\cite{Li2005}) attempt to address this issue, they lack the flexibility to independently prescribe the tangential momentum and 
thermal accommodation coefficients.

To bridge this gap, this study proposes a generalized kinetic boundary condition (GKBC) specifically designed for thermal non-equilibrium flows. 
This model innovatively combines the Maxwell scattering kernel with a decoupling mechanism that allows the thermal accommodation to be determined independently of the momentum 
accommodation $\sigma$. 
Furthermore, consistent with the two-temperature physical model employed in this study, the energy accommodation is explicitly split into two distinct parts: 
$\alpha_{tr}$ for the translational-rotational mode and $\alpha_{v}$ for the vibrational mode.

A local coordinate system is defined at the cell interface where the $x$-axis is aligned with the unit normal vector pointing into the fluid. The time-dependent gas 
distribution function for incident particles (hitting the wall, $u < 0$) is denoted as $f_{in}$. Based on the Chapman-Enskog expansion and the interior domain 
reconstruction, $f_{in}$ at the wall interface is expressed as:
\begin{equation}
    f_{in}(t, u, \xi) = g_0 \left[ 1 - \tau(\bar{a}u + \bar{A}) + t\bar{A} \right],
\end{equation}
where $g_0$ is the equilibrium state at the beginning of the time step, and $\bar{a}, \bar{A}$ are the microscopic slopes related to spatial and temporal derivatives.

The distribution function for reflected particles (leaving the wall, $u > 0$), denoted as $f_{ref}$, is modeled as a linear combination of specular reflection and a modified 
diffusive reflection:
\begin{equation}
    f_{ref}(t, u, \mathbf{\xi}) = (1 - \sigma) f_{s}(t, u, \mathbf{\xi}) + \sigma g_{d}(u, \mathbf{\xi}),
\end{equation}
where $\sigma \in [0, 1]$ is the tangential momentum accommodation coefficient. 

The specular reflection component $f_{s}$ describes particles bouncing elastically off the wall. In the local normal coordinate system, it relates to the incident function by 
reversing the normal velocity:
\begin{equation}
    f_{s}(t, u, \mathbf{\xi}) = f_{in}(t, -u, \mathbf{\xi}).
\end{equation}

Crucially, unlike the standard Maxwell model where the diffusive part assumes the wall temperature $T_w$, the diffusive reflection component $g_{d}$ here is assumed to follow 
a Maxwellian distribution defined by \textit{reflected temperatures} ($T_{d,tr}, T_{d,v}$) that are distinct from the wall temperature:
\begin{equation}
	\begin{aligned}
    g_{d} &= \rho_{d} \left(\frac{\lambda_{d,tr}}{\pi}\right)^{\frac{K_r+3}{2}} 
	 e^{-\lambda_{d,tr} 
	[(u-U_w)^2 + \xi_{r}^2]} \\
	&\quad \times \left(\frac{\lambda_{d,v}}{\pi}\right)^{\frac{K_v}{2}}
	 e^{- \lambda_{d,v} \xi_{v}^2} ,
	\end{aligned}
\end{equation}
where $U_w$ is the wall normal velocity (usually zero). Consequently, the complete time-dependent gas distribution function at the wall interface, $f_{wall}$, is constructed 
by the superposition:
\begin{equation}
    f_{wall}(t, u, \mathbf{\xi}) = f_{in}(t, u, \mathbf{\xi})\mathbb{H}(-u) + f_{ref}(t, u, \mathbf{\xi})
	\mathbb{H}(u).
\end{equation}

The unknown reflected temperatures $T_{d,tr}$ and $T_{d,v}$ are the key to decoupling the energy accommodation from the momentum accommodation, and they 
are solved using the definition of thermal accommodation coefficients.

\subsubsection{Thermal Accommodation and Decoupling}
In high-speed flows, the energy exchange efficiency varies across different internal modes. Consistent with the two-temperature model, distinct thermal accommodation 
coefficients are introduced: $\alpha_{tr}$ for the translational-rotational mode and $\alpha_{v}$ for the vibrational mode. 

The reflected state is determined by interpolating the specific internal energies based on the definition of these coefficients, effectively decoupling them from $\sigma$.
For the vibrational mode, $\alpha_v$ is defined as:
\begin{equation}
    \alpha_v = \frac{e_{v,in} - e_{v,d}}{e_{v,in} - e_{v,w}},
\end{equation}
where $e_{v,w} = \frac{K_v}{2} R T_{v,w}$ is the specific vibrational energy at the wall, and $e_{v,in}$ is calculated from the incident flux. From this, the specific energy of 
diffusely reflected particles $e_{v,d}$ and the corresponding $T_{d,v}$ are obtained.

Similarly, for the translational-rotational mode, $\alpha_{tr}$ is defined as:
\begin{equation}
    \alpha_{tr} = \frac{e_{tr,in} - e_{tr,d}}{e_{tr,in} - e_{tr,w}},
\end{equation}
where $e_{tr,w} = \frac{K_r+3}{2} R T_{tr,w}$. 
It is important to note that $e_{tr,in} = e_{in}-e_{v,in}$, assuming the kinetic energy of the incident flow is primarily exchanged with the translational-rotational mode of 
the reflected particles. This formulation allows for the precise determination of $T_{d,tr}$ independent of the momentum accommodation $\sigma$.

\subsubsection{Mass Balance}
The no-penetration condition imposes a constraint on the net mass flux. Since $f_s$ preserves mass flux, the condition simplifies to balancing the diffusive part against the 
incident flux:
\begin{equation}
    \int_0^{\Delta t} \int_{u > 0} u g_{d} \mathrm{d}\Xi \mathrm{d}t = - F_{in}^{\rho},
\end{equation}
where $F_{in}^{\rho}$ is the total mass flux of incident particles:
\begin{equation}
    F_{in}^{\rho} = \int_0^{\Delta t} \int_{u < 0} u f_{in}(t, u, \mathbf{\xi}) 
	\mathrm{d}\Xi \mathrm{d}t.
\end{equation}
The reflected density $\rho_{d}$ is thus uniquely determined by:
\begin{equation}
    \rho_{d} = \frac{-F_{in}^{\rho}}{\Delta t} \cdot 2\sqrt{\pi \lambda_{d, tr}}.
\end{equation}
With $\rho_{d}, T_{d, tr},$ and $T_{d, v}$ uniquely determined by $\sigma, \alpha_{tr},$ and $\alpha_v$, the boundary distribution function is fully defined, providing a robust 
boundary treatment for non-equilibrium flow simulations.

In the subsequent numerical simulations, the accommodation coefficients are prescribed as $\sigma = 0.85$, $\alpha_{tr} = 0.85$, and $\alpha_{v} = 0.001$. These values are 
identical to those used in the reference CFD studies~\cite{Candler2003} to ensure a consistent comparison. The choice of such a low value for $\alpha_{v}$, as adopted in the 
reference, reflects the physical consideration of the much slower relaxation process and inefficient energy exchange associated with the vibrational mode at the surface.

\subsection{Evaluation of Surface Quantities}
Once the time-dependent gas distribution function at the wall boundary, $f_{wall}$, is determined, 
the macroscopic aerothermodynamic quantities on the surface can be evaluated directly from the fluxes 
crossing the interface. 
A distinct advantage of the GKS framework is that it calculates surface quantities based on the moments 
of the particle distribution function, without assuming linear constitutive relationships 
(such as Newton's law of viscosity or Fourier's law of heat conduction). 
Consequently, GKS is capable of capturing non-equilibrium effects in the Knudsen layer more accurately 
than traditional macroscopic solvers, which rely on first-order derivatives of macroscopic variables.

In the local coordinate system defined at the cell interface, where the $x$-axis aligns with the wall 
normal direction $\mathbf{n}$ and the $y, z$-axes lie in the tangential plane, the numerical flux vector 
through the wall is computed as:
\begin{equation}
    \mathbf{F}_{wall} = \int u \boldsymbol{\psi} f_{wall}(t, u, \mathbf{\xi}) \mathrm{d}\Xi,
\end{equation}
where $\boldsymbol{\psi} = (1, u, v, w, \frac{1}{2}(u^2+v^2+w^2+\xi^2), \frac{1}{2}\xi_v^2)^T$.
Due to the no-penetration condition, the net convective mass flux ($F_{\rho}$) is zero. The remaining flux 
components directly correspond to the surface dynamic and thermodynamic quantities:

\begin{enumerate}
\item \textbf{Wall Pressure / Normal Stress:}
    The flux of normal momentum ($F_{\rho u}$), corresponding to the 2nd component of the flux vector, 
	represents the rate of normal momentum transfer to the wall. 
    Strictly speaking, this term corresponds to the \textbf{total normal stress} exerted by the fluid 
	on the surface, which includes both the thermodynamic pressure $p$ and the viscous normal stress 
	component $\tau_{xx}$:
    \begin{equation}
        F_{\rho u} = \int u^2 f_{wall} \mathrm{d}\Xi = p - \tau_{xx}.
    \end{equation}
    In the context of aerodynamic force evaluation, this total normal stress is the effective pressure 
	acting on the body surface used for calculating the pressure coefficient ($C_p$) and lift/drag forces.

    \item \textbf{Wall Shear Stress ($\boldsymbol{\tau}_w$):}
    The fluxes of tangential momenta ($F_{\rho v}$ and $F_{\rho w}$) correspond to the 3rd and 4th 
	components. These represent the tangential forces (shear stresses) exerted by the fluid on the wall 
	in the local $y$ and $z$ directions, respectively:
    \begin{equation}
		\begin{aligned}
        F_{\rho v} &= \int uv f_{wall} \mathrm{d}\Xi=\tau_{xy}, \\
        F_{\rho w} &= \int uw f_{wall} \mathrm{d}\Xi=\tau_{xz} .
		\end{aligned}
    \end{equation}
    The total wall shear stress magnitude is $\tau_w = \sqrt{\tau_{xy}^2 + \tau_{xz}^2}$.

\item \textbf{Total Energy Flux / Wall Heat Flux:}
    The total energy flux ($F_{\rho E}$), corresponding to the 5th component of the flux vector, 
	represents the total rate of energy transfer per unit area from the fluid to the wall.
    It is crucial to note that in the slip flow regime, the fluid velocity at the wall is non-zero 
	(velocity slip). Consequently, the work done by the shear stress (friction work) contributes to 
	the total energy exchange.
    Mathematically, the total energy flux calculated by the GKS moment integral can be decomposed into 
	the conductive heat flux $q_w$ and the rate of work done by the tangential shear stresses:
    \begin{equation}
		\begin{aligned}
        F_{\rho E} &= \int \frac{1}{2}u (u^2+v^2+w^2+\xi^2) f_{wall} \mathrm{d}\Xi \\
		&= q_w + V \tau_{xy} + W \tau_{xz},
		\end{aligned}
    \end{equation}
    where $V$ and $W$ denote the tangential slip velocity components at the wall surface, and $\tau_{xy}, 
	\tau_{xz}$ are the wall shear stresses. 
    Therefore, $F_{\rho E}$ captures the total aerodynamic heating load acting on the surface.
    Similarly, the vibrational heat flux is given directly by the 6th component, $F_{\rho E_v}$, which 
	represents the diffusive transport of vibrational energy.
\end{enumerate}

\section{Results and Discussion}

This section presents a comprehensive numerical investigation of hypersonic non-equilibrium flows involving complex shock-boundary layer interactions. The study is 
organized into four subsections to logically progress from baseline code validation to the assessment of flow sensitivities under varying conditions.

To establish the fundamental reliability of the present three-dimensional two-temperature gas-kinetic scheme, the first two subsections focus on validation against canonical 
benchmark experiments conducted in the CUBRC LENS hypervelocity shock tunnels~\cite{Holden2003, Holden2004}. 
Specifically, the sharp double cone (Run 35) and the hollow cylinder-flare (Run 11) configurations are selected, which serve as standard test cases for high-enthalpy flows.

Following the baseline validation, the subsequent subsections further evaluate the solver's accuracy in capturing the flow response to specific parameter variations. By 
comparing with the corresponding experimental measurements, the investigation examines the influence of rarefaction effects (via varying free-stream densities) and 
three-dimensional effects (via a non-zero angle of attack) on the flow structure and aerothermal loads.

\subsection{Validation Case I: Sharp Double Cone}

\subsubsection{Flow Conditions and Geometry}
The computational model replicates the experimental configuration tested in the LENS hypervelocity wind tunnels, specifically corresponding to the Run 35 test condition. 
The geometry consists of a first cone with a half-angle of $25^\circ$ and a second cone with a half-angle of $55^\circ$. The coordinate system is defined with the origin 
at the nose tip, the $x$-axis aligned with the symmetry axis, and the $y$-axis in the radial direction.

The detailed free-stream conditions are listed in Table~\ref{tab:all_conditions} (referred to as Run 35). A distinct feature of this 
case is the significant thermal non-equilibrium state of the incoming flow, where the vibrational temperature ($T_{v,\infty} = 2562$ K) is more than an order of magnitude 
higher than the translational temperature ($T_{\infty} = 98.27$ K). This phenomenon results from the rapid expansion in the wind tunnel nozzle, causing the vibrational energy 
mode to freeze.

\begin{table*}[hbt!]
	\centering
	\caption{Summary of free-stream conditions for all simulation cases. The test gas is Nitrogen ($N_2$).}
	\label{tab:all_conditions}
	\setlength{\tabcolsep}{5pt} % 调整列间距
	\renewcommand{\arraystretch}{1.2} % 稍微增加行高，避免单位和横线太挤
	\begin{tabular}{llccccccc}
		\hline
		\hline
		\multirow{2}{*}{Case} & \multirow{2}{*}{Description} & $Ma_{\infty}$ & $U_{\infty}$ & $\rho_{\infty}$ & $p_{\infty}$ & $T_{\infty}$ & $T_{v,\infty}$ & $T_{w}$ \\
		% 单位单独一行
		& & (-) & (m/s) & ($\times 10^{-4}$ kg/m$^3$) & (Pa) & (K) & (K) & (K) \\
		\hline
		Run 35 & Baseline Double Cone & $12.59$ & $2545$ & $5.848$ & $17.06$ & $98.3$ & $2562$ & $296.1$ \\
		Run 11 & Hollow Cylinder-Flare & $12.27$ & $2485$ & $5.866$ & $17.18$ & $98.7$ & $2497$ & $293.0$ \\
		Run 6 & $0.5\rho$ Double Cone & $15.57$ & $2139$ & $3.670$ & $4.95$ & $45.4$ & $1894$ & $296.1$ \\
		Run 7 & $0.33\rho$ Double Cone & $15.56$ & $2092$ & $1.730$ & $2.23$ & $43.5$ & $1810$ & $296.1$ \\
		AoA & $2^\circ$ Angle of Attack & $12.58$ & $2502$ & $6.190$ & $17.47$ & $95.1$ & $2498$ & $296.1$ \\
		\hline
		\hline
	\end{tabular}
\end{table*}

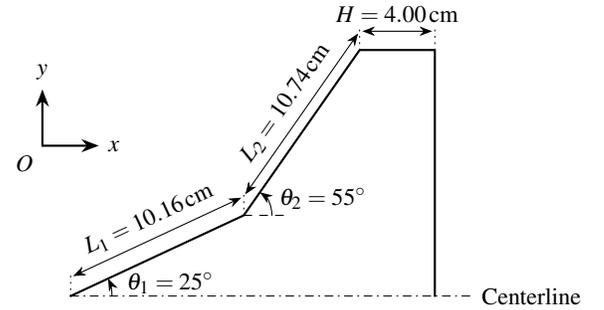
\begin{figure}[htp]
	\centering
\begin{tikzpicture}[
    % 尺度依然适配双栏，但字体稍大一些
    scale=0.25,
    font=\small, % 稍大的字体
    line join=round, line cap=round,
    >=Stealth,
    % 标注线样式：直接连到文字
    dim/.style={thin, <->, >=Stealth, shorten <=1pt, shorten >=1pt} 
]

% =====================
% 参数定义 (cm)
% =====================
\def\Lone{10.16}
\def\Ltwo{10.74}
\def\Hlen{4.00}
\def\thetaone{25}
\def\thetatwo{55}

% =====================
% 坐标计算
% =====================
\coordinate (O) at (0,0);
% 上半部分
\coordinate (A) at (\thetaone:\Lone);
\coordinate (C) at ($(A) + (\thetatwo:\Ltwo)$);
\coordinate (E) at ($(C) + (\Hlen,0)$);
% 下半部分
\coordinate (F) at (E |- O);

% =====================
% 1. 几何轮廓 (黑线白底)
% =====================
\draw[thick] (O) -- (A) -- (C) -- (E) -- (F);

% =====================
% 2. 中心线 (点划线)
% =====================
\draw[dash dot, thin] (O) -- ($(E|-O)+(2,0)$) node[right] {Centerline};

% =====================
% 3. 尺寸标注 (直接在标注线上写，不加辅助线)
% =====================
% L1
\draw[dim] ($(O)+(0,1.0)$) -- ($(A)+(0,1.0)$) 
    node[midway, above, sloped] {$L_1=10.16\,\mathrm{cm}$};
% L1 辅助线
\draw[thin, dotted] (O) -- ($(O)+(0,1.3)$);
\draw[thin, dotted] (A) -- ($(A)+(0,1.3)$);
% L2
\draw[dim] ($(A)+(0,1.0)$) -- ($(C)+(0,1.0)$) 
    node[midway, above, sloped] {$L_2=10.74\,\mathrm{cm}$};
% L2 辅助线
\draw[thin, dotted] (C) -- ($(C)+(0.0,1.3)$);
% H
\draw[dim] ($(C)+(0,1.0)$) -- ($(E)+(0,1.0)$) 
    node[midway, above] {$H=4.00\,\mathrm{cm}$};
% H 辅助线
\draw[thin, dotted] (E) -- ($(E)+(0,1.3)$);
	
% =====================
% 4. 角度标注
% =====================
% Theta 1
\draw[thin, ->] (2.2,0) arc[start angle=0, end angle=\thetaone, radius=2.2];
\node at (5.3, 0.7) [] {$\theta_1=25^\circ$};

% Theta 2
% 这里的辅助线可以保留，它是指明角度的起点
\draw[thin, dashed] (A) -- ++(2.2,0); 
\draw[thin, ->] ($(A)+(1.5,0)$) arc[start angle=0, end angle=\thetatwo, radius=1.5];
\node at ($(A)+(4.2, 0.9)$) [] {$\theta_2=55^\circ$};

% =====================
% 5. 坐标轴 (左下角，单箭头)
% =====================
\coordinate (AxisOrigin) at (-1.5, 8); % 位置微调
\draw[->, thick] (AxisOrigin) -- ++(3,0) node[right] {$x$};
\draw[->, thick] (AxisOrigin) -- ++(0,3) node[above] {$y$};
\node[below left] at (AxisOrigin) {$O$};
\end{tikzpicture}
	\caption{Schematic of the double cone geometry.}
	\label{fig:geo_double_cone}
\end{figure}

\subsubsection{Computational Mesh and Grid Independence Study}
\begin{figure}[htp]
	\centering
	% 第一个子图：拐角处
	\subfloat[Cone-cone junction]{%
		\includegraphics[width=0.45\textwidth]{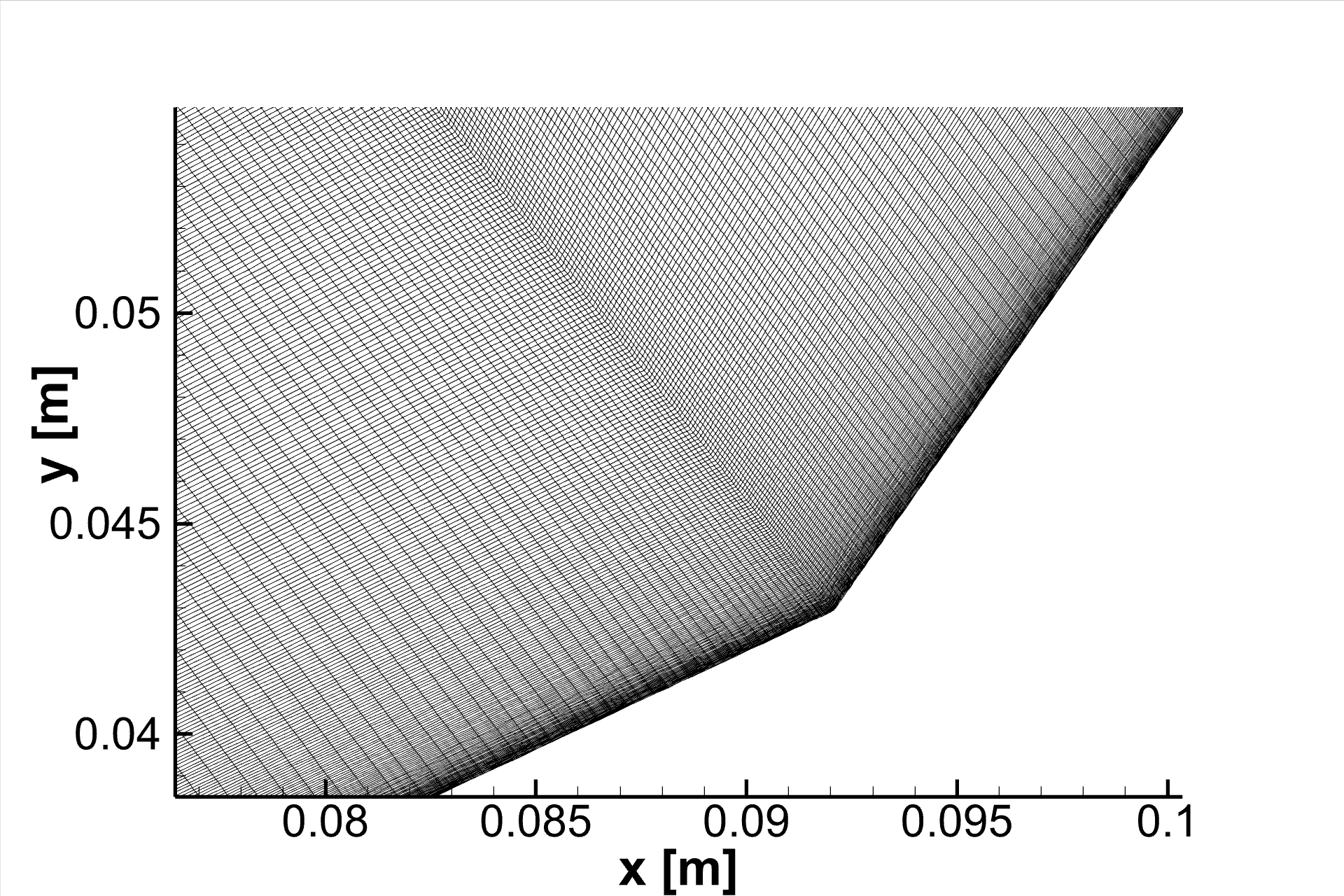} % 请替换为拐角网格图文件名
		\label{fig:mesh_junction}
	}
	\hfill % 换行或者并排
	% 第二个子图：再附点处
	\subfloat[Reattachment region]{%
		\includegraphics[width=0.45\textwidth]{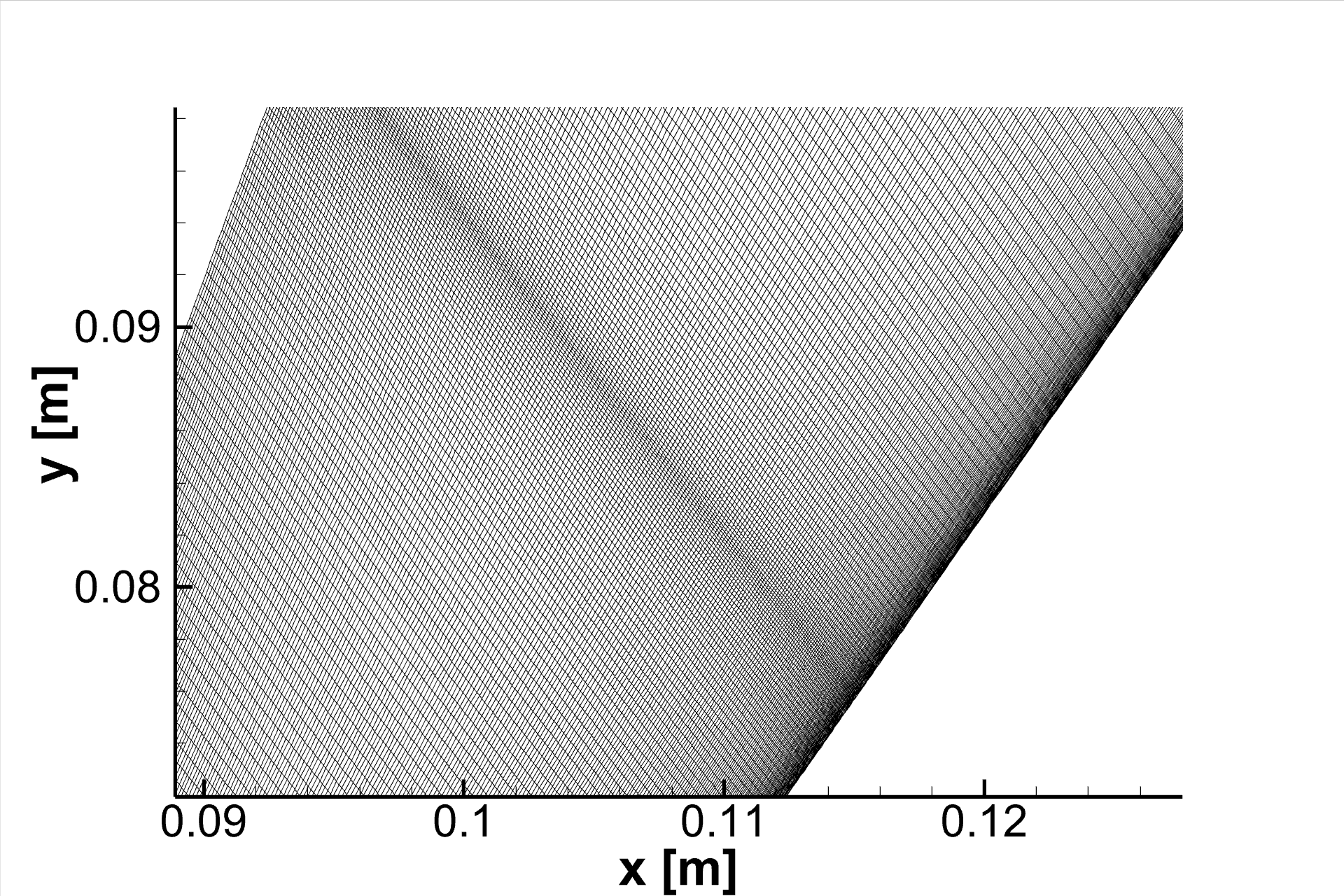} % 请替换为再附点网格图文件名
		\label{fig:mesh_reattachment}
	}
	\caption{Close-up views of the medium grid ($500 \times 250$) distribution: (a) near the cone-cone junction and (b) near the flow reattachment region.}
	\label{fig:mesh_details}
\end{figure}

To ensure numerical independence from spatial discretization, a grid convergence study is conducted using three resolution levels: coarse, medium, and fine. 
The computational grids consist of approximately $300 \times 125$, $500 \times 250$, and $1000 \times 500$ cells in the streamwise and wall-normal directions, 
respectively. 
To accurately resolve the complex shock-boundary layer interaction, the computational mesh is specifically refined in significant regions. 
In the wall-normal direction, the grid is clustered near the solid surface with a first cell height of $1 \times 10^{-5}$ m. This spacing maintains a cell Reynolds number of 
order unity, which is critical for accurately resolving the heat flux within the viscous boundary layer. 
In the streamwise direction, the grid density is increased near the cone-cone junction and the flow reattachment zone to capture the separation bubble and shock impingement. 
The detailed grid topology in these two critical regions is illustrated in Figure~\ref{fig:mesh_details}.

\begin{figure*}[htp]
	\centering
	\subfloat[Surface pressure]{%
		\includegraphics[width=0.45\textwidth]{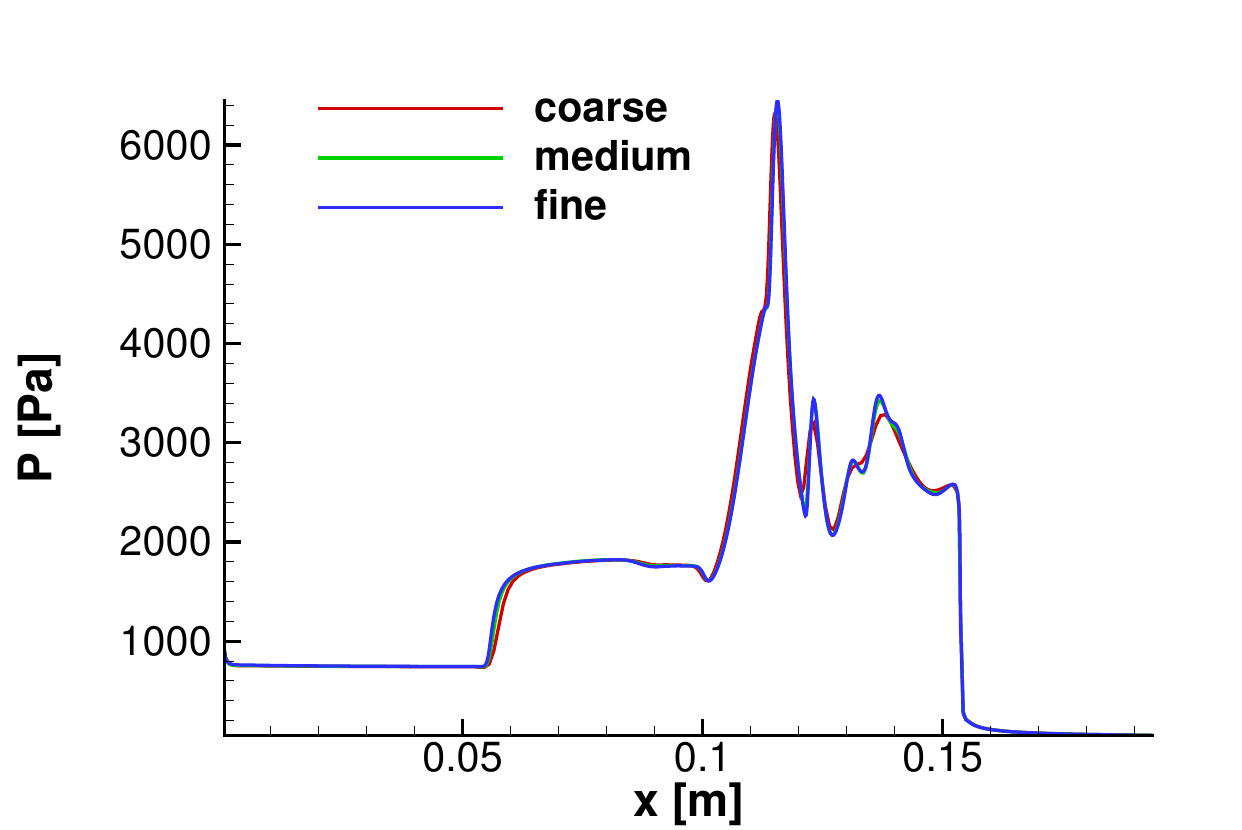}%
	}%
	\hspace{0.05\textwidth}
	\subfloat[Wall shear stress]{%
		\includegraphics[width=0.45\textwidth]{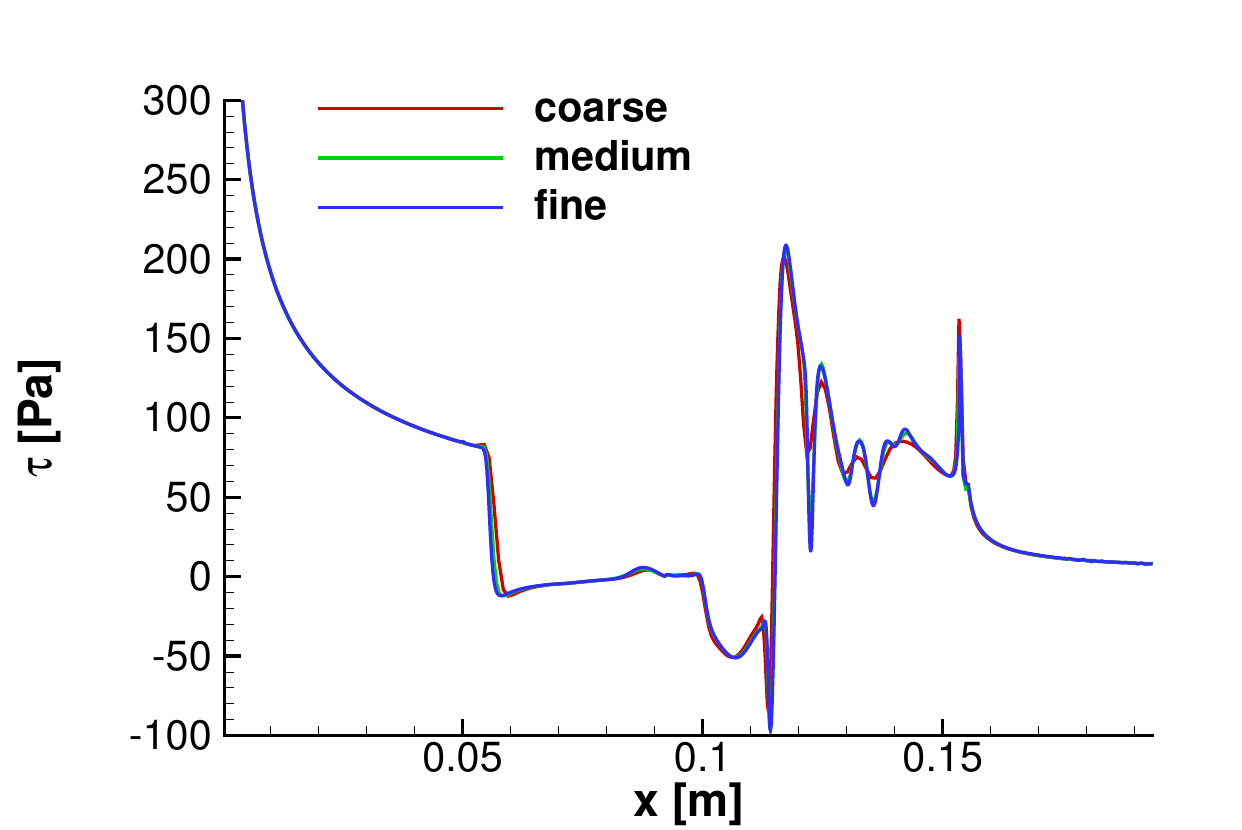}%
	}%
	
	\subfloat[Surface heat flux]{%
		\includegraphics[width=0.45\textwidth]{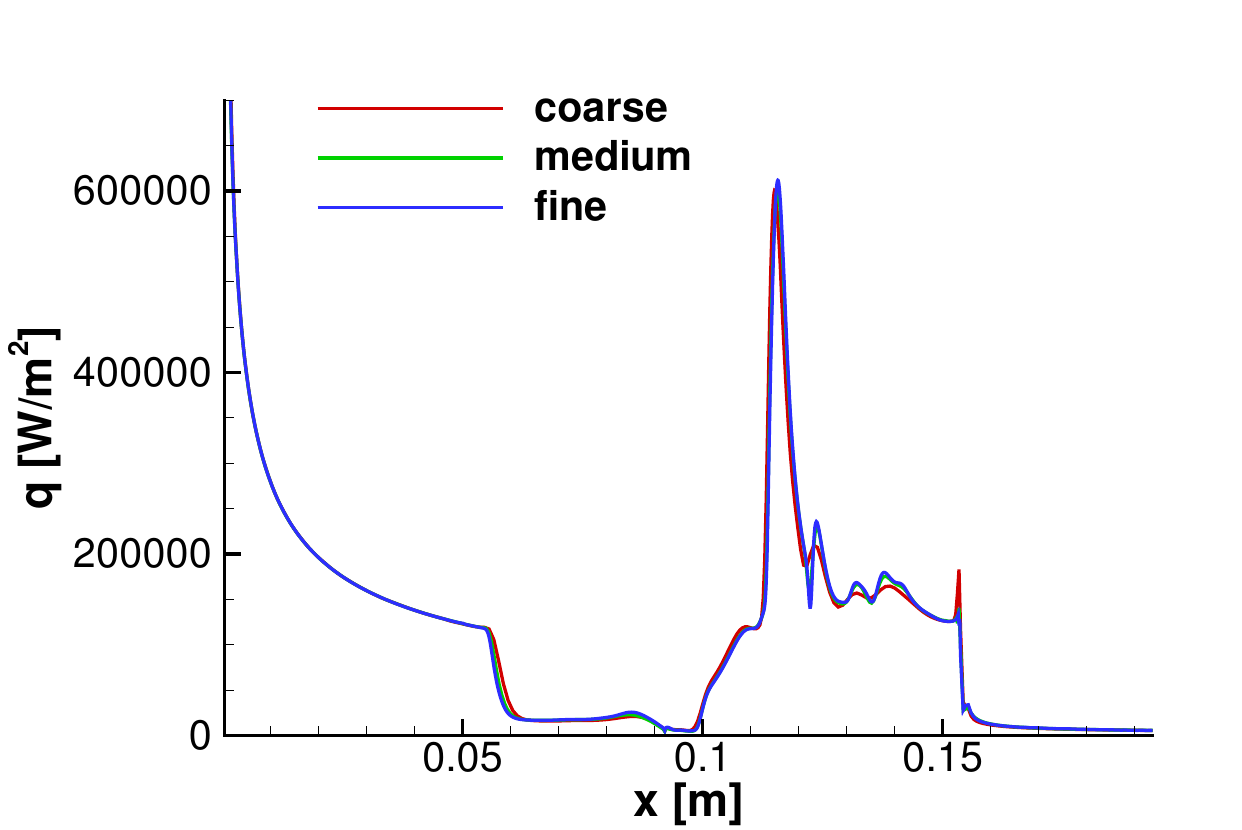}%
	}%
	\caption{Grid convergence results on three resolution levels: (a) surface pressure, (b) wall shear stress, and (c) surface heat flux distributions.}
	\label{grid_indep}
\end{figure*}

Figure~\ref{grid_indep} presents the quantitative comparison of surface flow properties along the cone surface obtained from the three grid levels. Panels 
(a), (b), and (c) correspond to the distributions of surface pressure ($p_w$), wall shear stress ($\tau_w$), and surface heat flux ($q_w$), respectively. As observed, 
the results from the coarse grid exhibit slight deviations, particularly in the extent of the separation region and the location of peak heating. However, the profiles 
obtained from the medium and fine grids are virtually indistinguishable, demonstrating excellent consistency in predicting the separation shock impingement and reattachment 
peak values. Consequently, the medium grid resolution is deemed sufficient for the subsequent analysis, striking an optimal balance between computational efficiency and 
numerical accuracy.

\subsubsection{Flow Structure}

\begin{figure}[htp]  
	\centering  
	\includegraphics[width=0.5\textwidth]{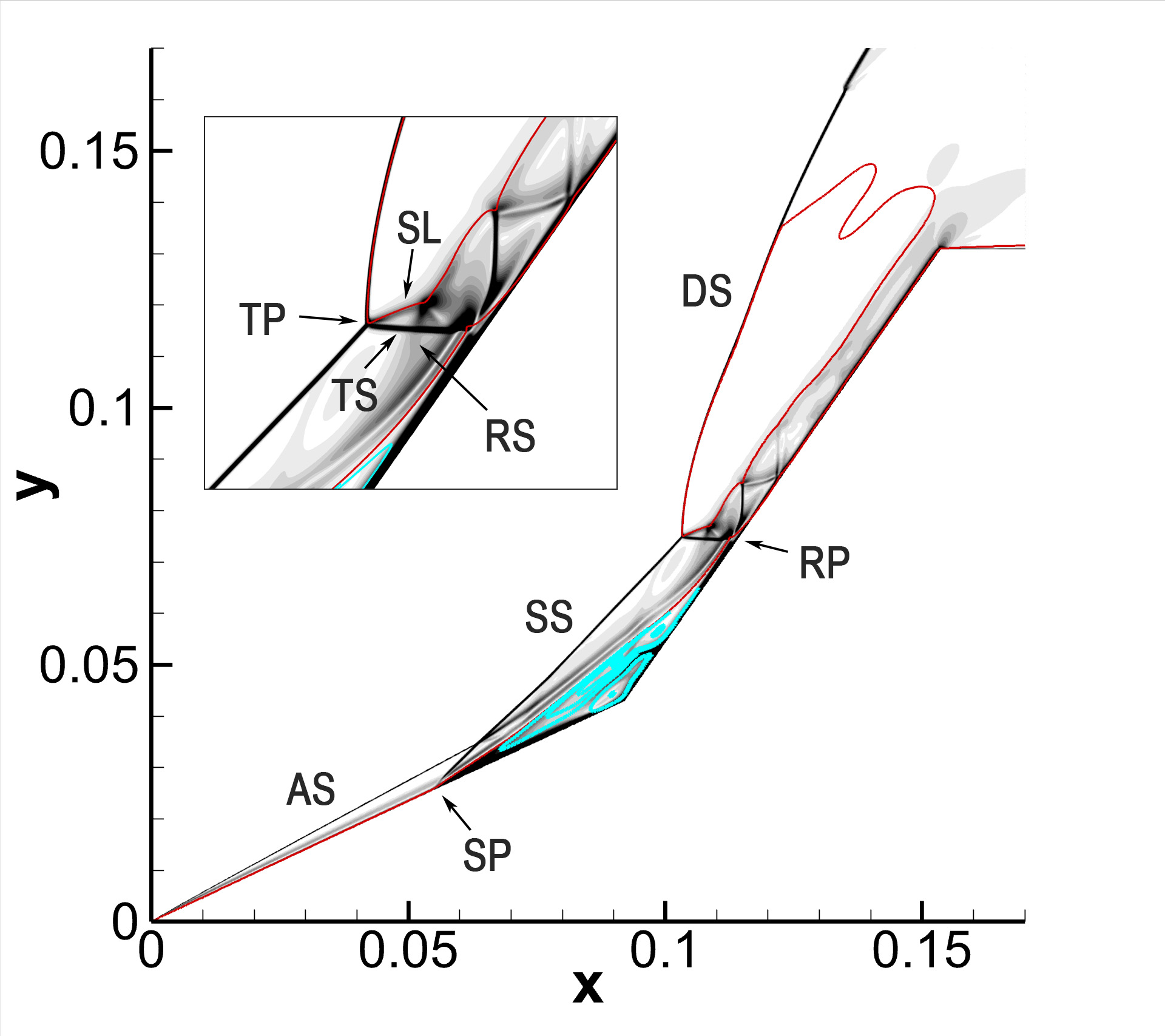}  
	\vspace{-4mm}  
	\caption{Flow structure over the sharp double-cone configuration computed using the fine grid. The cyan region indicates the separation vortex, and the red solid line in 
	the zoomed-in view marks the sonic line.}  
	\label{schematic_cone}  
\end{figure} 

The flow field exhibits a hierarchical shock interaction pattern involving multiple interference types. Figure~\ref{schematic_cone} illustrates the flow structure obtained 
from the fine grid computation. Initially, an attached oblique shock (AS) originates from the tip of the first cone. Due to the adverse pressure gradient imposed by the steeper 
second cone, the laminar boundary layer on the first cone separates, forming a recirculation region highlighted in cyan. This separation generates a separation shock (SS).

As the flow develops downstream, the separation shock (SS) intersects with the initial attached shock (AS). This interaction between two weak shocks of the same family is 
classified as Edney Type VI interference, resulting in a merged, stronger incident shock layer. Subsequently, this merged shock structure intersects with the strong detached 
bow shock (DS) of the second cone at a triple point (TP). This primary interaction is identified as a classic Edney Type IV interference.

From the triple point, a transmitted shock (TS) emanates downwards, and a slip line (SL) is generated, separating the supersonic jet from the subsonic region (indicated by 
the red sonic line). Concurrently, the reattachment of the separated shear layer on the second cone surface generates an oblique reattachment shock (RS). The reattachment 
shock propagates upward and intersects with the incoming transmitted shock (TS). This event constitutes a regular intersection of shocks of opposite families. 

This interaction bifurcates into two resulting transmitted shocks. The upper branch impinges upon the slip line, causing a deflection of the shear layer. The lower branch 
impinges directly onto the surface of the second cone. It is this specific shock impingement that leads to the severe peak localized heat flux and pressure loads observed 
in the surface distributions. Capturing the size of the separation bubble and the precise topology of these sequential shock interactions is highly sensitive to the numerical 
dissipation, the modeling of vibrational relaxation, and the implementation of wall boundary conditions.

\begin{figure}[htp]  
	\centering  
	\includegraphics[width=0.45\textwidth]{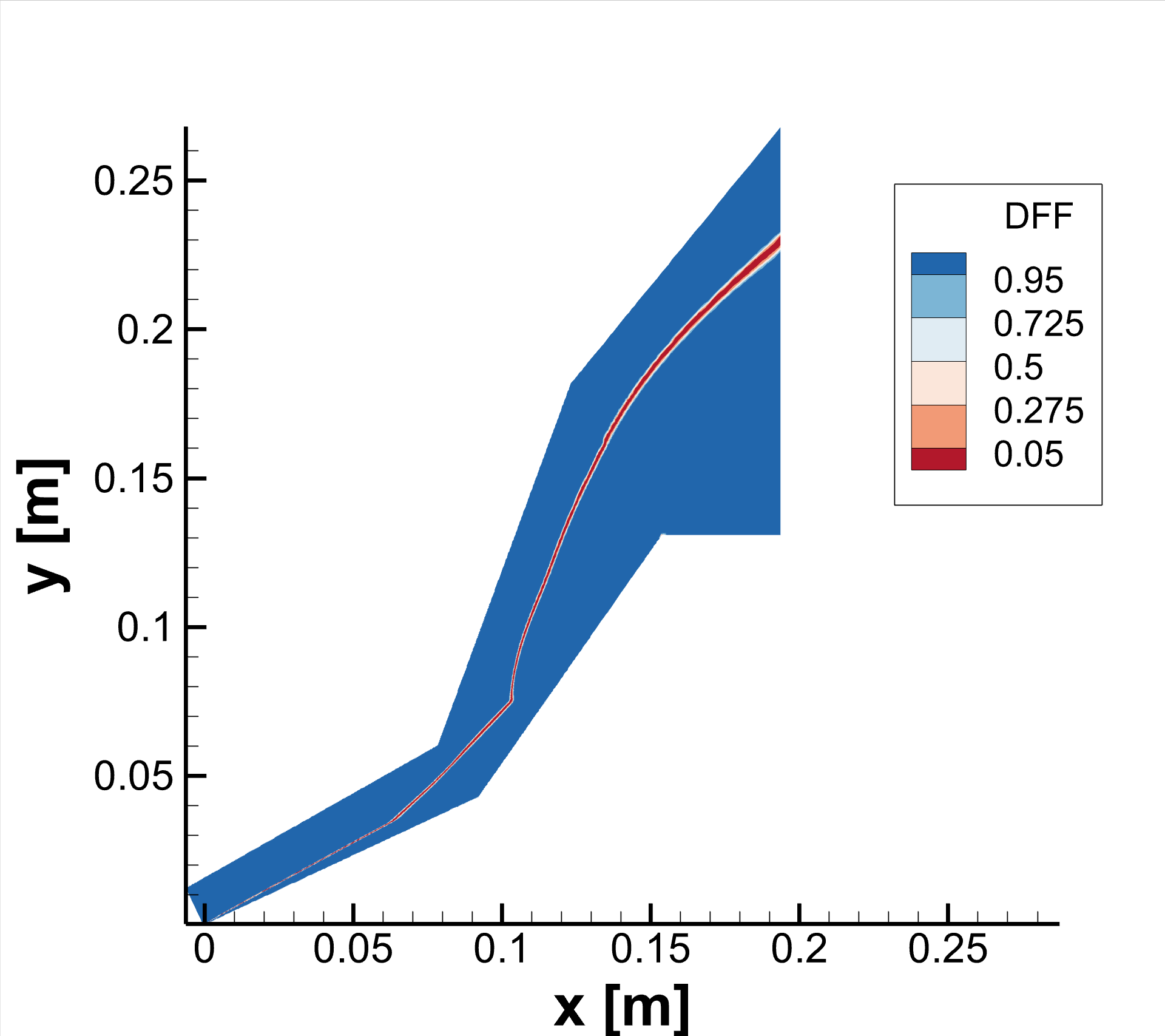}  
	\vspace{-4mm}  
	\caption{Contours of the Discontinuity Feedback Factor (DFF) for Run35.}  
	\label{fig:dff_cone}  
\end{figure}

To further verify the numerical fidelity in capturing these steep gradients without compromising viscous accuracy, Figure~\ref{fig:dff_cone} presents the distribution of the 
Discontinuity Feedback Factor (DFF) limiter. 
It is evident that the DFF is precisely activated ($\alpha < 1$, indicated by the red regions) only at the locations of strong discontinuities, sharply tracing the strong 
shock structures. 
Crucially, the DFF remains completely inactive ($\alpha = 1$, blue region) within the boundary layer and along the entire wall surface. This selective activation ensures 
that the high-order spatial reconstruction is fully preserved in the smooth viscous regions, preventing artificial numerical dissipation from contaminating the prediction 
of surface aerothermal loads.

\subsubsection{Validation of Surface Quantities}

\begin{figure}[htp]  
	\centering  
	\includegraphics[width=0.45\textwidth]{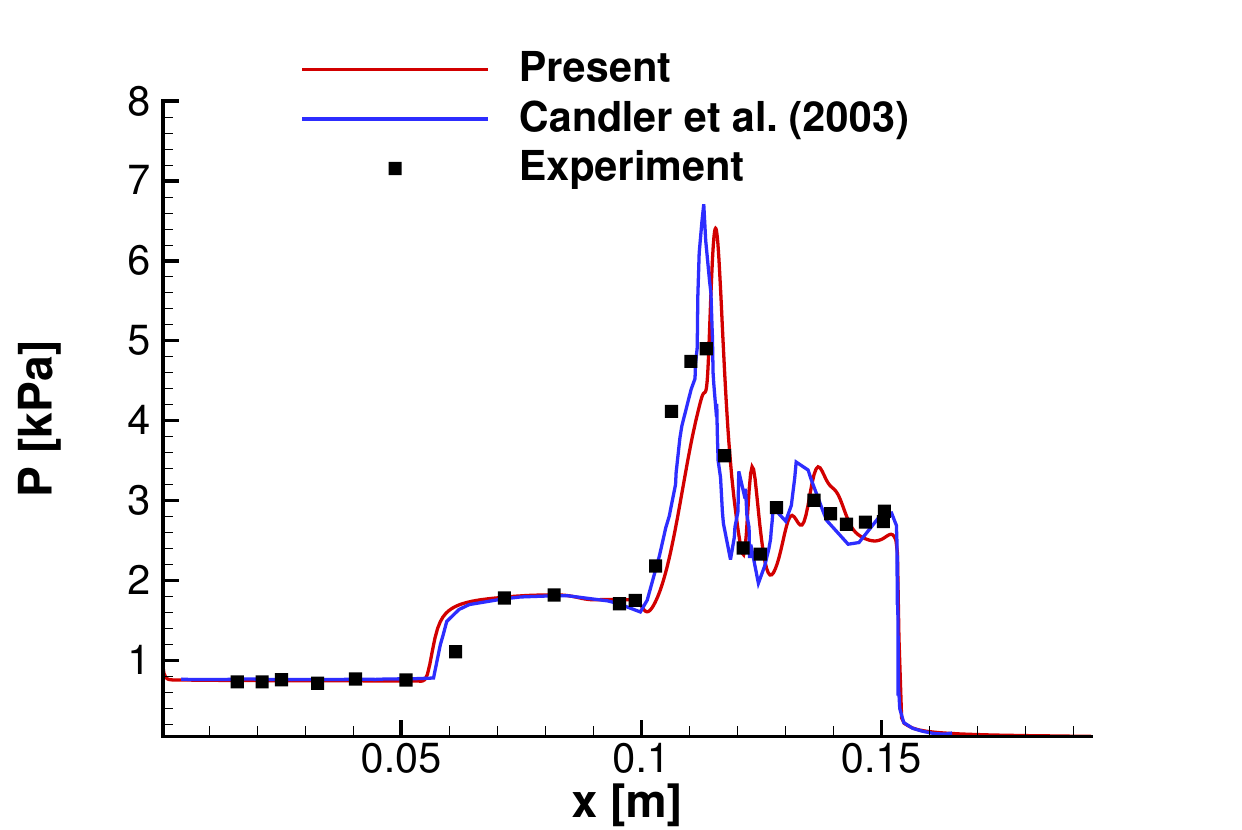}  
	\includegraphics[width=0.45\textwidth]{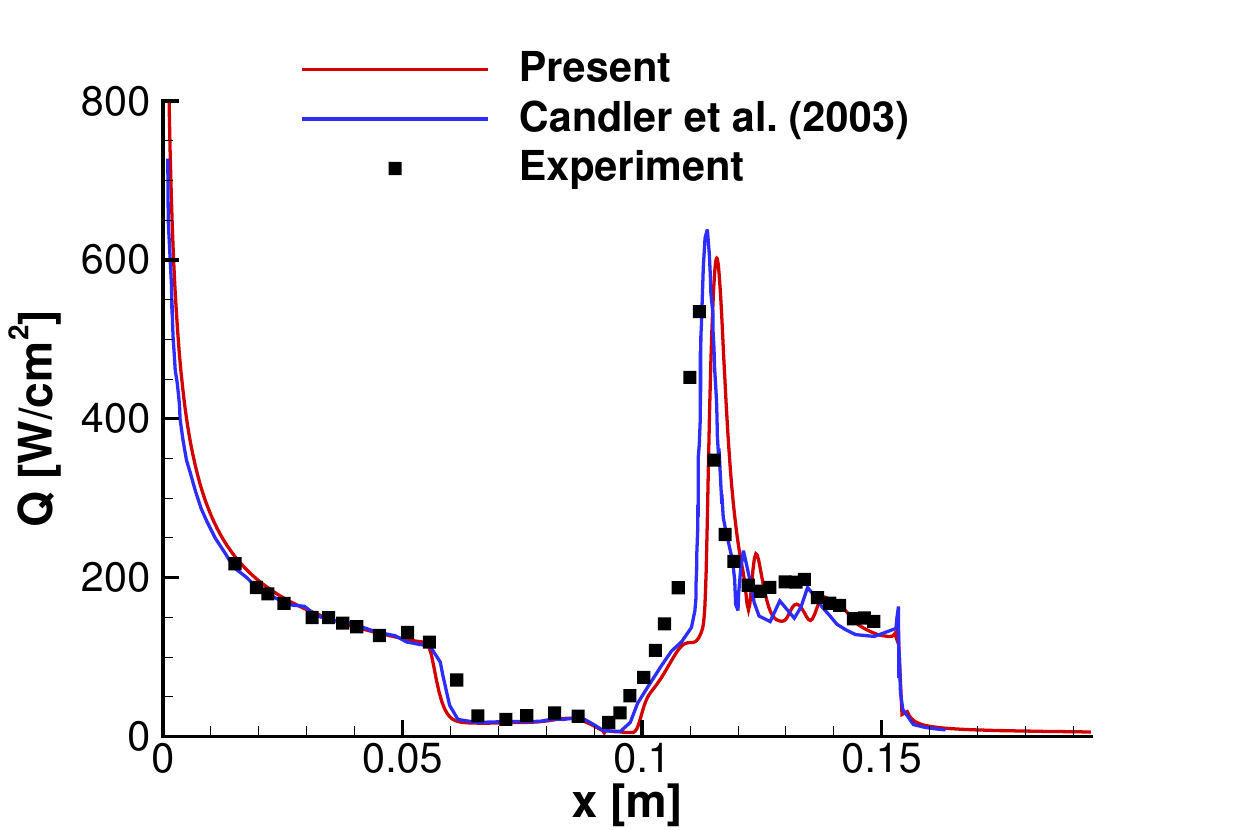}  
	\vspace{-4mm}  
	\caption{Comparison of surface pressure and heat flux with experimental data and reference results for sharp double cone Run 35.}  
	\label{surface-run35}  
\end{figure} 

To validate the accuracy of the current numerical framework, the computed surface pressure and heat flux distributions are compared with experimental measurements for the Run 35 
condition and the reference numerical results reported by Candler et al.~\cite{Candler2003}. 
The reference CFD results were obtained using the same two-temperature model and accommodation coefficients as the present work. 
Figure~\ref{surface-run35} displays the comparison of these surface quantities.

While the current solver captures the overall wave structures and load distributions, a quantitative comparison reveals a slightly larger deviation from the experimental data 
compared to the results of Candler et al. Specifically, the present simulation predicts a more extensive separation region. This is evidenced by an earlier rise in surface 
pressure and a delayed drop in the heat flux profile relative to both the experimental and reference data. 

Despite the overprediction of the recirculation zone size (characterized by earlier separation and later reattachment), the peak values of pressure and heat flux at the shock 
impingement location are predicted with high fidelity. The discrepancy in the separation extent is a known sensitivity in hypersonic double-cone flows, often attributed to subtle 
differences in numerical dissipation and the specific implementation of vibrational relaxation models. Given that the critical peak thermal and mechanical loads are resolved 
accurately, the overall results are considered reliable for the purposes of the present study.

\subsubsection{Assessment of the Generalized Kinetic Boundary Condition}

\begin{figure*}[htp]
	\centering
	\subfloat[Surface pressure]{
		\includegraphics[width=0.45\textwidth]{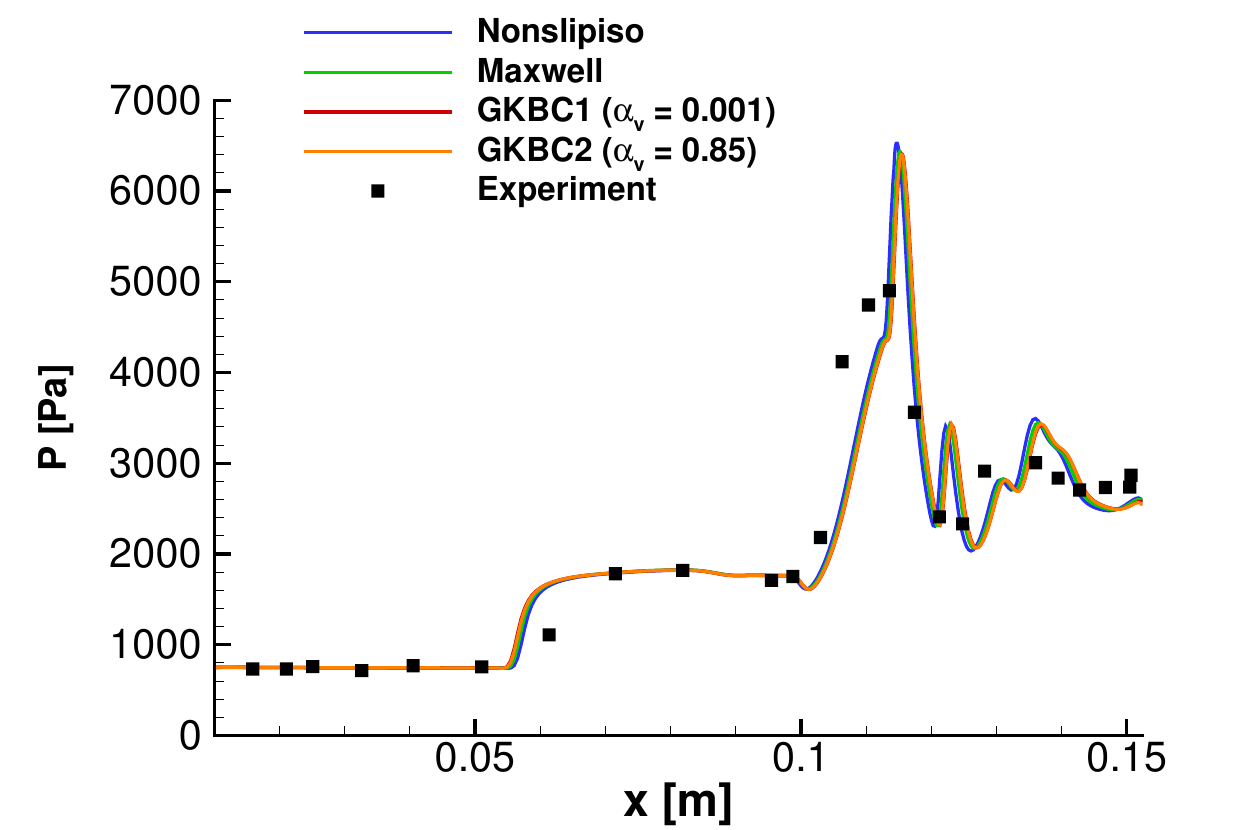}
	}
	\hfill
	\subfloat[Surface heat flux]{
		\includegraphics[width=0.45\textwidth]{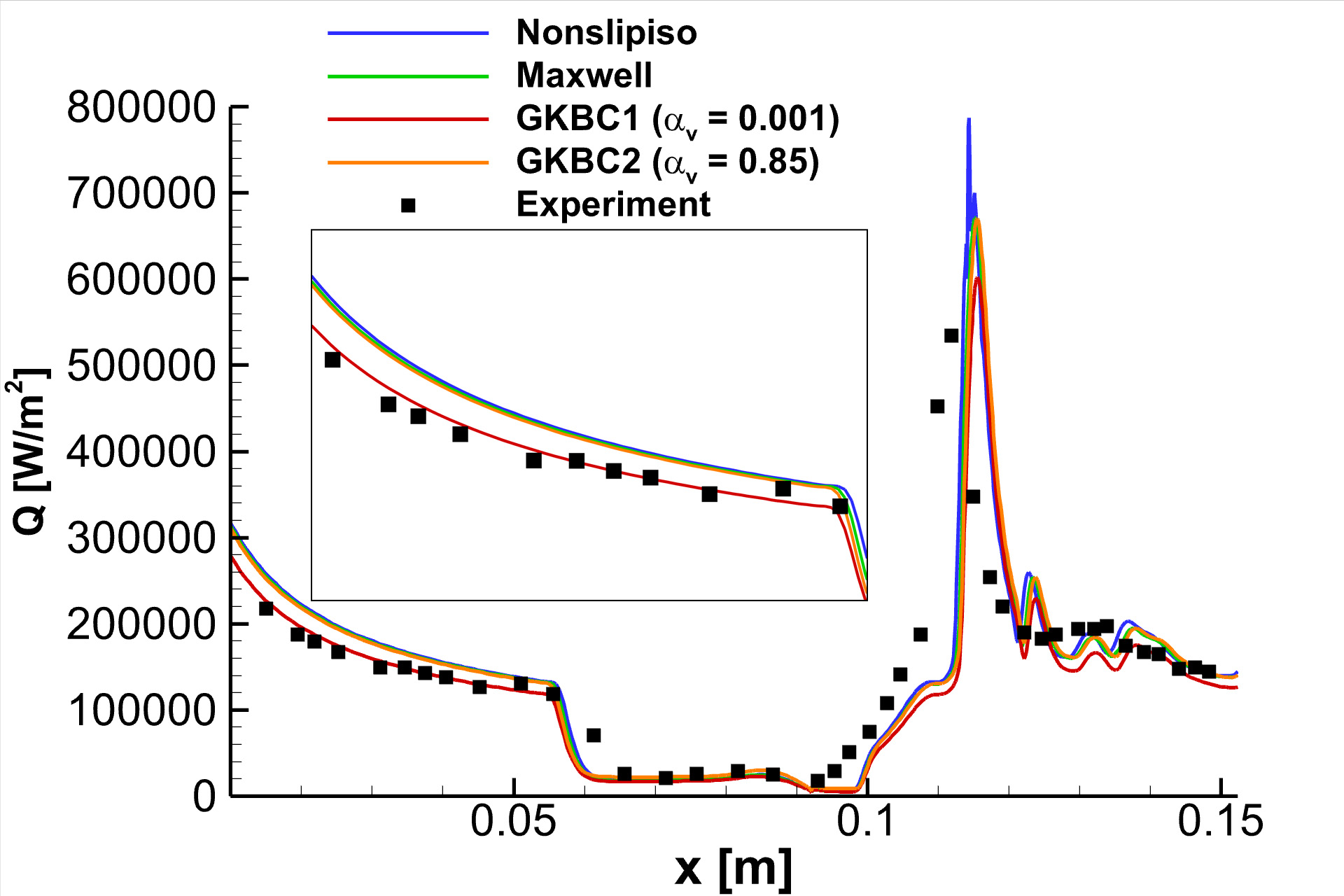}
	}
	\caption{Comparison of surface properties using different boundary conditions. The inset in (b) shows a magnified view of the heat flux on the first cone.}
	\label{fig:bc_comparison}
\end{figure*}

To verify the effectiveness of the proposed Generalized Kinetic Boundary Condition (GKBC) and to elucidate the specific impact of the vibrational thermal accommodation 
coefficient, comparative simulations are performed using four distinct boundary modeling strategies:
(1) The traditional no-slip isothermal condition;
(2) The standard Maxwell boundary condition with full accommodation ($\sigma = \alpha = 1$);
(3) A comparative GKBC case with high vibrational accommodation ($\sigma = \alpha_{tr} = \alpha_v = 0.85$);
(4) The present GKBC model employed in this study, with parameters set to $\sigma = \alpha_{tr} = 0.85$ and $\alpha_v = 0.001$.
Figure~\ref{fig:bc_comparison} presents the surface pressure and heat flux distributions obtained from these four models against the experimental measurements.

The comparison reveals that all four boundary conditions predict virtually identical flow structures, with the surface pressure curves being nearly indistinguishable. 
However, significant discrepancies are observed in the heat flux profiles, particularly on the first cone surface. 
As shown in Figure~\ref{fig:bc_comparison}b, the results from the comparative GKBC case ($\alpha_v = 0.85$) align closely with those of the traditional no-slip and 
standard Maxwell conditions. All three models markedly overpredict the thermal load. This confirms that when the vibrational accommodation coefficient is 
high (close to unity), the boundary condition implicitly forces the high-energy vibrational modes frozen in the free stream ($T_{v,\infty} = 2562$ K) to 
equilibrate rapidly with the cold wall ($T_w \approx 296$ K). This results in a massive and unphysical release of vibrational energy.
In stark contrast, the present GKBC, by explicitly assigning a low vibrational accommodation coefficient ($\alpha_v = 0.001$), correctly captures the 
near-elastic reflection of the vibrational mode consistent with its slow relaxation time scale. This configuration yields a heat flux prediction that matches 
the experimental measurements with high accuracy. 

It is also worth noting that non-physical oscillations appear near the heat flux peak in the no-slip isothermal result. In the GKS framework, surface heat flux 
is evaluated directly from the moments of the time-dependent distribution function at the cell interface (as detailed in Section III.G). 
This formulation requires high spatial accuracy in the 
boundary reconstruction. The no-slip condition imposes a rigid constraint on the interface velocity, which triggers slope limiters and degrades local reconstruction 
accuracy, resulting in numerical jitters. The kinetic boundary conditions alleviate this stiffness and yield smoother profiles.

To further investigate the independent influence of momentum and energy accommodation coefficients on the flow field and aerothermal loads, a parametric study is conducted 
using the proposed GKBC. Table~\ref{tab:gkbc_cases} summarizes the four sets of accommodation coefficients tested. GKBC1 serves as the baseline strategy. GKBC2 represents 
the high vibrational accommodation case discussed above. GKBC3 and GKBC4 are designed to isolate the effects of tangential momentum accommodation ($\sigma$) and 
translational-rotational thermal accommodation ($\alpha_{tr}$), respectively.

\begin{table*}[hbt!]
	\centering
	\caption{Summary of accommodation coefficients for GKBC parametric study.}
	\label{tab:gkbc_cases}
	\setlength{\tabcolsep}{8pt} 
	\begin{tabular}{lccccl}
		\hline
		Case & $\sigma$ & $\alpha_{tr}$ & $\alpha_{v}$ & Description \\
		\hline
		GKBC1 & 0.85 & 0.85 & 0.001 & Baseline (Present Model) \\
		GKBC2 & 0.85 & 0.85 & 0.85  & High Vibrational Accommodation \\
		GKBC3 & 0.50 & 0.85 & 0.001 & Low Momentum Accommodation \\
		GKBC4 & 0.85 & 0.50 & 0.001 & Low Trans-Rot Accommodation \\
		\hline
	\end{tabular}
\end{table*}

\begin{figure}[htp]
	\centering
	\includegraphics[width=0.45\textwidth]{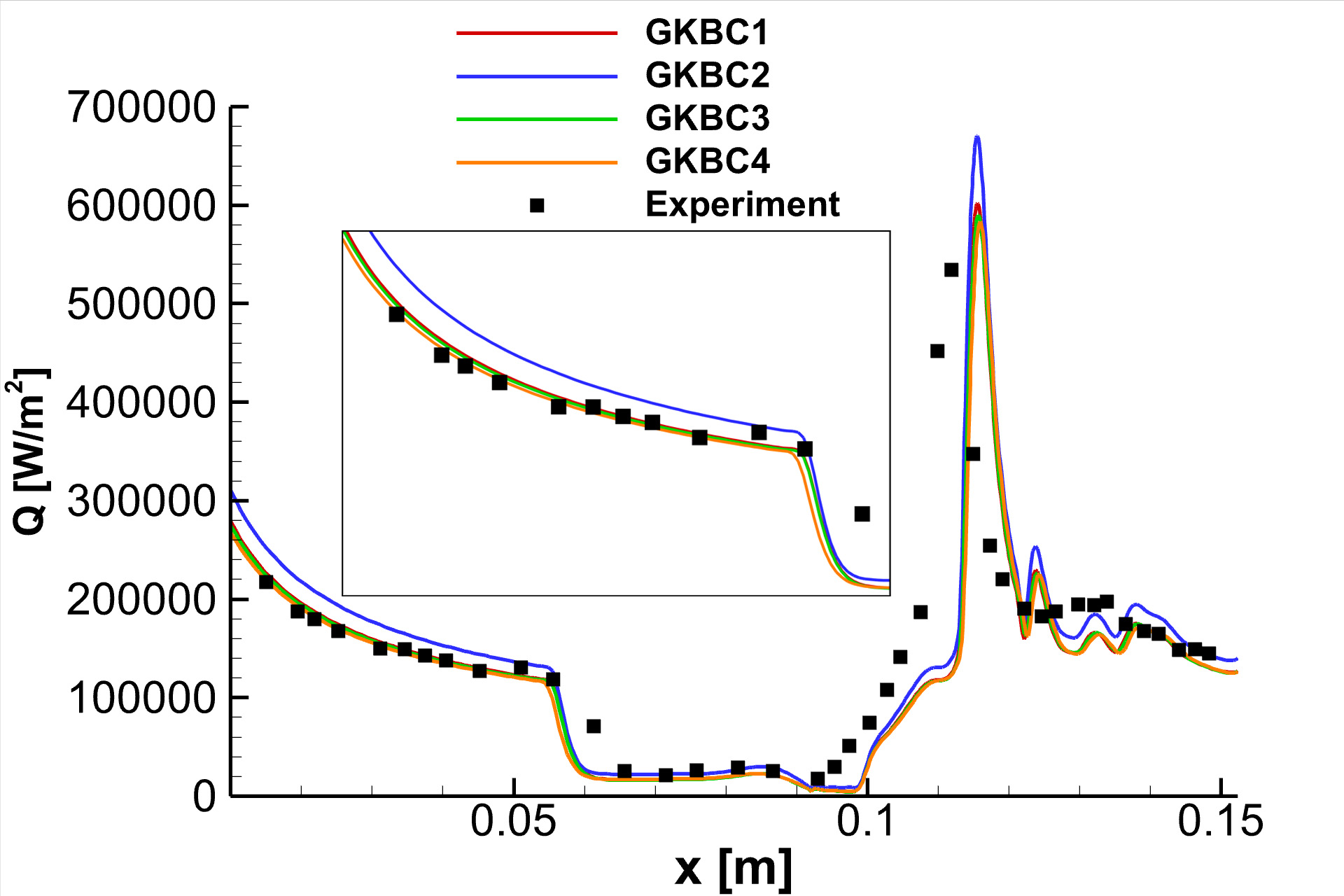} 
	\caption{Comparison of surface heat flux distributions using GKBC with different accommodation coefficients.}
	\label{fig:gkbc_parametric}
\end{figure}

Figure~\ref{fig:gkbc_parametric} compares the surface heat flux distributions predicted by these four parameter sets. 
Comparing GKBC3 with the baseline GKBC1, where $\sigma$ is reduced from 0.85 to 0.50 while keeping thermal coefficients constant, the results are virtually identical. 
This insensitivity confirms that the velocity slip has a negligible impact on the macroscopic thermal loads for the current flow condition ($Kn_\infty \approx 0.0014$), 
which falls within the near-continuum regime. This observation validates the previous assumption that the discrepancies observed in Figure~\ref{fig:bc_comparison} were 
not caused by slip effects but by thermal non-equilibrium.

In contrast, GKBC4, which reduces $\alpha_{tr}$ from 0.85 to 0.50, exhibits a noticeable deviation from the baseline. The surface heat flux shows a global reduction, 
and the separation point shifts slightly upstream. This demonstrates that even when velocity slip is negligible, the thermal accommodation of the translational-rotational 
mode remains a critical factor governing the energy exchange efficiency at the wall.

In summary, this parametric analysis highlights the unique capability of the proposed GKBC to physically decouple the accommodation mechanisms. It proves that the reliability of the present method relies on the correct modeling of the slow vibrational relaxation ($\alpha_v$), while the flow topology remains robust against variations in momentum accommodation ($\sigma$) in this regime.

\subsection{Validation Case II: Hollow Cylinder-Flare}

\subsubsection{Flow Conditions and Geometry}
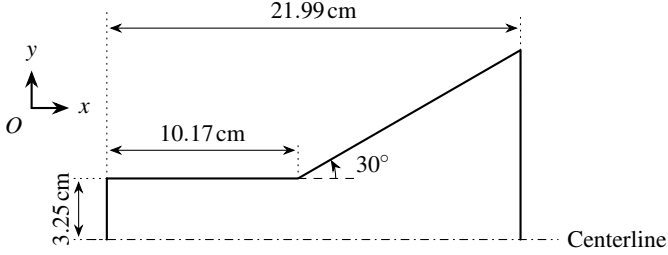
\begin{figure}[htp]
	\centering
	\begin{tikzpicture}[
	    scale=0.25, 
	    font=\small,
	    line join=round, line cap=round,
	    >=Stealth,
	    dim/.style={thin, <->, >=Stealth, shorten <=1pt, shorten >=1pt} 
	]
	
	% =====================
	% 参数定义 (cm)
	% =====================
	\def\Lcyl{10.17}       
	\def\Ltotal{21.999}     
	\def\R_LE{3.2506}      % Leading Edge Radius (User specified)
	\def\ThetaFlare{30}    
	
	% =====================
	% 坐标计算
	% =====================
	\coordinate (O) at (0,0); 
	\coordinate (Start) at (0, \R_LE); 
	\coordinate (Hinge) at (\Lcyl, \R_LE); 
	\coordinate (End) at (\Ltotal, {\R_LE + (\Ltotal-\Lcyl)*tan(\ThetaFlare)});
	
	% =====================
	% 1. 几何轮廓
	% =====================
	\draw[thick] (O) -- (Start) -- (Hinge) -- (End) -- (\Ltotal, 0);
	
	% =====================
	% 2. 中心线
	% =====================
	\draw[dash dot, thin] (-1,0) -- (\Ltotal+2,0) node[right] {Centerline};
	
	% =====================
	% 3. 尺寸标注
	% =====================
	% R_LE 标注
	\draw[dim] ($(O)+(-1.5,0.0)$) -- ($(Start)+(-1.5,0.0)$) 
	node[midway, above, sloped] {$3.25\,\mathrm{cm}$};
	\draw[thin, dotted] (Start) -- ++(-1.8,0.0);
	\draw[thin, dotted] (O) -- ++(-1.8,0.0);
	% \draw[dim] (0, 0) -- (Start) node[midway, left] {$3.2506\,\mathrm{cm}$};
	
	% L_cylinder
	\draw[dim] ($(Start)+(0,1.5)$) -- ($(Hinge)+(0,1.5)$) 
	    node[midway, above] {$10.17\,\mathrm{cm}$};
	\draw[thin, dotted] (Start) -- ++(0,9.0);
	\draw[thin, dotted] (Hinge) -- ++(0,1.8);
	\draw[thin, dotted] (End) -- ++(0,1.8);

	% L_total
	\draw[dim] ($(Start)+(0,8.0)$) -- ($(End|-Start)+(0,8.0)$) 
	    node[midway, above] {$21.99\,\mathrm{cm}$};
	\draw[thin, dotted] (End) -- ($(End|-Start)+(0,3.3)$);
	
	% Flare Angle
	\draw[thin, dashed] (Hinge) -- ++(3,0); 
	\draw[thin, ->] ($(Hinge)+(2,0)$) arc[start angle=0, end angle=\ThetaFlare, radius=2];
	\node at ($(Hinge)+(4.0, 0.8)$) [] {$30^\circ$};
	
	% =====================
	% 4. 坐标轴
	% =====================
	\coordinate (AxisOrigin) at (-4, 7);
	\draw[->, thick] (AxisOrigin) -- ++(2,0) node[right] {$x$};
	\draw[->, thick] (AxisOrigin) -- ++(0,2) node[above] {$y$};
	\node[below left] at (AxisOrigin) {$O$};
	
	\end{tikzpicture}
	\caption{Schematic of the hollow cylinder-flare geometry.}
	\label{fig:geo_cylinder}
\end{figure}

The second validation case involves a hollow cylinder-flare configuration, corresponding to the experimental Run 11 condition. The specific geometric profile and dimensions 
are illustrated in Figure~\ref{fig:geo_cylinder}. The coordinate system is defined with the origin at the center of the leading edge plane, with the $x$-axis along the symmetry 
axis. The corresponding free-stream parameters are summarized in Table~\ref{tab:all_conditions}. Similar to the double-cone case, the flow exhibits distinct thermal 
non-equilibrium characteristics.

\subsubsection{Computational Mesh}
\begin{figure}[htp]
\centering
% 请替换为圆柱-扩张段拐角处的局部网格放大图文件名
\includegraphics[width=0.45\textwidth]{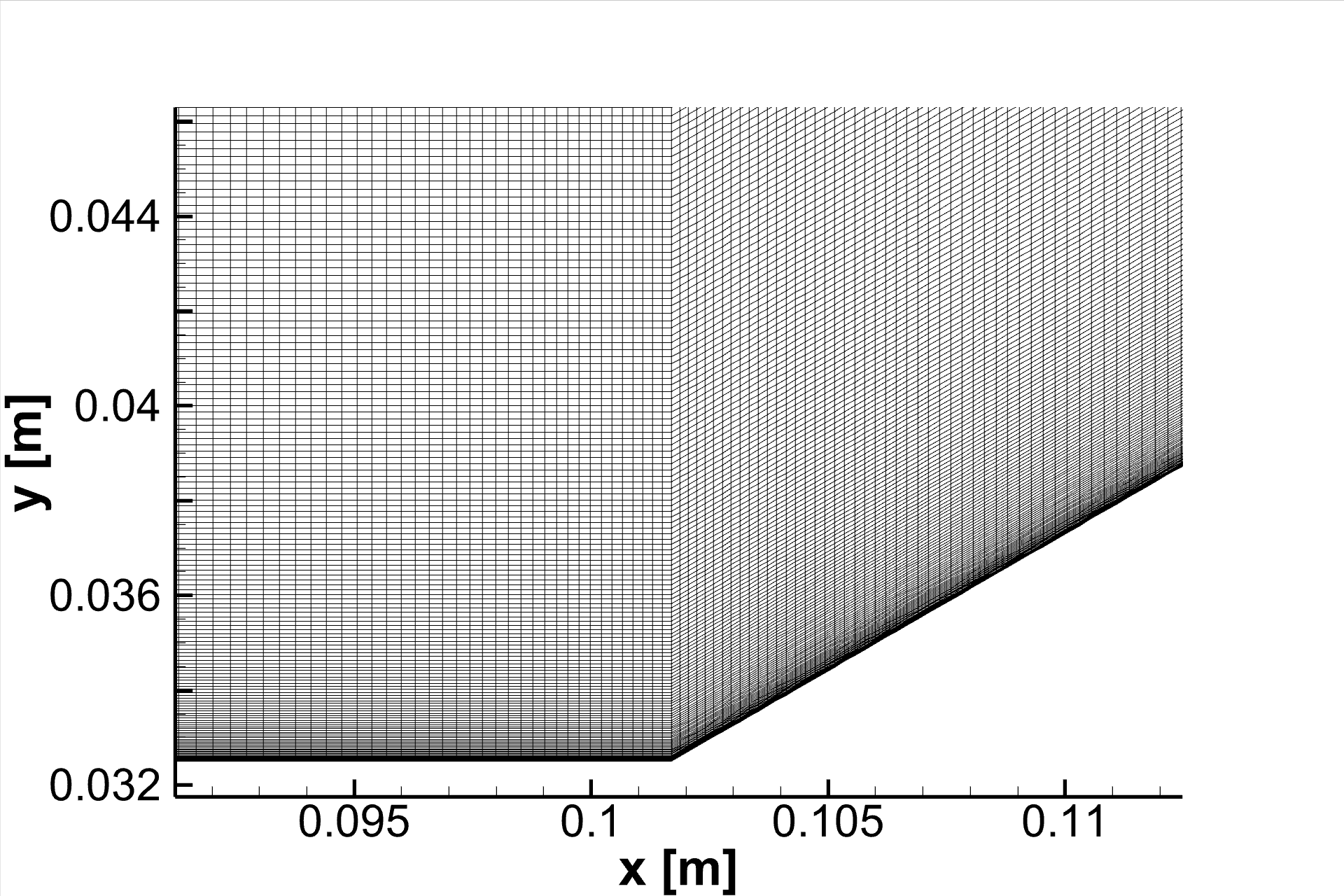} 
\caption{Close-up view of the computational mesh near the cylinder-flare junction.}
\label{fig:mesh_cylinder}
\end{figure}

Consistent with the grid verification strategy employed in the previous double-cone case, a grid independence study was conducted for the current hollow cylinder-flare 
configuration using three grid density levels. The medium-resolution grid, consisting of $500 \times 250$ cells in the streamwise and wall-normal directions, respectively, 
was confirmed to provide grid-converged solutions. For the sake of brevity, the specific comparisons of the grid convergence study are omitted here. 
To accurately resolve the flow separation and the interaction between the boundary layer and the geometric discontinuity, the computational mesh is specifically refined near 
the cylinder-flare junction. The local grid topology illustrating the clustering in this critical region is presented in Figure~\ref{fig:mesh_cylinder}.

\subsubsection{Flow Structure}

\begin{figure}[htp]  
	\centering  
	\includegraphics[width=0.45\textwidth]{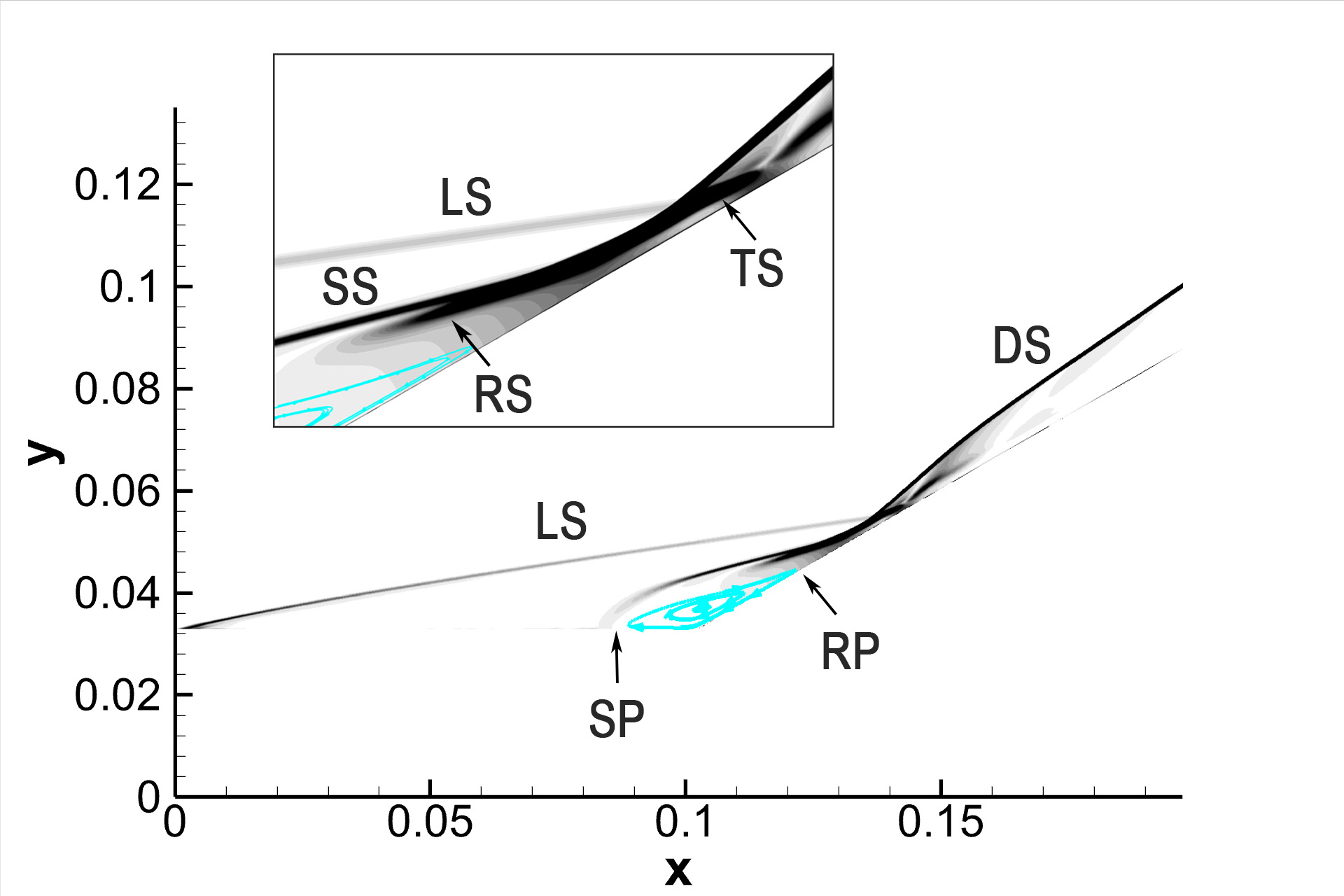} % 假设上图是压力
	\includegraphics[width=0.45\textwidth]{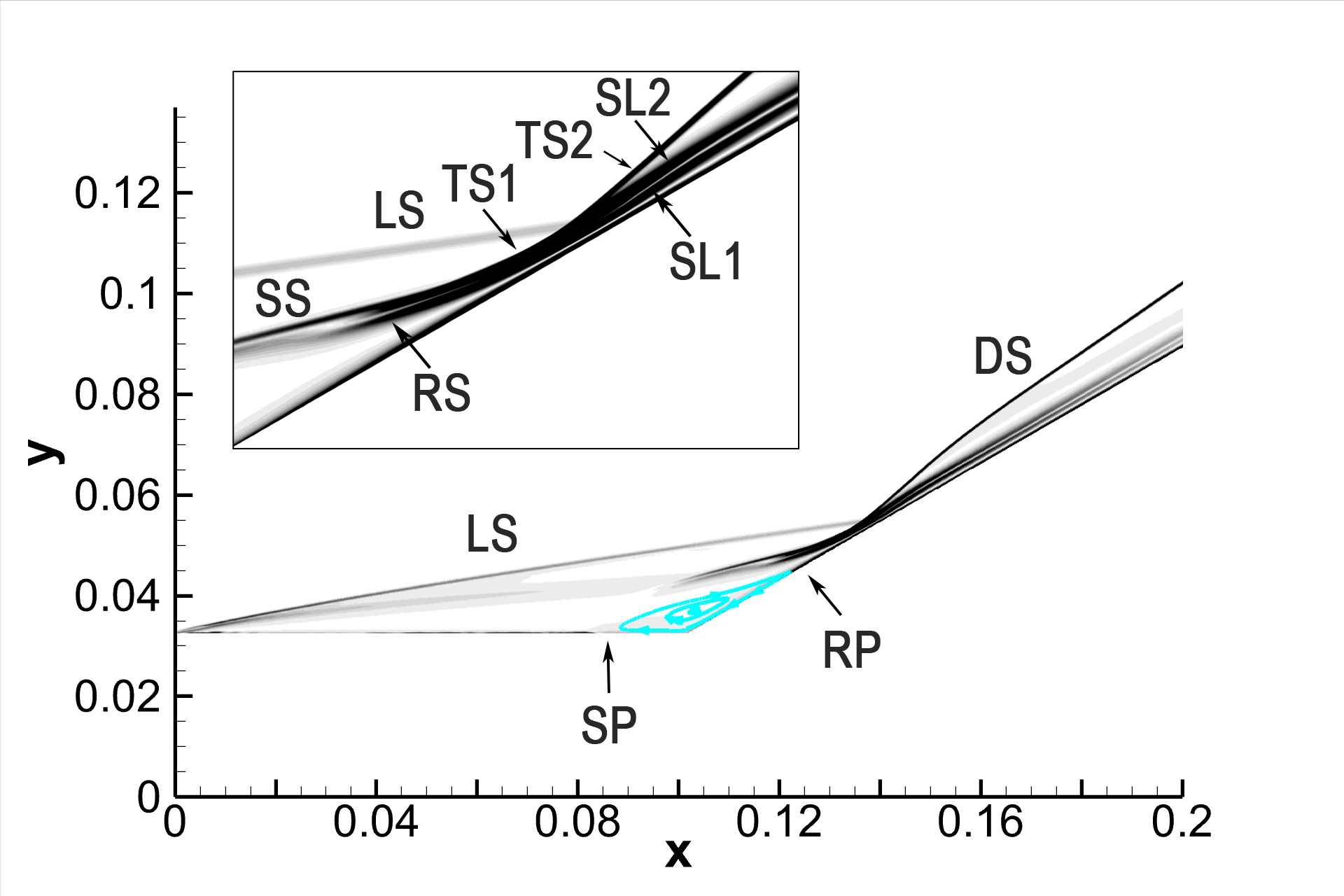}  % 假设下图是密度
	\vspace{-4mm}  
	\caption{Numerical schlieren images of the hollow cylinder-flare configuration. Top: Pressure gradient contours; Bottom: Density gradient contours. The cyan region 
	indicates the separation vortex.}  
	\label{schematic_cylinder}  
\end{figure} 

The flow field exhibits a complex multiple-shock interaction pattern. Figure~\ref{schematic_cylinder} illustrates the flow structure. To accurately distinguish between shock 
waves and shear layers, the analysis employs a comparative view using both pressure gradient (top) and density gradient (bottom) contours. While shock waves induce gradients 
in both fields, shear layers are identified as contact discontinuities featuring strong density gradients but vanishing pressure gradients.

The interaction process evolves through two distinct stages. First, in the lower region, the separation shock (SS) intersects with the reattachment shock (RS). This interaction, 
typical of Type VI interference, generates an upward-propagating transmitted shock and a distinct lower shear layer (SL1).

Subsequently, this upward transmitted shock intersects with the leading-edge shock (LS), resulting in the generation of two transmitted shocks separated by an upper shear 
layer (SL2). The outward-propagating branch merges to form the final strong bow shock. 

Observations from the pressure gradient contours reveal that the inward-deflected transmitted shock does not impinge on the wall. Instead, it becomes confined within a narrow 
channel, where it reflects internally and eventually dissipates downstream. The density gradient contours confirm the nature of this confinement channel: two prominent linear 
structures parallel to the wall are clearly visible. Since these structures exhibit strong density variations but continuous pressure profiles (as seen in the upper figure), they 
are identified as the upper and lower shear layers.

Since the inward-deflected shock is trapped and dissipated within this shear-layer corridor, there is no direct shock impingement on the flare surface. This mechanism effectively 
shields the wall, explaining why the peak surface pressure and heat flux in this case are significantly lower than those observed in the double-cone case under similar 
free-stream conditions.

\subsubsection{Validation of Surface Quantities}

\begin{figure}[htp]  
	\centering  
	\includegraphics[width=0.45\textwidth]{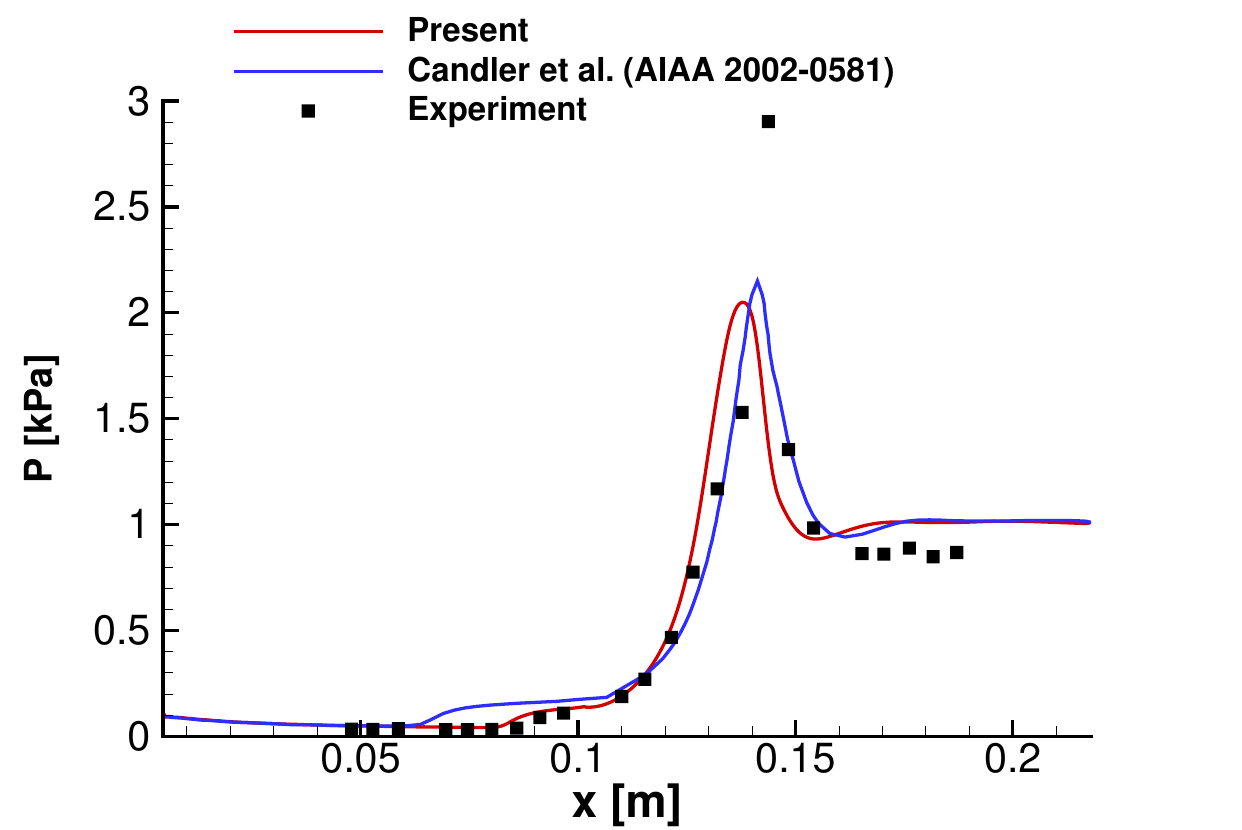}  
	\includegraphics[width=0.45\textwidth]{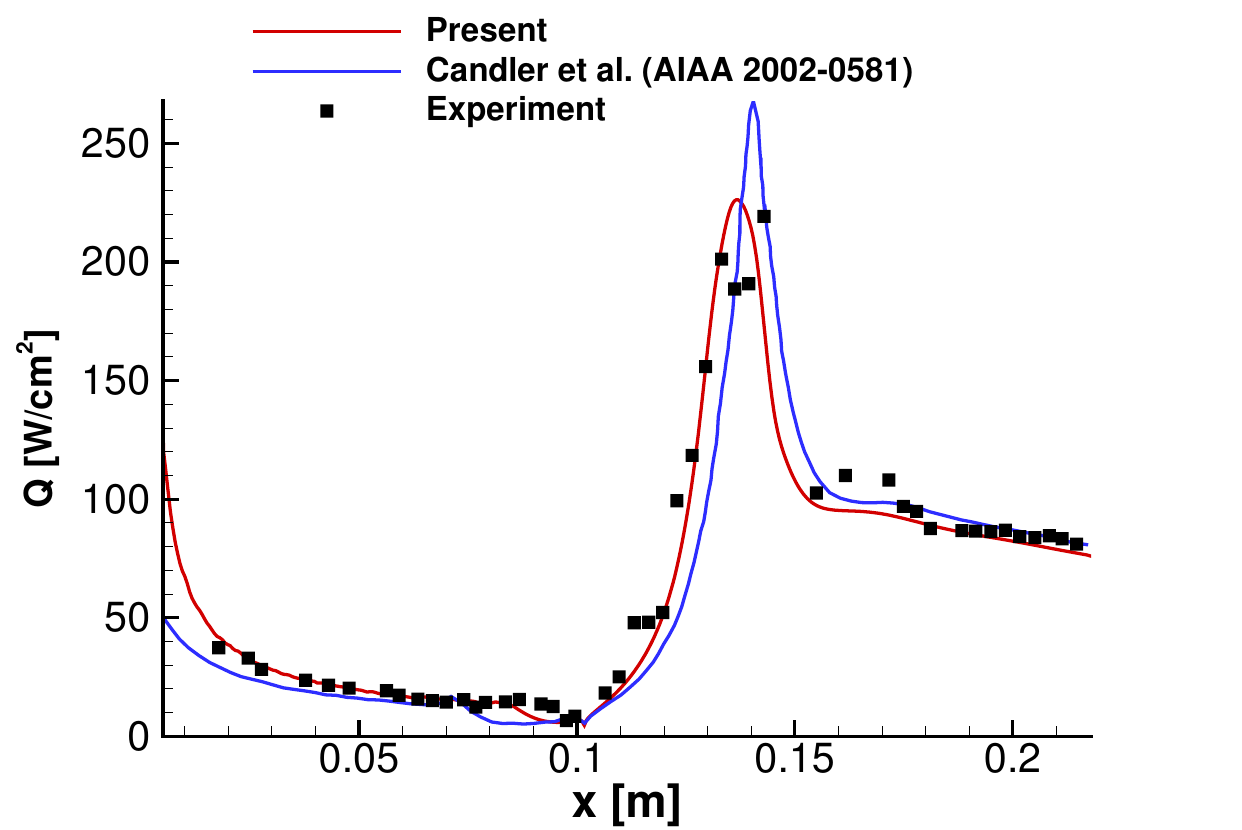}  
	\vspace{-4mm}  
	\caption{Comparison of surface pressure and heat flux with experimental data and reference results for hollow cylinder-flare Run 11.}  
	\label{surface-run11}  
\end{figure} 

Figure~\ref{surface-run11} compares the computed surface pressure and heat flux distributions with the experimental measurements and the numerical results reported by Candler 
et al.~\cite{Candler2003}. The reference CFD results were obtained using the same two-temperature model and accommodation coefficients as the present work.

A notable observation is that the present method exhibits significantly better agreement with the experimental data than the reference results, particularly regarding the 
separation characteristics. As indicated by the heat flux profile, the present method accurately predicts the separation location, whereas the reference results indicate 
a much earlier separation. Furthermore, the heat flux magnitude within the separation bubble and along the cylindrical forebody is predicted with higher accuracy. However, 
both methods show comparable deviations in predicting the peak pressure and the pressure levels downstream of the reattachment point.

The superior performance of the present method is attributed to its enhanced capability in modeling non-equilibrium slip effects, which are more pronounced in the hollow 
cylinder flow than in the cone case. Unlike Navier-Stokes solvers that rely on macroscopic gradients to determine slip quantities, the present method employs the generalized 
kinetic boundary condition detailed in the previous section. By characterizing the particle distribution function at the wall, this approach allows for a more precise 
evaluation of interface fluxes, thereby effectively capturing the non-equilibrium gas-surface interactions in this flow regime.

\subsection{Effects of Varying Freestream Density}

To evaluate the predictive capability of the numerical method for SBLI under varying flow conditions, simulations were performed for three cases selected from the experimental campaign by Holden et al. 
The baseline case corresponds to the experimental Run 35. The two additional cases correspond to Run 6 and Run 7, which represent flow environments with significantly lower freestream densities.
The detailed free-stream conditions for these cases are listed in Table~\ref{tab:all_conditions}.
Compared to the baseline case ($1.0\rho$), the free-stream densities of Run 6 and Run 7 are approximately $0.5\rho$ and $0.33\rho$, respectively.

\subsubsection{Flow Topology and Separation Characteristics}

\begin{figure*}[htp]
	\centering
	\subfloat[Baseline ($1.0\rho$)]{
		\includegraphics[width=0.45\textwidth]{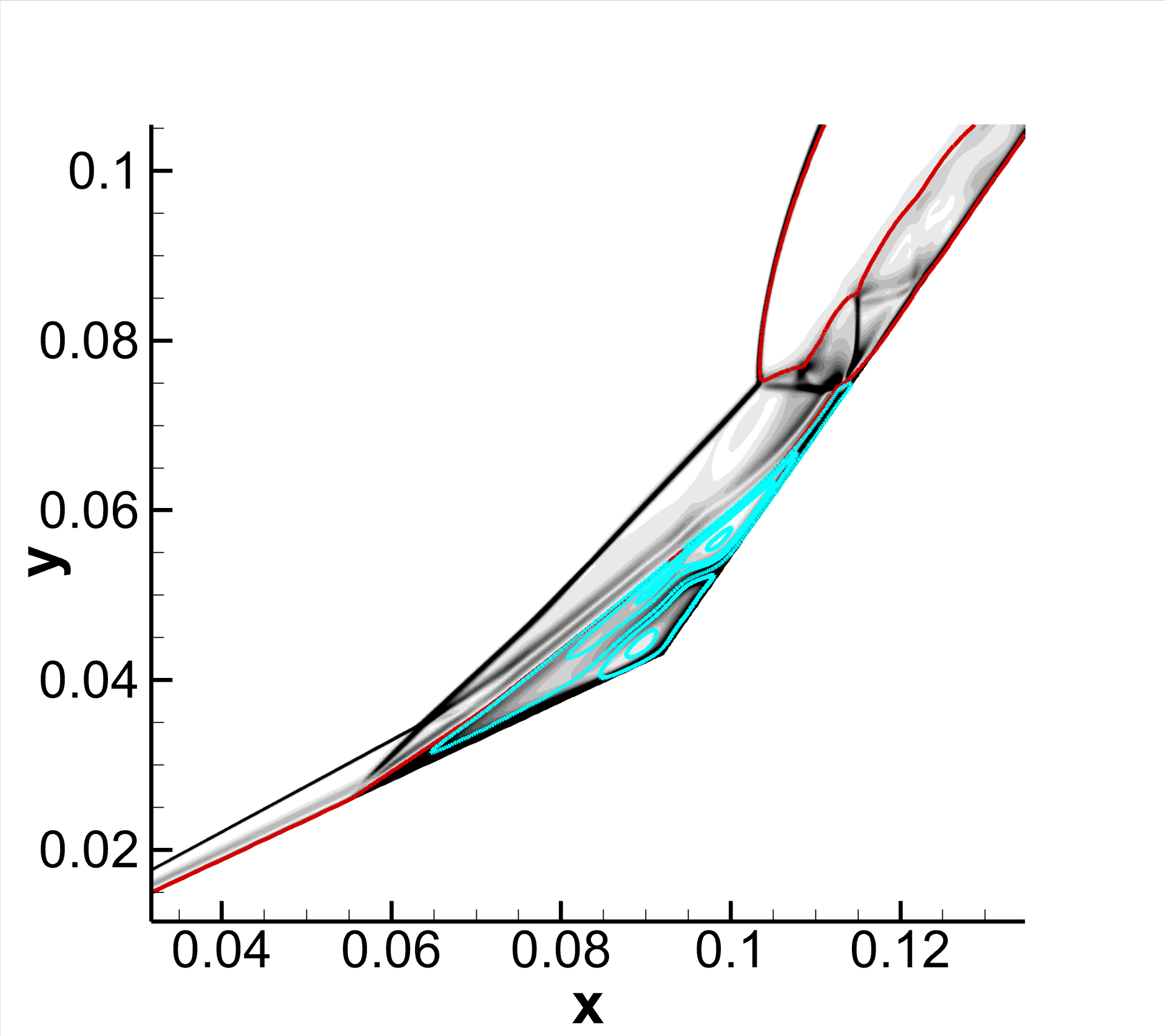}
		\label{fig:flow_baseline}
	}
	\subfloat[$0.5\rho$]{
		\includegraphics[width=0.45\textwidth]{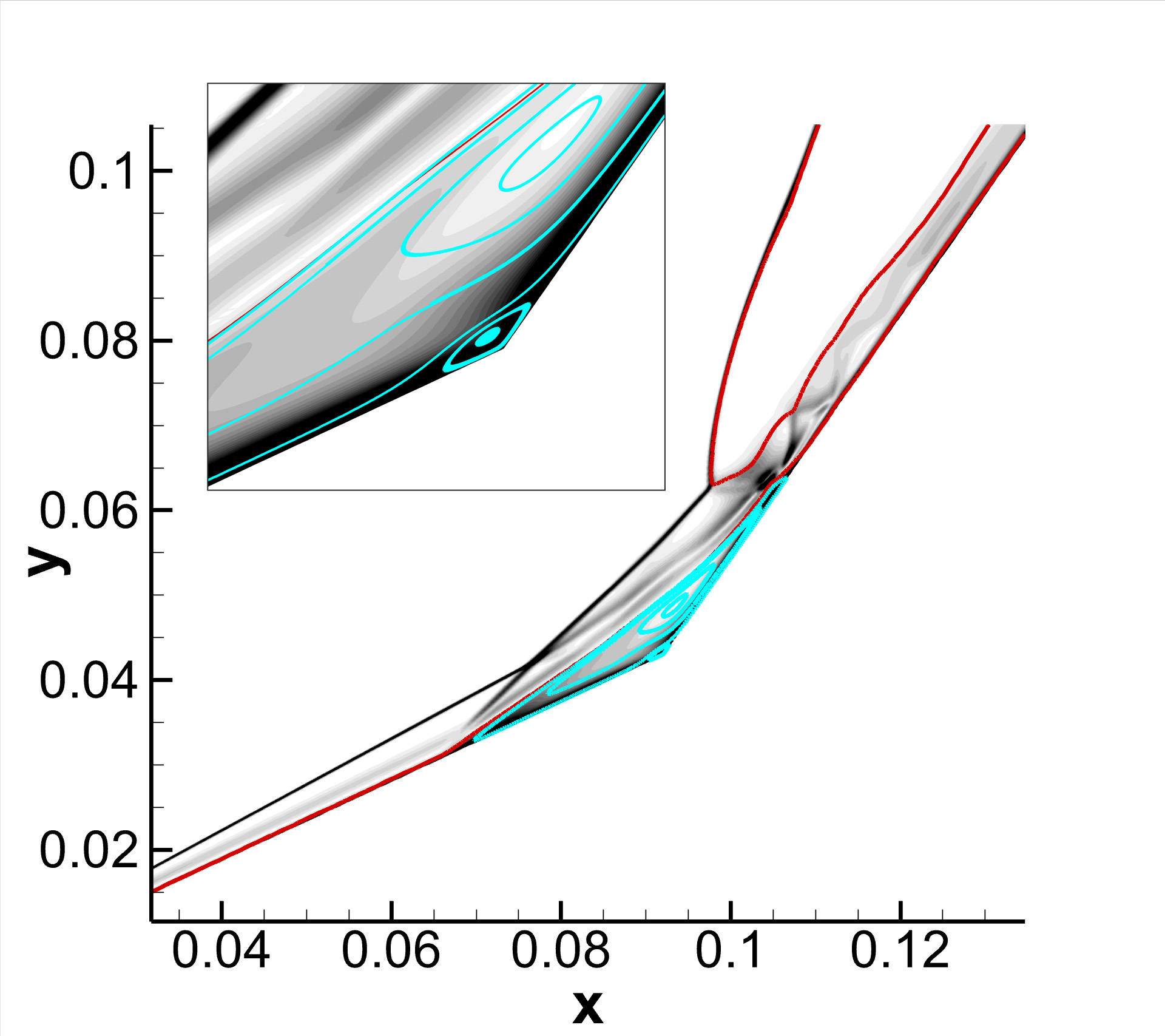}
		\label{fig:flow_0.5rho}
	}
	
	\subfloat[$0.33\rho$]{
		\includegraphics[width=0.45\textwidth]{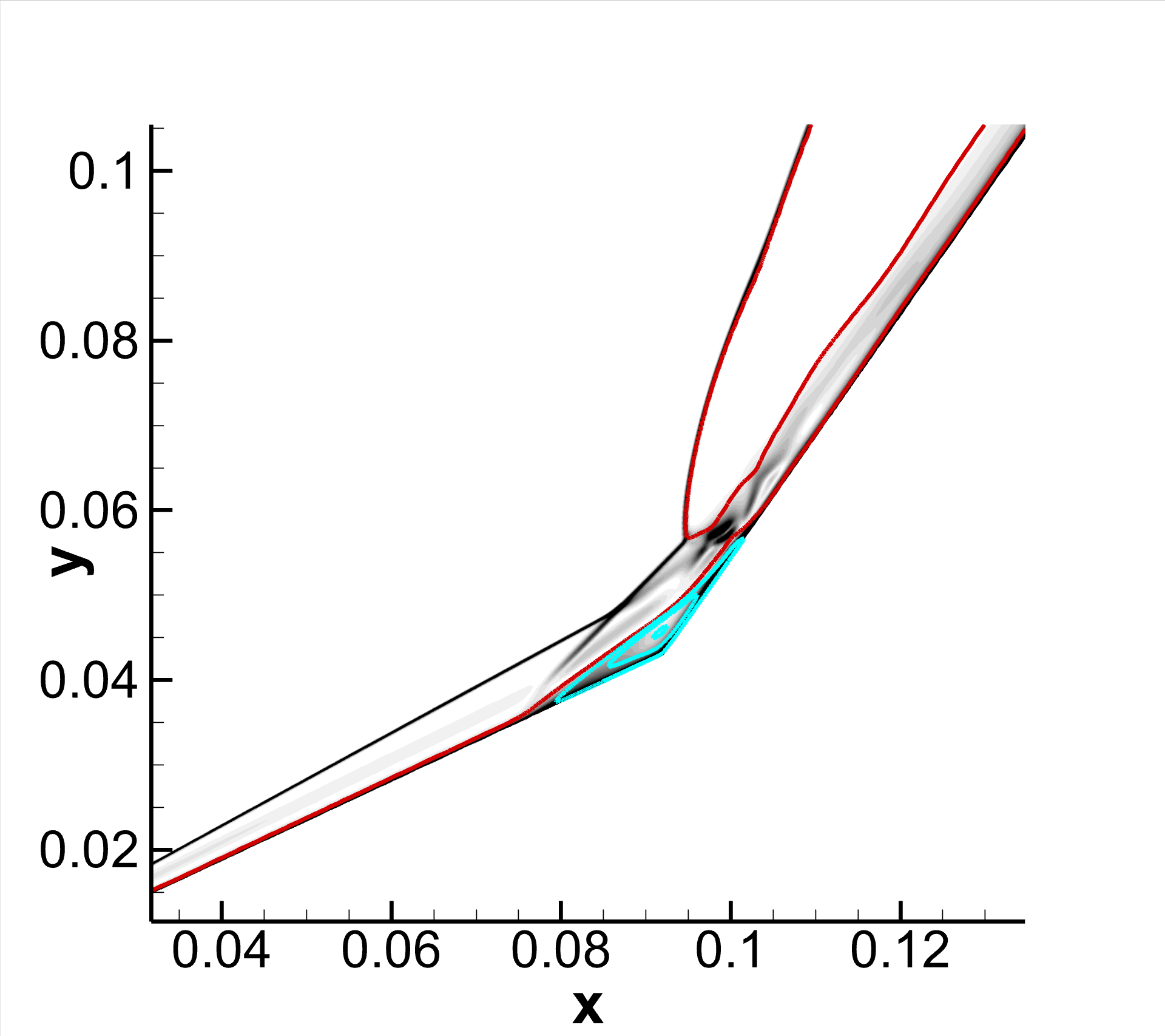}
		\label{fig:flow_0.33rho}
	}
	\caption{Contours of density gradient magnitude superimposed with streamlines for different density cases. The cyan regions indicate separation vortices.}
	\label{fig:rarefaction_flow}
\end{figure*}

Figure~\ref{fig:rarefaction_flow} presents the contours of density gradient magnitude superimposed with streamlines for the three cases. Since the general wave structures (Type VI interaction) are topologically consistent with the baseline case detailed in the previous section, the description here focuses on the variations in the separation region.

As the freestream density decreases, the separation region exhibits a monotonic reduction in streamwise extent. The most notable difference lies in the secondary separation structure near the corner.
For the baseline Run 35 ($1.0\rho$), a large secondary separation vortex is evident.
For the Run 6 ($0.5\rho$) case, the magnified view in Figure~\ref{fig:rarefaction_flow}b reveals that although the separation bubble has shrunk, a minute yet distinct secondary vortex persists near the corner.
In contrast, for the lowest density Run 7 ($0.33\rho$) case, the secondary separation is completely suppressed, and the recirculation zone simplifies into a single primary vortex.

\subsubsection{Surface Aerothermal Loads and Pressure Gradient Analysis}

To facilitate a quantitative analysis, the surface aerothermal loads presented in this section are expressed in terms of dimensionless coefficients. The skin friction coefficient ($C_f$), surface pressure coefficient ($C_p$), and Stanton number ($St$) are defined as follows:
\begin{equation}
	C_f = \frac{\tau_w}{0.5\rho_\infty u_\infty^2}, \quad 
	C_p = \frac{p_w-p_{\infty}}{0.5\rho_\infty u_\infty^2}, \quad 
	St  = \frac{q_w}{0.5\rho_\infty u_\infty^3}
\end{equation}
where $\tau_w$, $p_w$, and $q_w$ represent the wall shear stress, surface pressure, and surface heat flux, respectively, and $p_\infty$ denotes the freestream static pressure.
Additionally, the parameter $dC_p/ds$ denotes the streamwise gradient of the surface pressure coefficient along the model surface.

\begin{figure*}[htp]
	\centering
	\subfloat[Skin friction coefficient]{
		\includegraphics[width=0.45\textwidth]{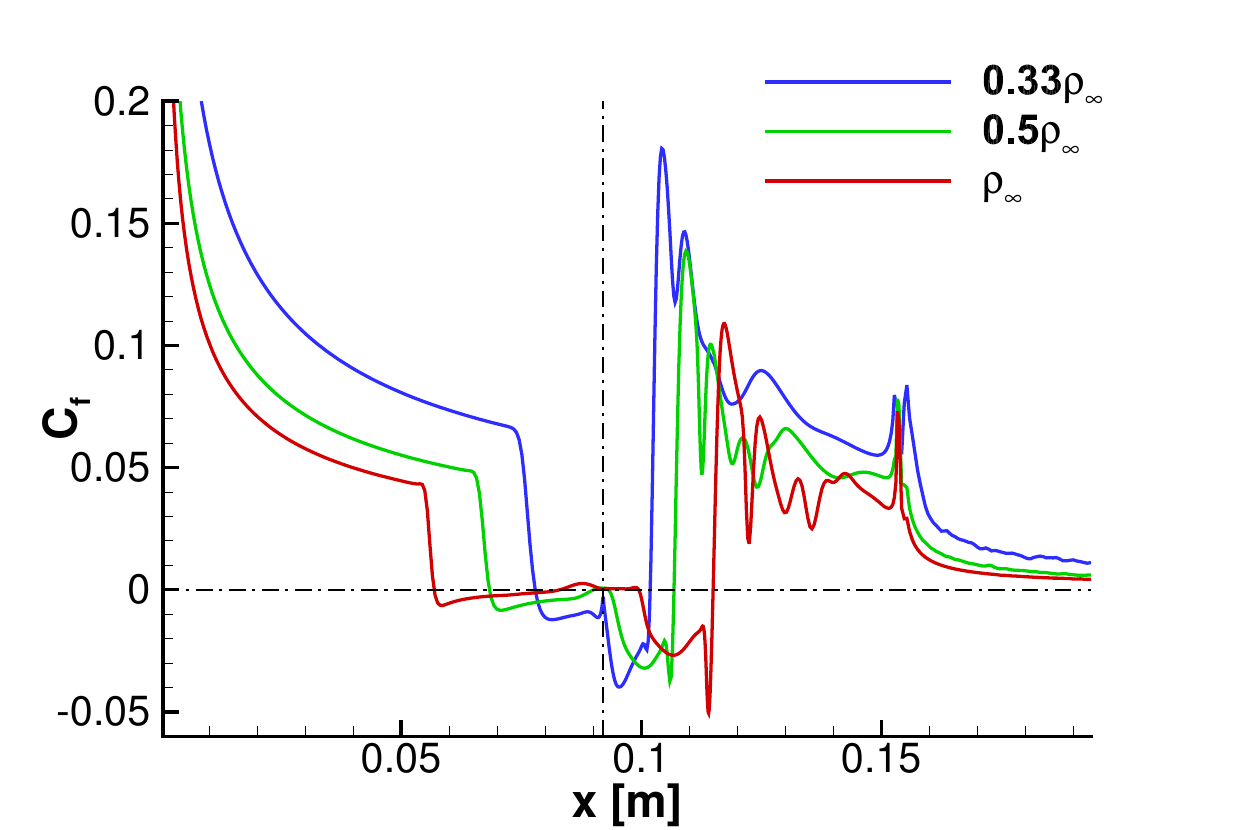}
		\label{fig:rare_Cf}
	}
	\hspace{0.05\textwidth}
	\subfloat[Enlarged view of $C_f$ near corner]{
		\includegraphics[width=0.45\textwidth]{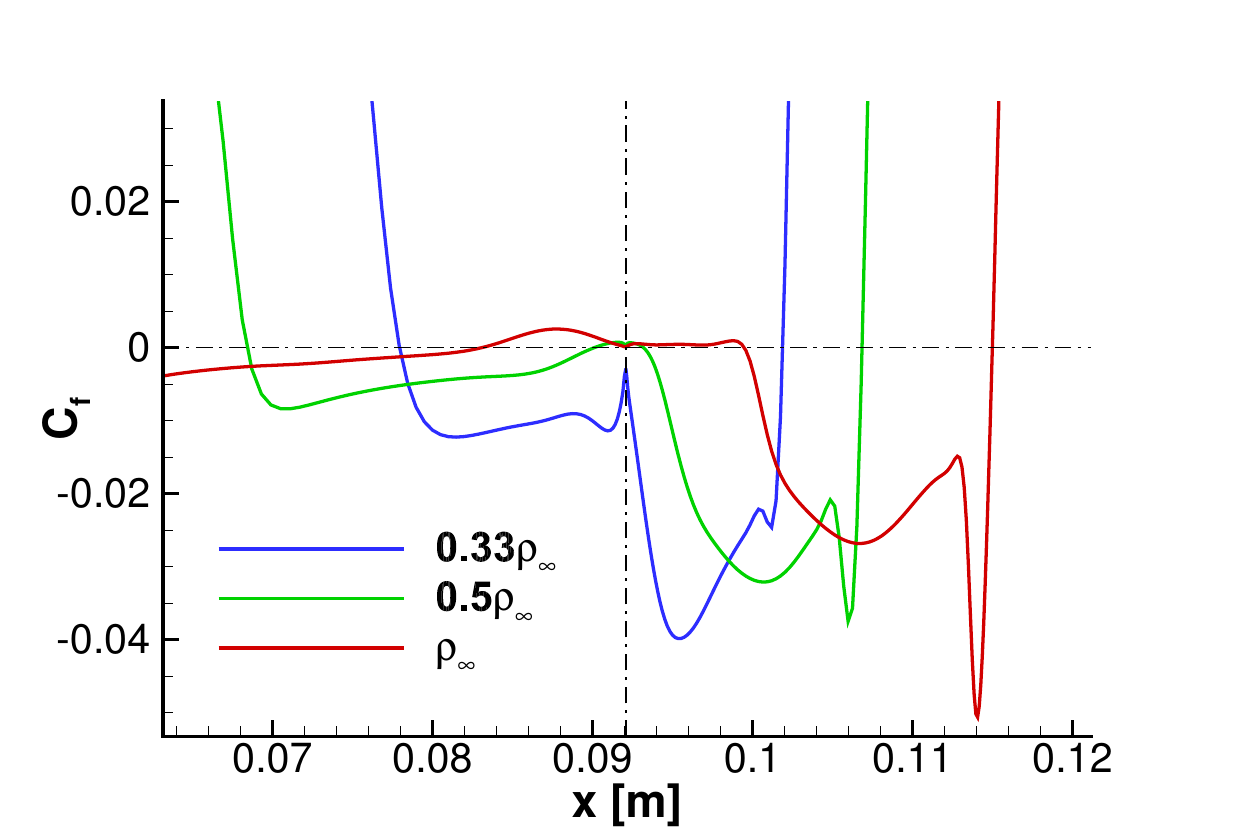}
		\label{fig:rare_Cf_zoom}
	}\\
	\vspace{0.3cm}
	\subfloat[Surface pressure coefficient]{
		\includegraphics[width=0.45\textwidth]{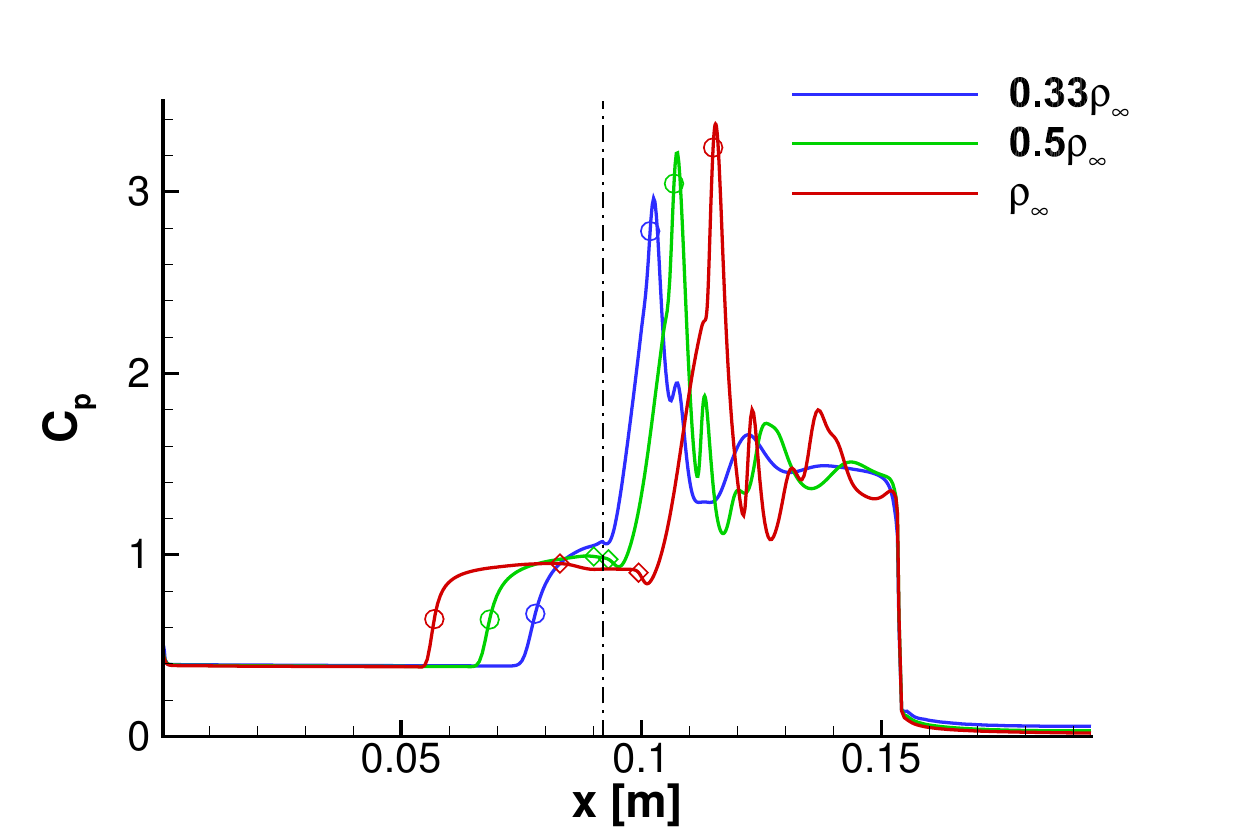}
		\label{fig:rare_Cp}
	}
	\hspace{0.05\textwidth}
	\subfloat[Surface Stanton number]{
		\includegraphics[width=0.45\textwidth]{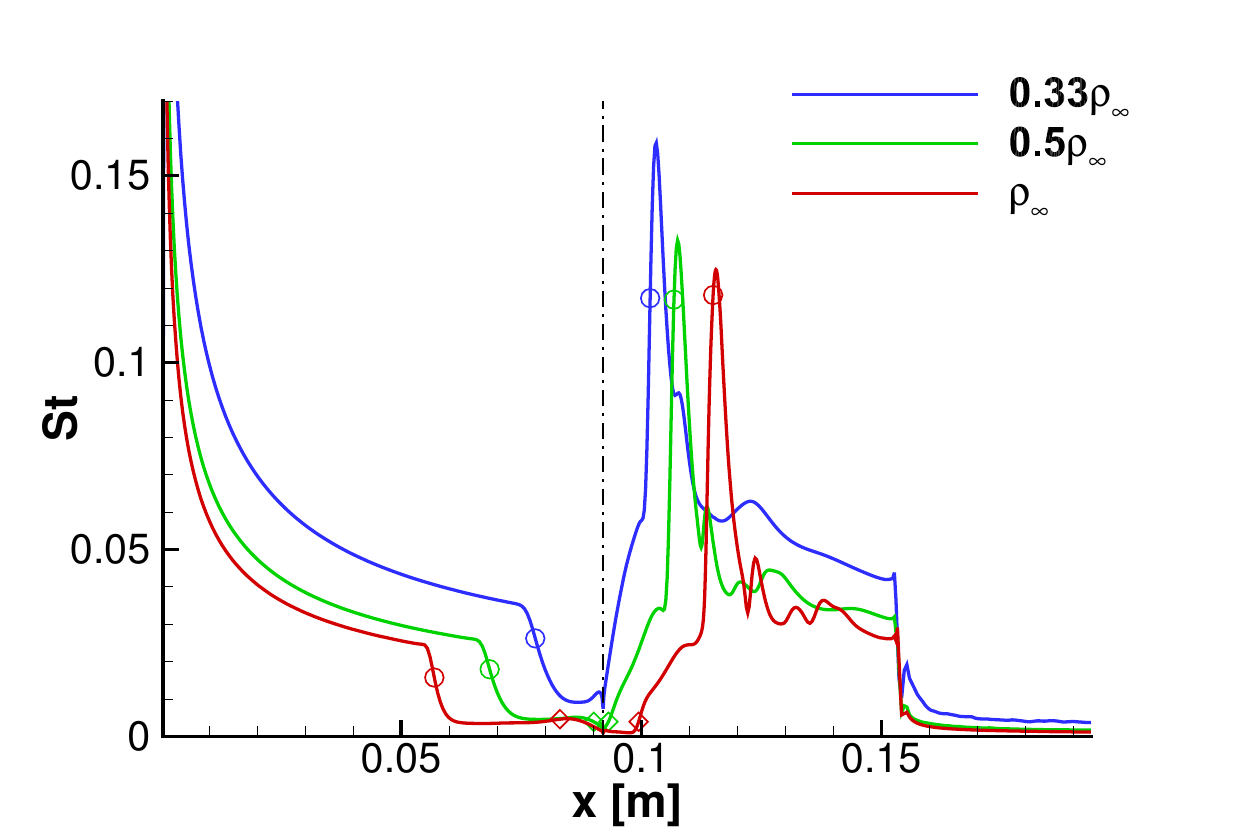}
		\label{fig:rare_St}
	}\\
	\vspace{0.3cm}
	\subfloat[Streamwise pressure gradient]{
		\includegraphics[width=0.45\textwidth]{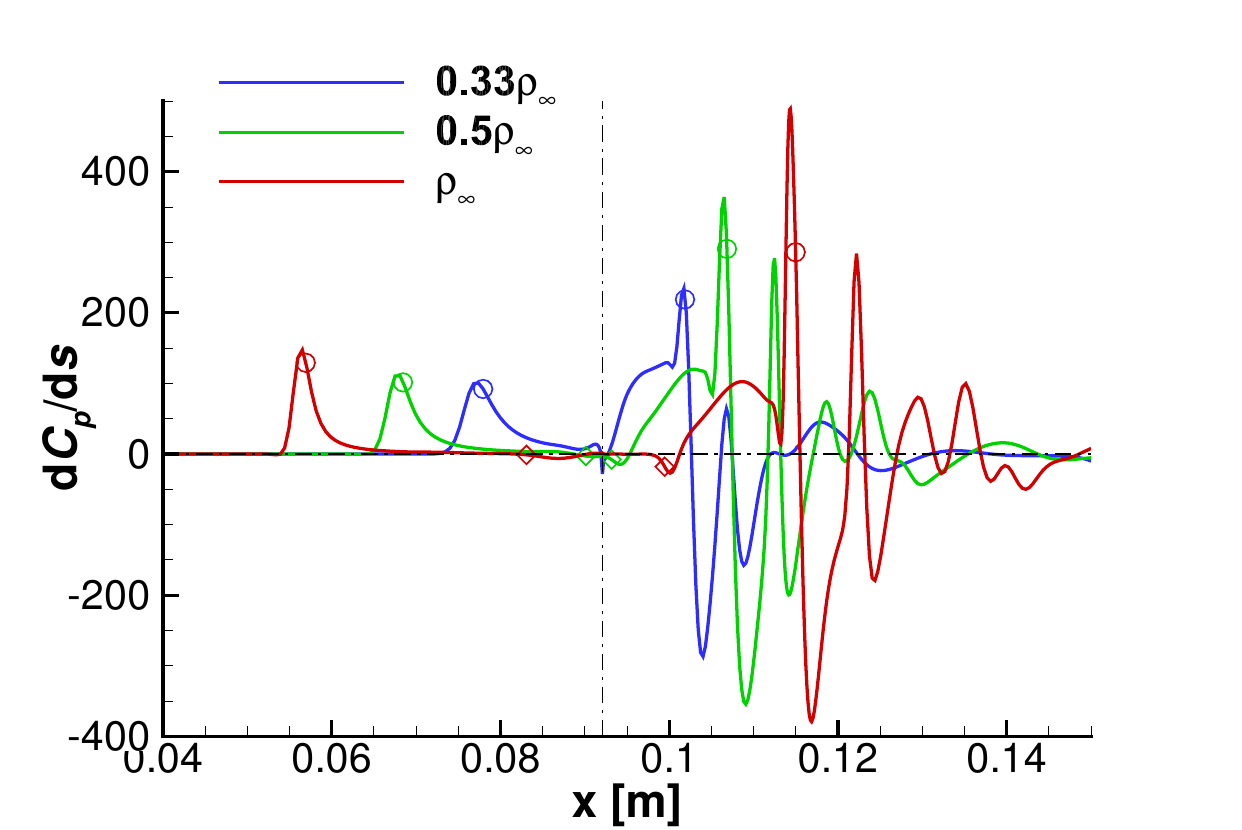}
		\label{fig:rare_dpds}
	}
	\hspace{0.05\textwidth}
	\subfloat[Enlarged view of $dC_p/ds$]{
		\includegraphics[width=0.45\textwidth]{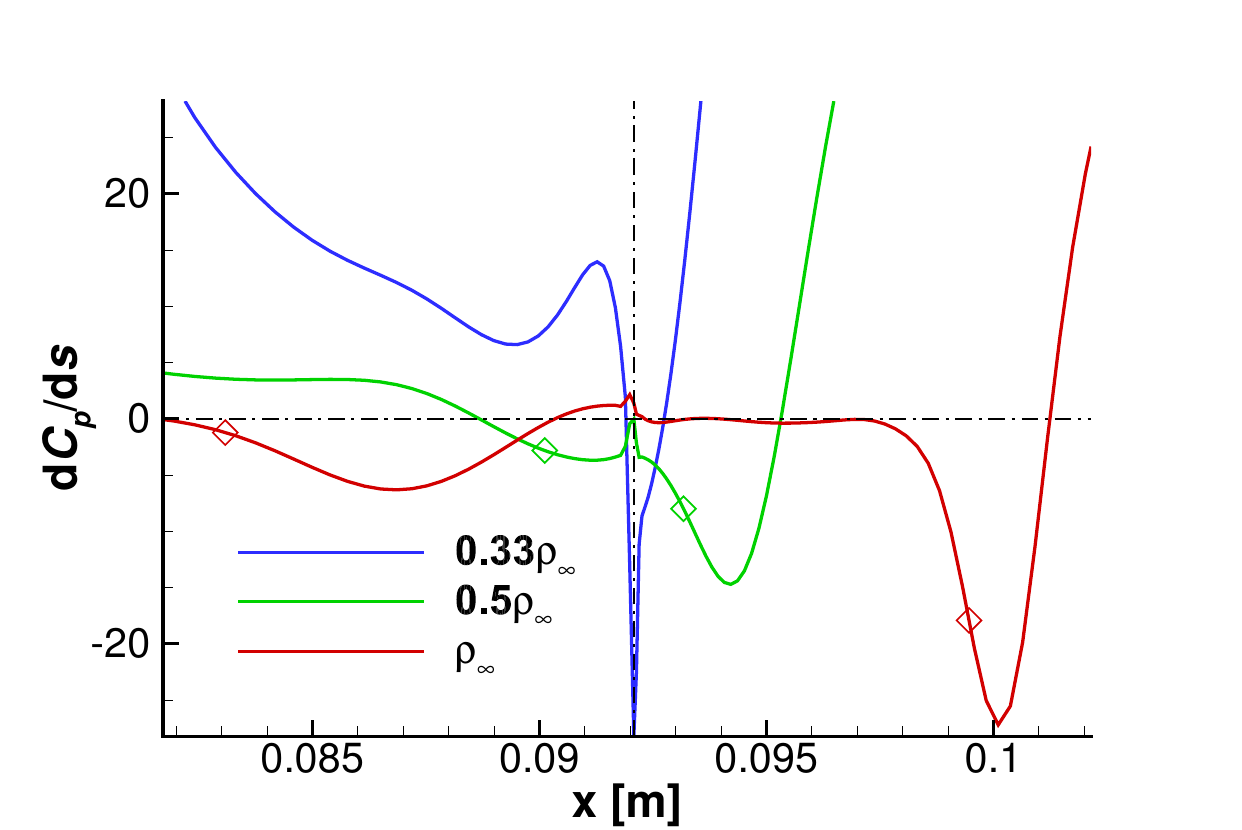}
		\label{fig:rare_dpds_zoom}
	}
	\caption{Comparisons of surface quantities for varying freestream densities. Left column shows the overall distributions; Right column shows the enlarged views near the corner. Open circles ($\circ$) indicate primary separation and reattachment points; open diamonds ($\diamond$) indicate secondary separation and reattachment points.}
	\label{fig:rarefaction_surface}
\end{figure*}

The variations in these surface quantities provide quantitative insight into the flow physics. Figure~\ref{fig:rarefaction_surface} presents the distributions of $C_f$, $C_p$, $q_w$ (represented by $St$), and $dC_p/ds$. In these plots, the separation and reattachment points of the primary bubble are marked by open circles ($\circ$), while those of the secondary bubble (if present) are marked by open diamonds ($\diamond$).

Figure~\ref{fig:rarefaction_surface}a displays the $C_f$ distributions. Upstream of separation, $C_f$ decreases as the laminar boundary layer develops along the first cone. Near the separation point, $C_f$ experiences a sudden decrease to a local minimum and then rises. Downstream of the corner, $C_f$ drops again to two successive local minima. The one closer to the reattachment point is caused by the impingement of the transmitted shock on the wall. Inside the separation region, the $C_f$ profile for the baseline case exhibits a positive peak near the corner, indicating the presence of secondary separation. For the $0.5\rho$ case, $C_f$ rises to approach zero near the corner, corresponding to the minute secondary vortex identified in the streamlines. For the $0.33\rho$ case, $C_f$ remains negative throughout the separation region.

The surface pressure distributions (Figure~\ref{fig:rarefaction_surface}b) show a characteristic plateau region associated with the separation. Near the separation point, $C_p$ begins to rise and enters the plateau. Subsequently, the pressure rises dramatically to its peak value near flow reattachment. A slight pressure `dip' can be observed at the end of the pressure plateau for all cases. In fact, this pressure `dip' is the footprint of the low pressure in the vortex core of the primary bubble. Physically, the recirculating flow within the separation bubble requires an inward pressure gradient to provide the necessary centripetal force; consequently, the pressure decreases towards the vortex core, manifesting as a local depression on the adjacent wall surface.

The surface heat flux (Figure~\ref{fig:rarefaction_surface}c) behaves similarly to the skin friction in the incoming boundary layer. The Stanton number enters a valley region downstream of separation and increases to a peak value near the reattachment point. With decreasing freestream density, the pressure coefficient exhibits an overall decreasing trend, whereas both the skin friction coefficient and the Stanton number increase.

The behavior of the pressure `dip' is further demonstrated in Figure~\ref{fig:rarefaction_surface}d, which plots the distributions of $dC_p/ds$. Downstream of the separation point, there is a nearly zero-pressure-gradient region corresponding to the pressure plateau. Near the corner, a distinct local minimum of the streamwise pressure gradient is induced by the pressure `dip'. When the freestream density exceeds a certain threshold, the reverse boundary layer developing from the primary reattachment point cannot resist this local adverse pressure gradient, causing it to separate again and form a secondary bubble. Inside this secondary separation region, another zero-pressure-gradient region can be observed for the baseline case.

\subsubsection{Validation against Experiments and Reference CFD}

\begin{figure*}[htp]
	\centering
	\subfloat[Run 6: Surface pressure coefficient]{
		\includegraphics[width=0.45\textwidth]{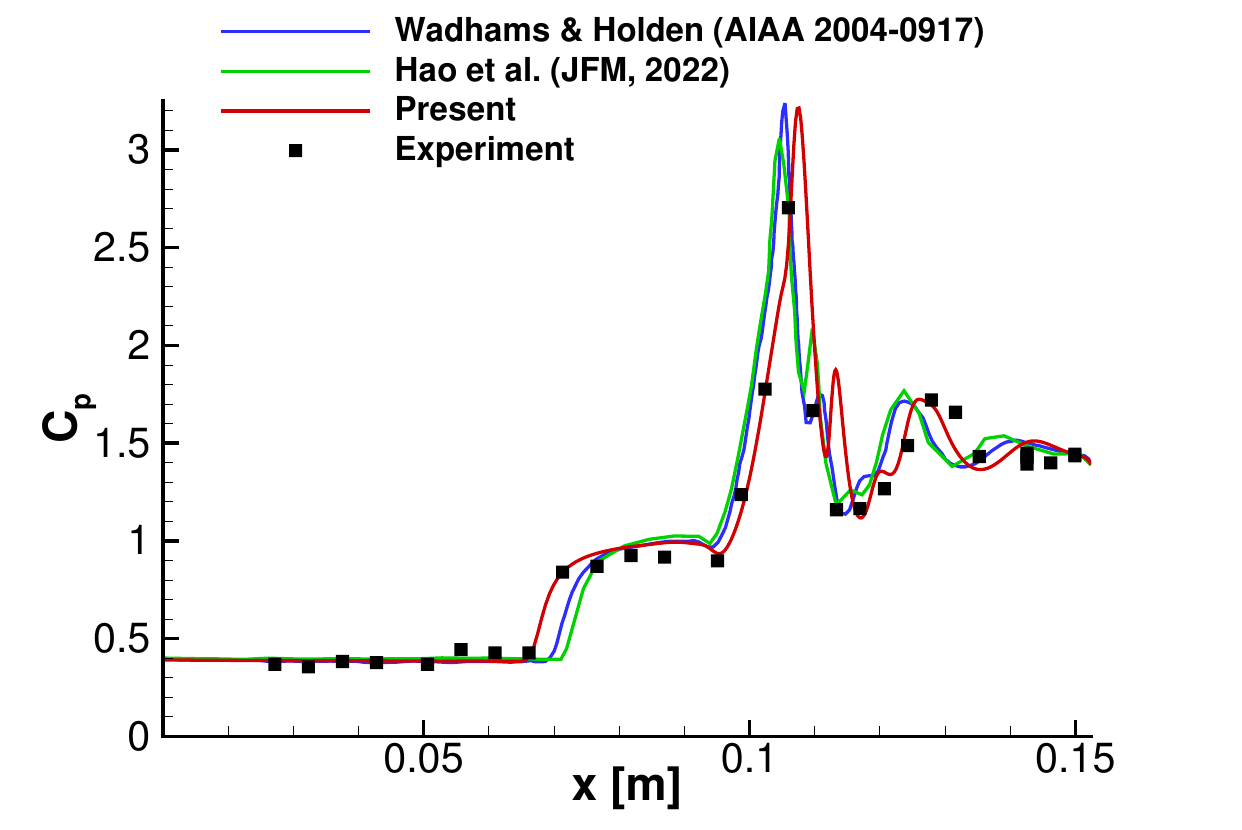}
	}
	\hfill
	\subfloat[Run 6: Surface Stanton number]{
		\includegraphics[width=0.45\textwidth]{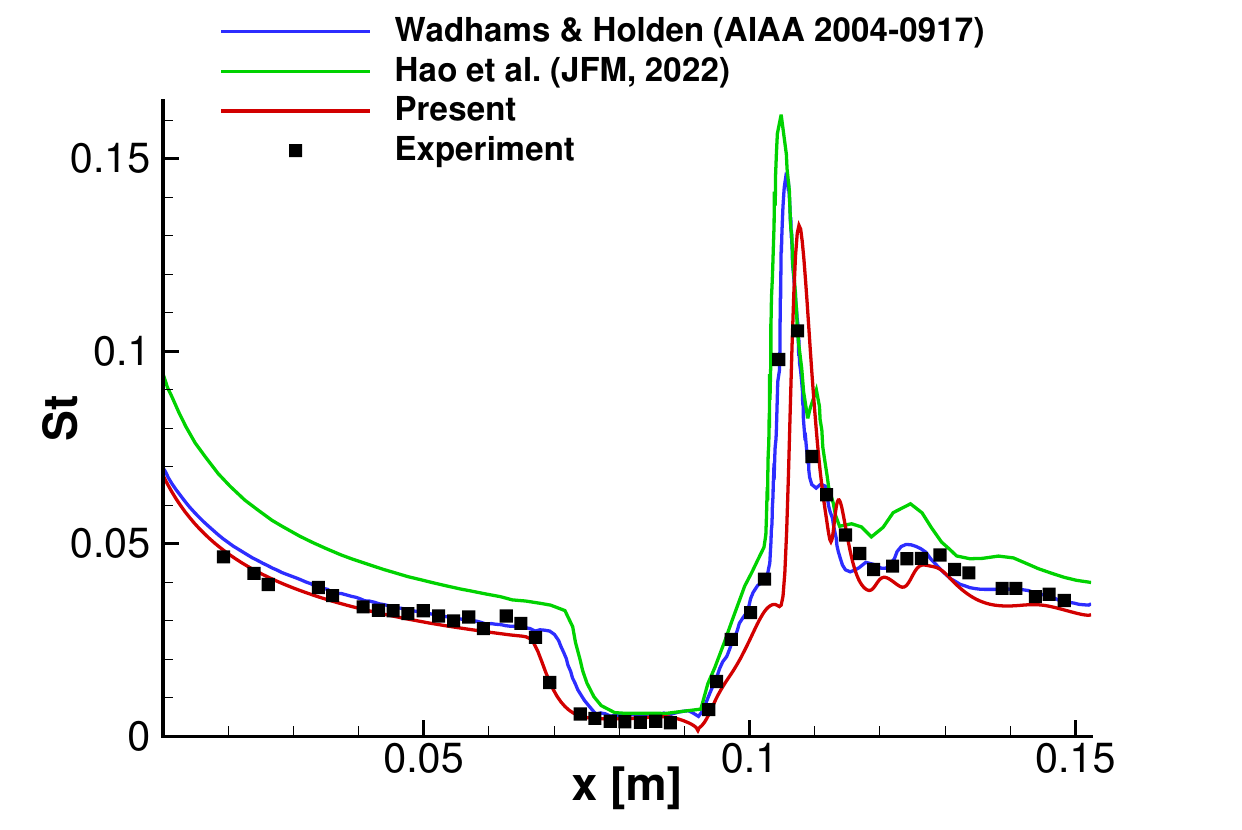}
	}
	
	\vspace{0.5em}
	
	\subfloat[Run 7: Surface pressure coefficient]{
		\includegraphics[width=0.45\textwidth]{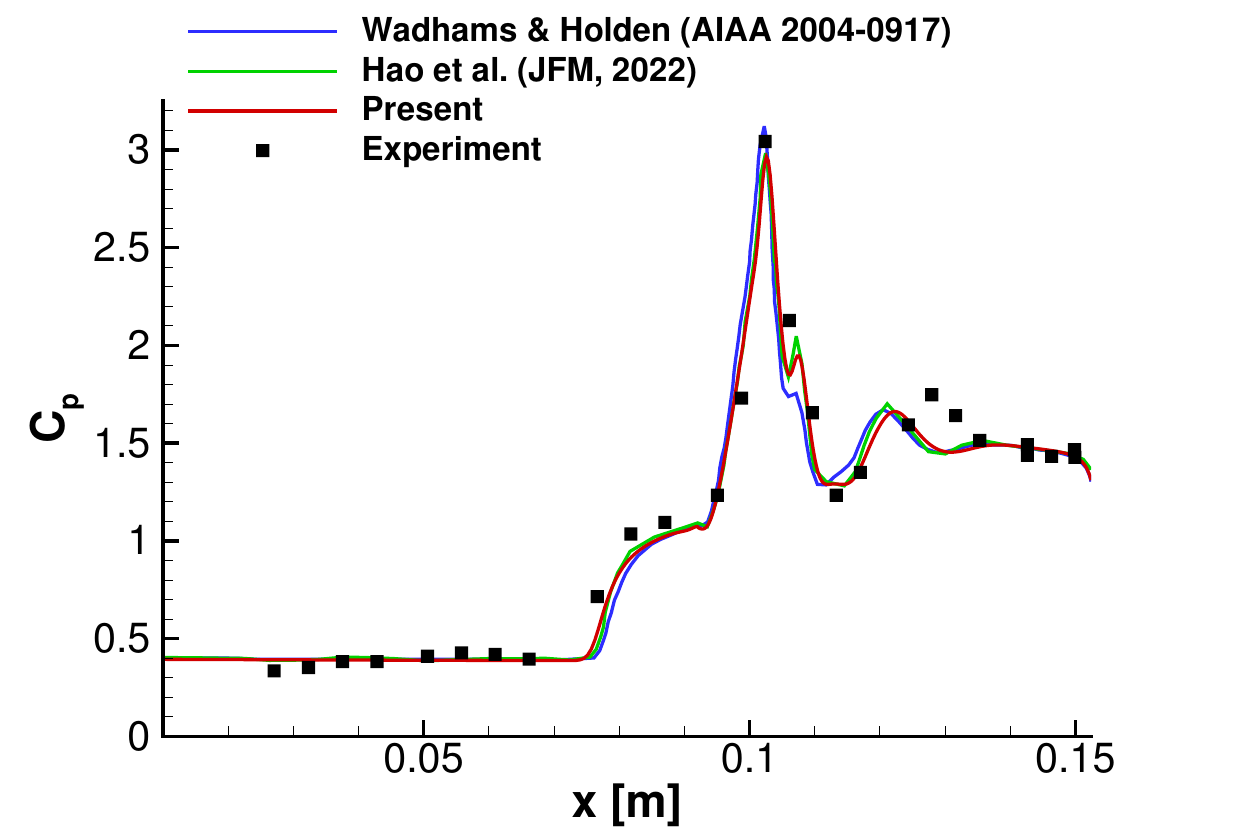}
	}
	\hfill
	\subfloat[Run 7: Surface Stanton number]{
		\includegraphics[width=0.45\textwidth]{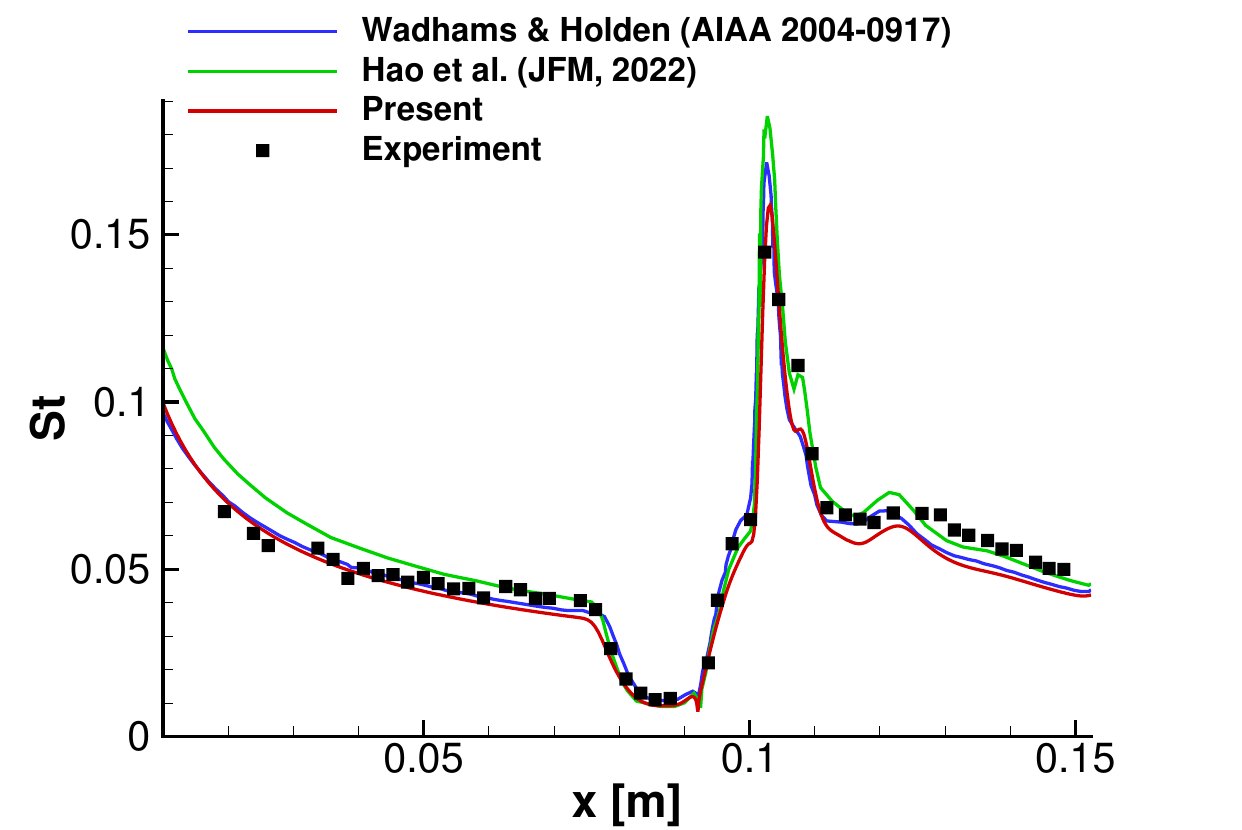}
	}
	\caption{Comparison of surface pressure coefficient and Stanton number with experimental data and reference CFD results for Run 6 ($0.5\rho$) and Run 7 ($0.33\rho$).}
	\label{fig:validation_low_density}
\end{figure*}

To further assess the quantitative accuracy of the present method, the computed surface pressure coefficient and Stanton number distributions for the lower density cases are 
compared with the experimental measurements~\cite{Holden2003,Holden2004} and reference CFD solutions provided by Holden et al.~\cite{Holden2004} and Hao et al.~\cite{Hao2022}.
It is crucial to note that Holden's results were obtained using the identical two-temperature model and wall accommodation coefficients as the present study, whereas Hao's 
results were based on a one-temperature model combined with a no-slip isothermal wall. The comparisons are presented in Figure~\ref{fig:validation_low_density}.

Figures~\ref{fig:validation_low_density}a and \ref{fig:validation_low_density}b display the results for the Run 6 case. The present method demonstrates superior accuracy in 
predicting the separation onset, yielding a separation point that aligns closely with the experimental data and is notably more accurate than the reference CFD results. On 
the first cone, the predicted Stanton number agrees well with the measurements. Furthermore, the peak pressure, peak heat flux, and the recovery profiles downstream of reattachment 
are all captured within a reasonable range.

For the lower density Run 7 case (Figures~\ref{fig:validation_low_density}c and \ref{fig:validation_low_density}d), all three numerical approaches predict similar separation 
and reattachment locations, which are in good agreement with the experiment. The peak loads predicted by the different methods are also comparable. The present method continues 
to provide accurate Stanton number predictions on the first cone surface.

A distinct disparity is observed in the Stanton number magnitude, where Hao's results are consistently higher than both the present and Holden's results. This significant 
overprediction is directly attributed to the deficiency of the one-temperature model in neglecting the vibrational relaxation process, whereas the consistent agreement 
between the present work and Holden's confirms the necessity of the two-temperature formulation.

\subsection{Three-dimensional Effects at Angle of Attack}

Finally, to demonstrate the capability of the proposed 3D solver in handling complex flow separation, a simulation was performed for the double-cone configuration at 
a $2^\circ$ angle of attack, designated herein as Case AoA. The specific flow parameters are listed in Table~\ref{tab:all_conditions}. Aside from the non-zero 
incidence, the free-stream conditions are similar to the Run 35 baseline.

\begin{figure*}[htp]
	\centering
	\subfloat[Windward]{
		\includegraphics[width=0.45\textwidth]{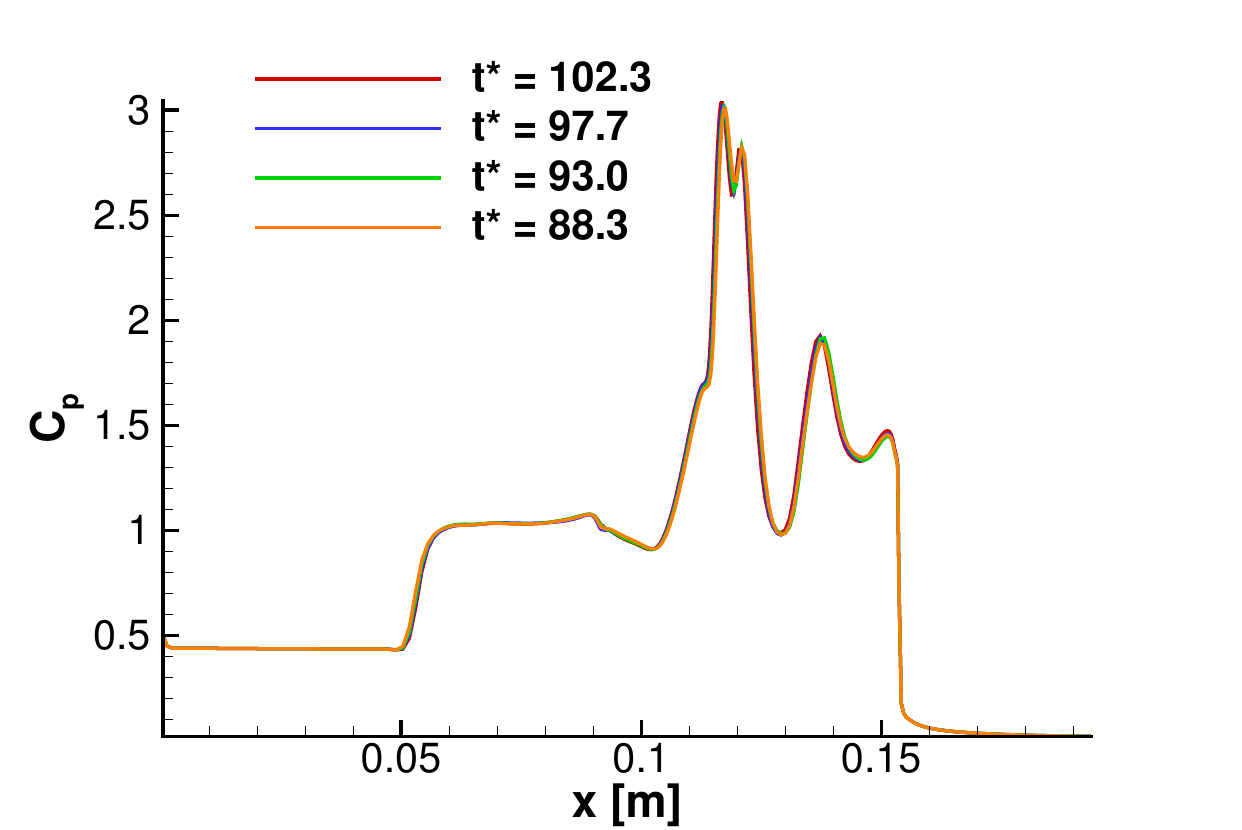}
	}
	\hfill
	\subfloat[Leeward]{
		\includegraphics[width=0.45\textwidth]{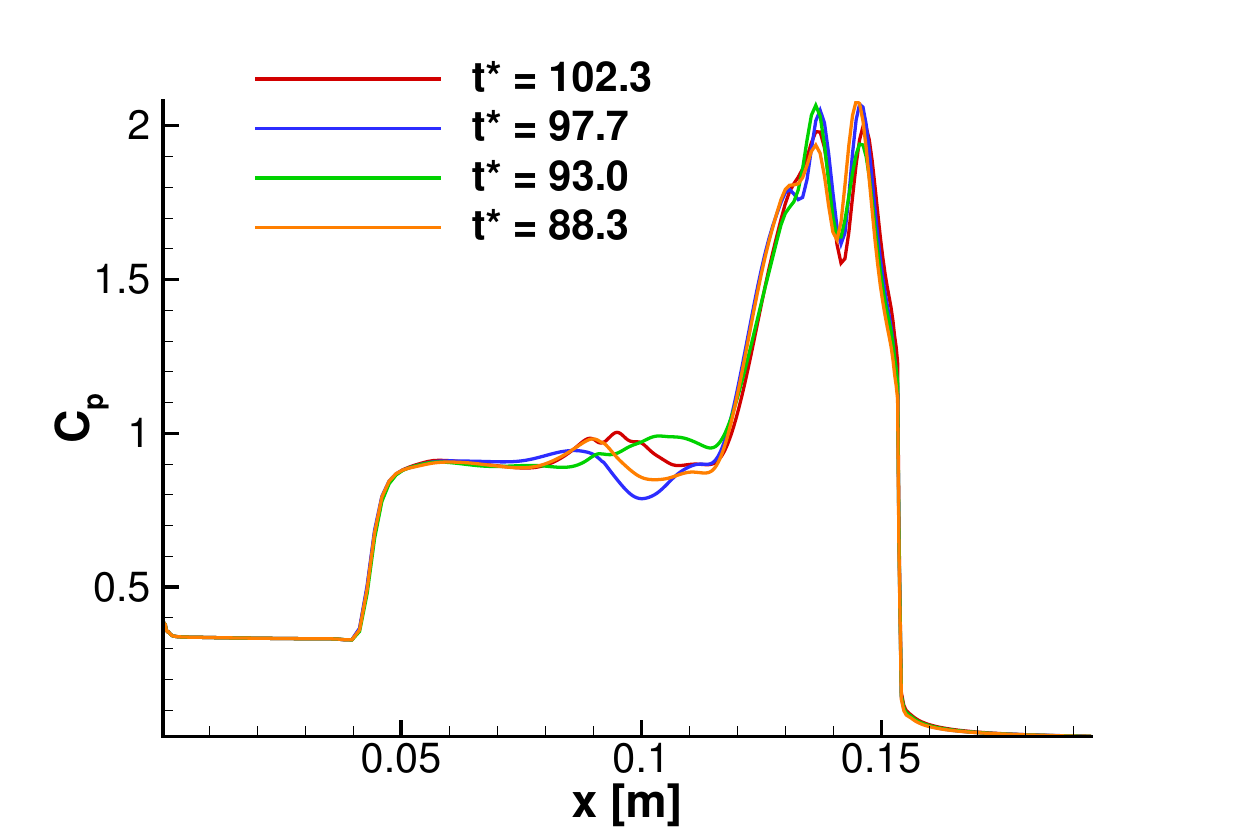}
	}
		\caption{Instantaneous surface pressure coefficient distributions on the windward and leeward sides at different normalized time instances ($t^*$) for Case AoA.}
	\label{fig:unsteady_Cp}
\end{figure*}

Previous stability analyses by Hao et al.~\cite{Hao2022} demonstrated that the flow under Run 35 conditions intrinsically supports three-dimensional global instabilities. 
Consequently, physical unsteadiness is anticipated for the present case. 
Figure~\ref{fig:unsteady_Cp} depicts the instantaneous surface pressure coefficient distributions along the windward and leeward centerlines at various time instants. 
The physical time is normalized by the characteristic flow time, defined as $t^* = t/t_c$. As evident from the plots, the windward flow field remains essentially stable. 
Conversely, the flow in the leeward separation region exhibits distinct unsteady periodic fluctuations. To analyze the dominant topological features, the results presented 
in the following sections correspond to a representative instantaneous snapshot taken at $t^* = 97.7$.

\subsubsection{Three-dimensional Flow Structure}

\begin{figure}[htp]  
	\centering  
	\includegraphics[width=0.45\textwidth]{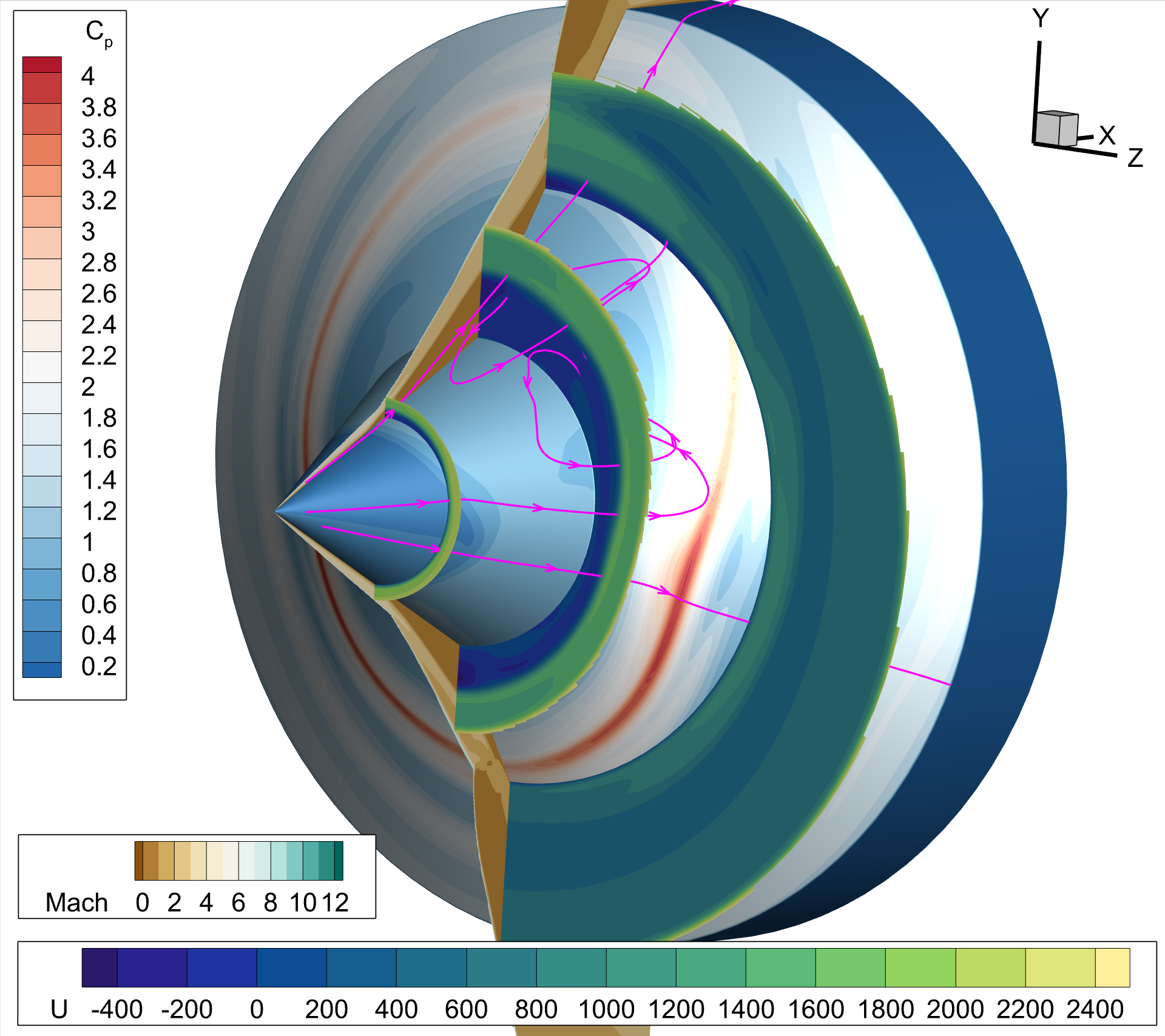}
	\vspace{-4mm}  
	\caption{Composite visualization of the flow field for Case AoA. The plot displays Mach number on the symmetry plane, streamwise velocity ($U$) on axial slices, surface pressure coefficient ($C_p$), and representative three-dimensional streamlines.}
\label{fig:3d_flow_structure}
\end{figure}

Figure~\ref{fig:3d_flow_structure} presents a composite visualization of the flow field for Case AoA. 
The macroscopic asymmetry is clearly revealed by the Mach number contours on the symmetry plane ($y=0$). On the windward side (bottom), 
the separation region is suppressed by the higher compression. In contrast, the leeward side (top) exhibits a significantly enlarged separation zone due to the 
expansion effect.

The streamwise evolution of the separation bubble is captured by the $U$-velocity contours at three axial stations ($x=0.05, 0.09208, 0.12$ m). At $x=0.05$ m, the flow remains 
attached with positive near-wall velocities. At $x = 0.09208$ m (corner), a distinct region of negative velocity (blue) appears, marking the core of the recirculation zone. 
By $x=0.12$ m, the flow begins to reattach and accelerate.
Crucially, the surface pressure coefficient ($C_p$) contours show that the high-pressure reattachment zone is distinctly non-circular. The peak pressure band is distorted 
along the circumference, indicating significant three-dimensional modulation of the reattachment line.

This complex topology is further elucidated by the three-dimensional streamlines (purple lines). The streamlines passing outside the separation region remain smooth and 
exhibit quasi-two-dimensional behavior. However, the streamline trapped inside the separation zone ($x = 0.09208$ m) displays highly disordered twisting and 
spiraling motion. This stark contrast confirms that while the external flow is dominated by inviscid shock dynamics, the internal separation region is governed by strong 
three-dimensional crossflow and vortical interactions, which are responsible for the observed flow unsteadiness.

\begin{figure*}[htp]
	\centering
	\subfloat[Side view ($x$-$y$ plane)]{
		\includegraphics[width=0.45\textwidth]{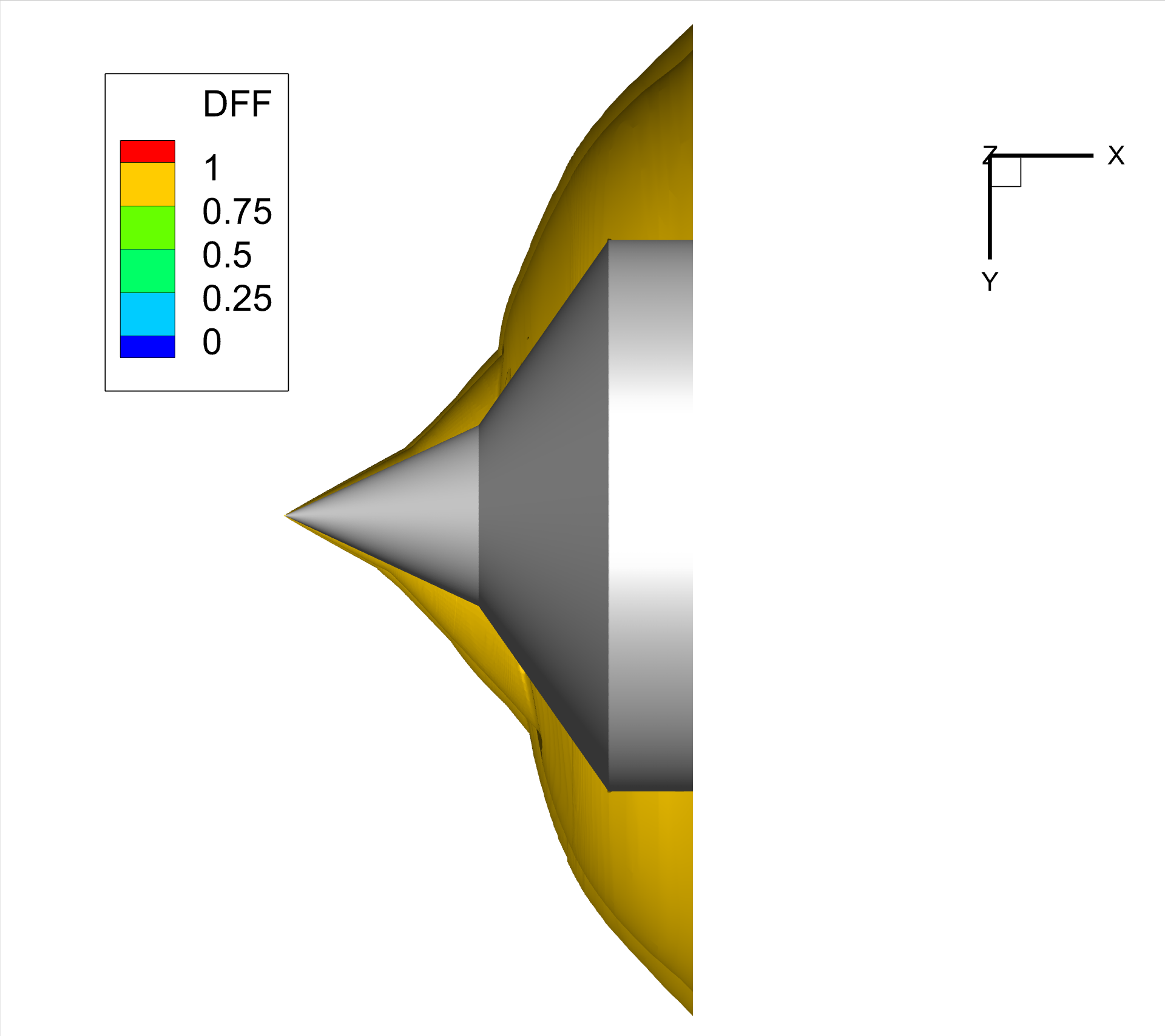}
	}
	\hfill
	\subfloat[Perspective view (rotated $40^\circ$ about $y$-axis)]{
		\includegraphics[width=0.45\textwidth]{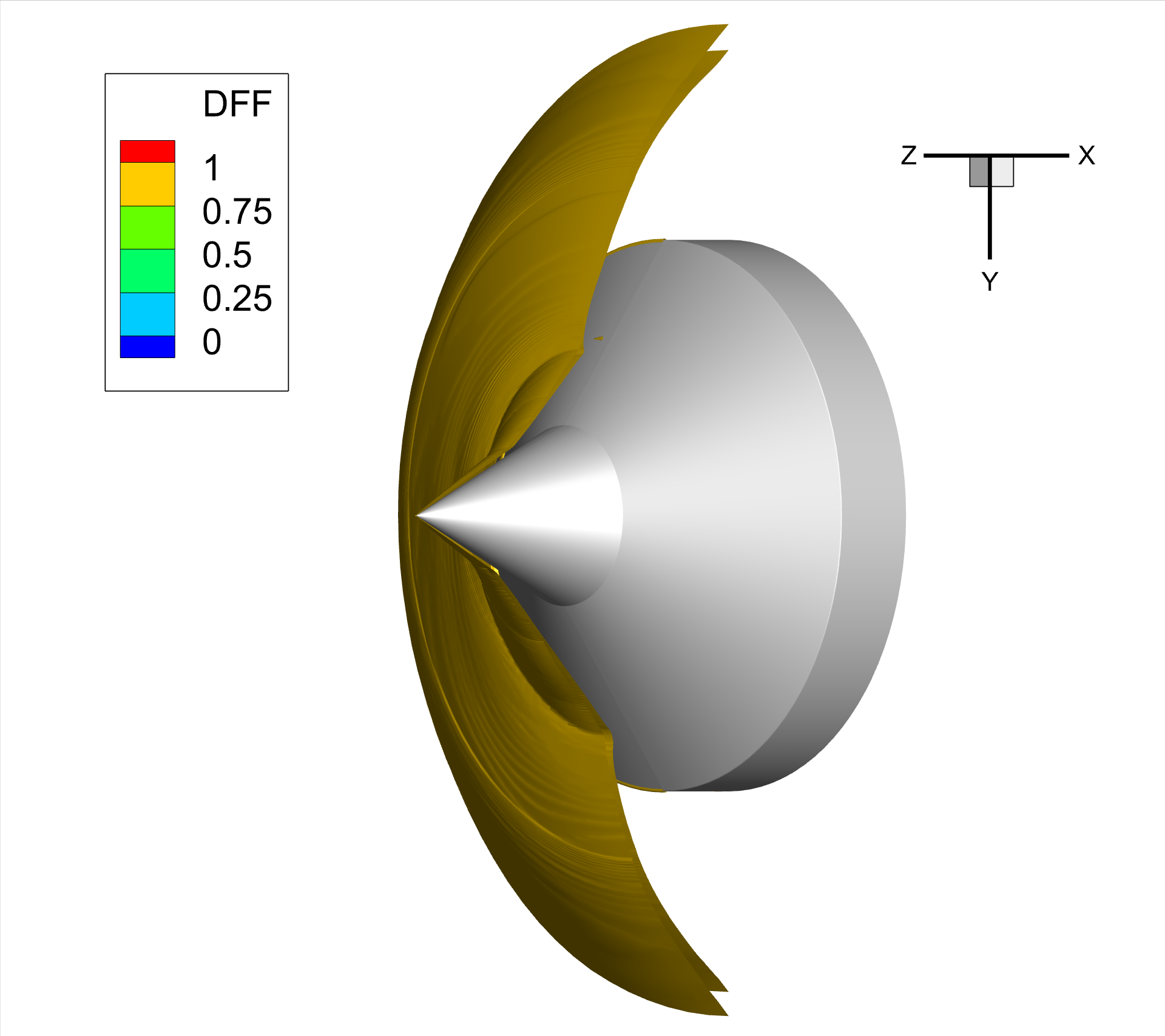}
	}
	\caption{Iso-surfaces of DFF = 0.85 (yellow) for Case AoA.}
	\label{fig:3d_dff}
\end{figure*}

To further confirm the numerical fidelity in resolving steep gradients within this complex three-dimensional environment, the activation status of the slope limiter is 
examined. Figure~\ref{fig:3d_dff} presents the iso-surfaces of the Discontinuity Feedback Factor (DFF) at a value of 0.85 (colored in yellow) from two different perspectives. 
Consistent with the observations in the axisymmetric cases, the DFF is selectively triggered only at the locations of strong discontinuities, clearly outlining the 
three-dimensional shock structures. Crucially, even under the influence of strong crossflow and secondary vortex interactions, the limiter 
remains inactive near the wall surface. This selective behavior ensures that the physical viscous gradients within the boundary layer are preserved without contamination 
from artificial numerical dissipation.

\subsubsection{Surface Topology and Vortex Organization}

\begin{figure}[htp]  
	\centering  
	\includegraphics[width=0.45\textwidth]{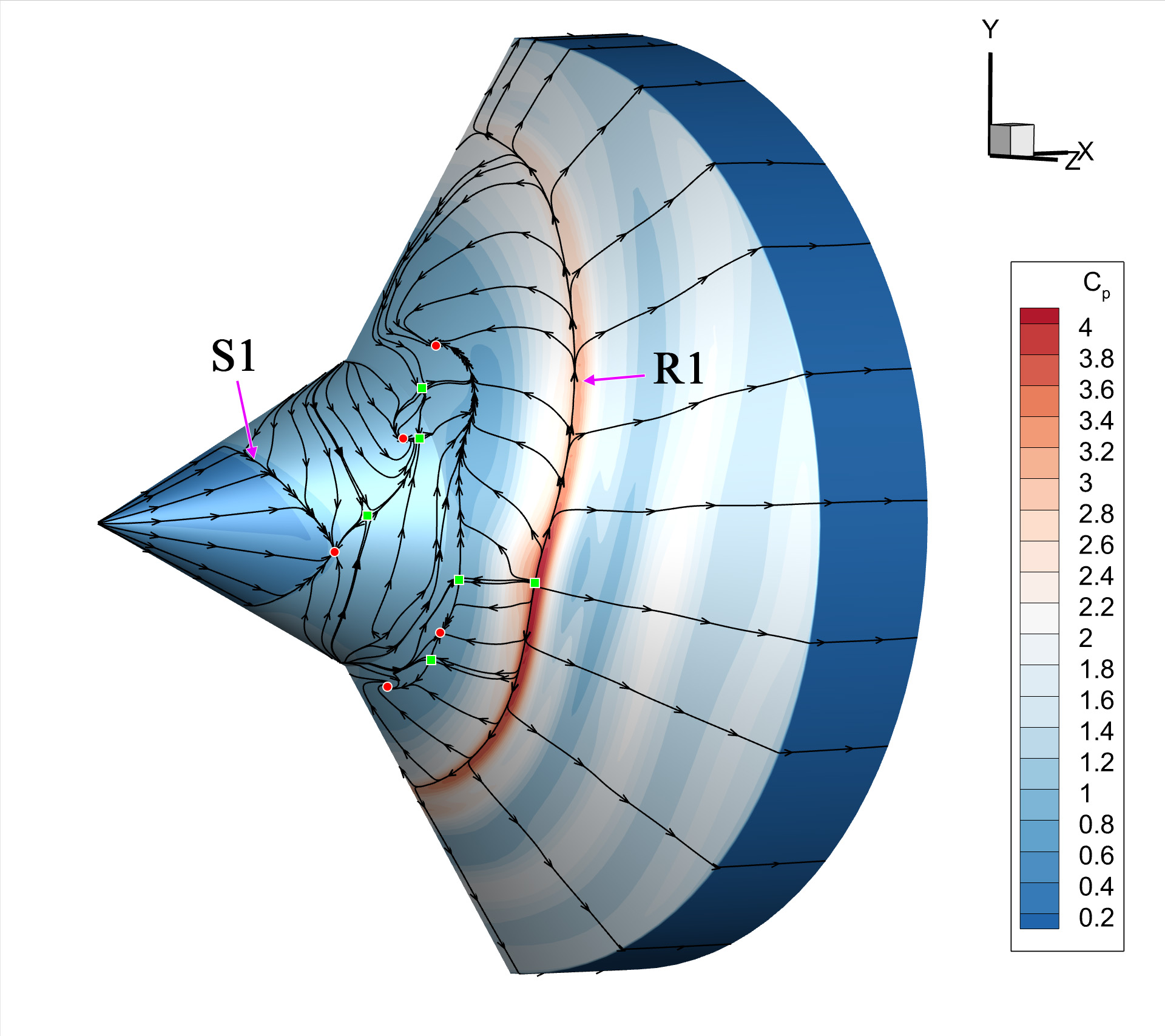}
	\vspace{-4mm}  
	\caption{Surface skin-friction line topology for Case AoA. The primary separation line ($S_1$) and reattachment line ($R_1$) are explicitly marked. 
	Key topological singularities are highlighted: red circles ($\bullet$) denote stable foci, and green squares ({\scriptsize$\blacksquare$}) denote saddle points.}
\label{fig:3d_friction_line}
\end{figure} 

Figure~\ref{fig:3d_friction_line} presents the surface skin friction lines, which reveal a distinct topological hierarchy between the primary and secondary flow structures.
The primary separation ($S_1$) and reattachment ($R_1$) lines retain a global continuity, manifesting as continuous lines traversing from the windward to the leeward 
side. This indicates that the primary interaction is driven by the strong inviscid shock forcing, which maintains a quasi-two-dimensional topology despite the geometric 
asymmetry.

In stark contrast, the secondary flow structure inside the separation bubble undergoes a topological disintegration. Instead of the continuous secondary separation line 
observed in axisymmetric cases, the topology within the separation zone fragments into a complex pattern of discrete singularities. This structure is characterized by a 
series of stable foci and saddle points~\cite{Tobak1981}. 
Topologically, a stable focus serves as a sink where skin-friction lines spiral inward and lift off the surface, whereas a saddle point acts as a separator where streamlines 
converge from two directions and diverge in the others.
This transition from "line-type" to "point-type" topology is physically fundamental. In axisymmetric flows, the rotational invariance constrains the separation to form 
closed rings (singular lines). However, the angle of attack introduces a strong circumferential crossflow. This crossflow destroys the two-dimensional stability of the 
secondary vortex tube, forcing it to fracture and lift off from the surface. The footprints of these lifted vortices manifest as the spiral foci on the wall, marking the 
roots of three-dimensional "tornado-like" vortices~\cite{Lighthill1963}. 
This phenomenon confirms that the present solver correctly captures the transition from closed (2D) to open (3D) separation topology~\cite{Wang1972}.

\subsubsection{Quantitative Comparison}
\begin{figure*}[hbt!]
	\centering
	% 第一行：背风面 (Leeward)
	\subfloat[Leeward side: Surface pressure]{
		\includegraphics[width=0.45\textwidth]{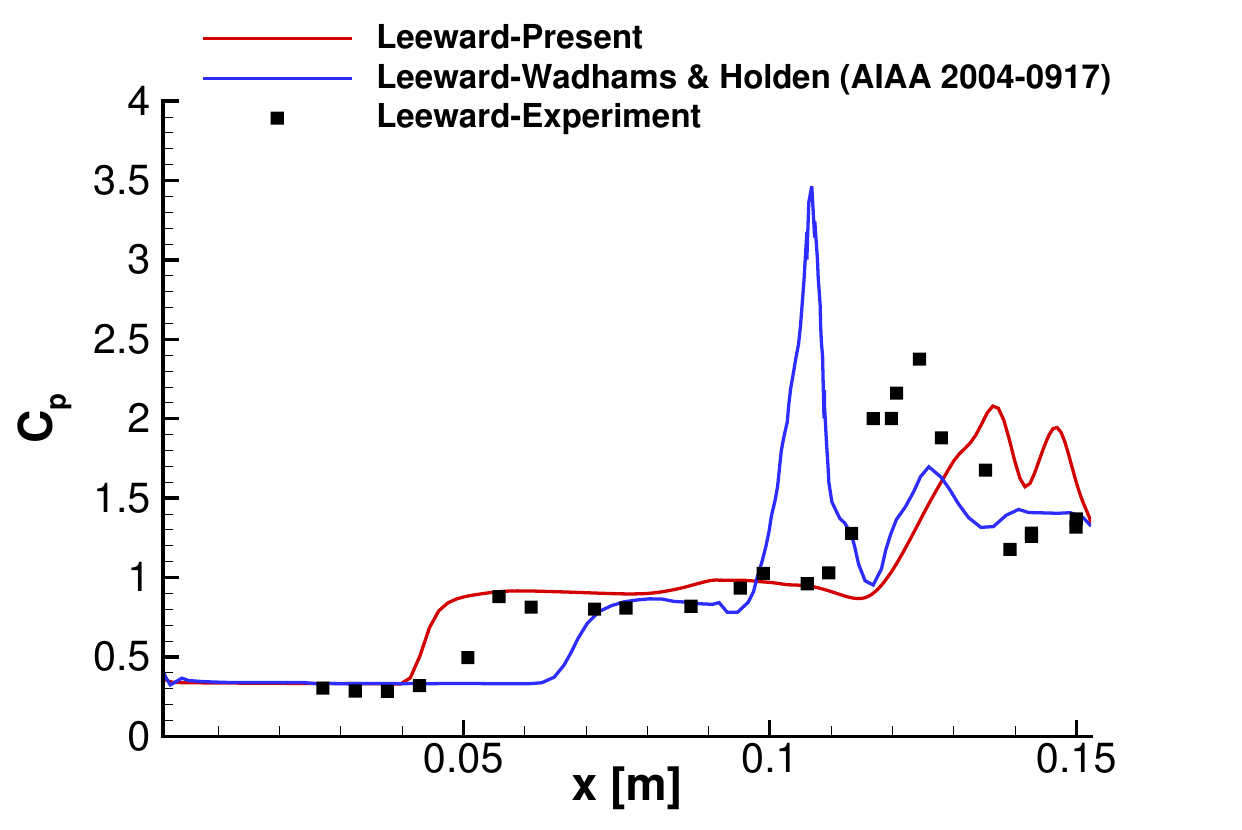}
	}
	\hfill
	\subfloat[Leeward side: Surface Stanton number]{
		\includegraphics[width=0.45\textwidth]{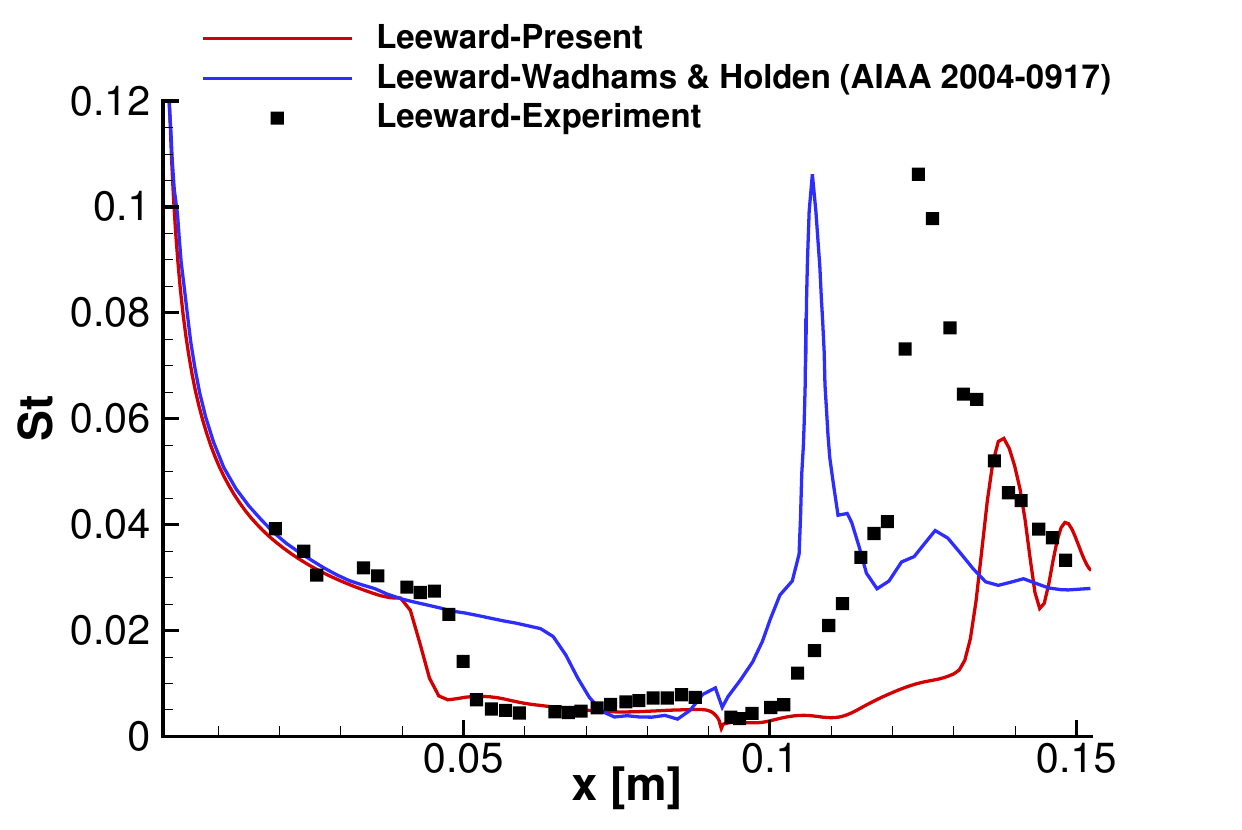}
	}
	
	\vspace{0.5em} % 行间距
	
	% 第二行：迎风面 (Windward)
	\subfloat[Windward side: Surface pressure]{
		\includegraphics[width=0.45\textwidth]{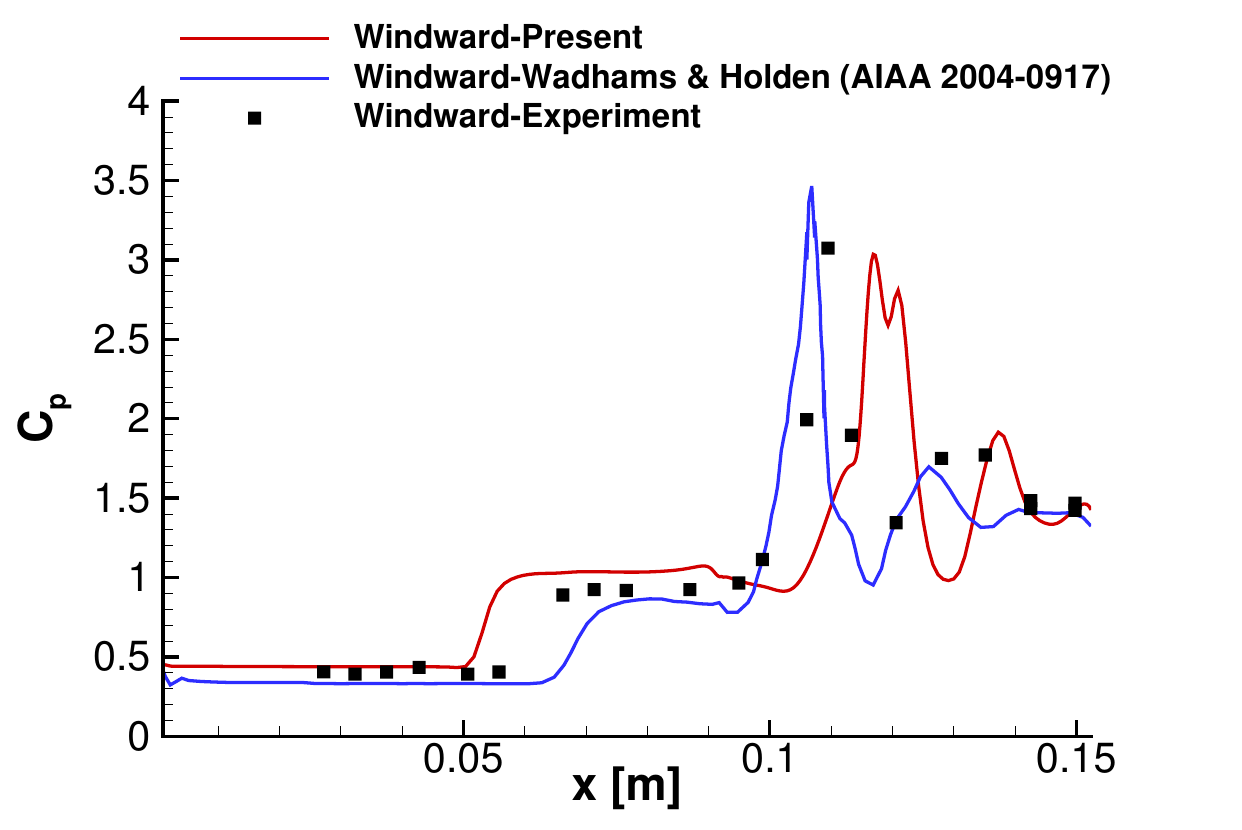}
	}
	\hfill
	\subfloat[Windward side: Surface Stanton number]{
		\includegraphics[width=0.45\textwidth]{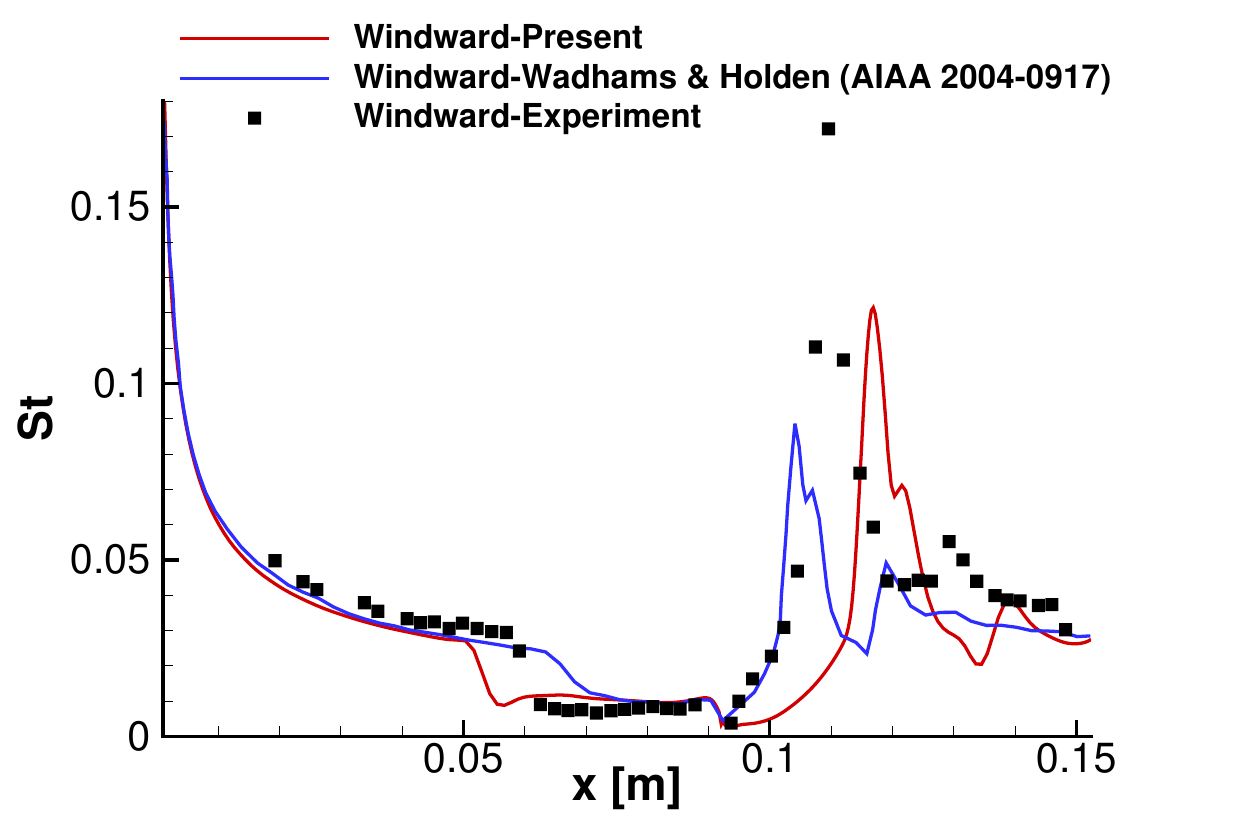}
	}
	
	\caption{Quantitative comparison of surface pressure coefficient and Stanton number with experimental data~\cite{Holden2003} and reference CFD results for Case AoA.}
	\label{fig:3d_validation}
\end{figure*}

Figure~\ref{fig:3d_validation} compares the computed surface pressure and heat flux distributions on the windward ($\phi=0^\circ$) and leeward ($\phi=180^\circ$) centerlines 
with experimental measurements and reference CFD results. It should be noted that the reference CFD results are cited as a private communication in the experimental 
report~\cite{Holden2003}; consequently, specific details regarding their numerical methods and grid resolution are unavailable.

Compared to the axisymmetric baseline, the deviations between numerical predictions and experimental data are more pronounced for this complex three-dimensional case. 
Generally, the present method tends to predict a larger separation region (earlier separation and later reattachment) compared to the experiment, whereas the reference 
CFD results exhibit the opposite trend, predicting a smaller separation zone.

On the leeward side, the present method predicts the flow separation location more accurately than the reference solution. Regarding the pressure peak at reattachment, 
the present simulation slightly underpredicts the magnitude, while the reference results overpredict it, with both showing comparable deviations from the experiment. 
However, the leeward heat flux peak is notably underpredicted by the present method, showing a larger discrepancy than the reference results. 
This underprediction is likely attributed to two factors: the insufficient resolution of the intense vortex-boundary layer interaction on the medium grid, 
and the inherent limitation of the steady-state solver which tends to time-average the peak heating associated with unsteady fluctuations.

On the windward side, both numerical methods exhibit similar deviations in predicting the separation onset. The peak pressure is captured well within the experimental range. 
For the peak heat flux, although both methods show an underprediction, the present results are closer to the experimental measurements than the reference solution.

In summary, despite the quantitative discrepancies inherent to such complex unsteady flows, the proposed method successfully captures the dominant three-dimensional flow 
features. Overall, it provides a prediction capability that is comparable to, and in certain aspects (such as separation region size and windward heating), superior to the 
reference CFD results.

\section{Conclusions}

In this work, a three-dimensional two-temperature gas-kinetic scheme (3D 2T-GKS) on unstructured meshes is developed to simulate hypersonic thermochemical non-equilibrium 
flows. A Generalized Kinetic Boundary Condition (GKBC) is proposed to physically decouple the thermal accommodation of vibrational energy from the translational-rotational 
mode. The main findings are summarized as follows:

The GKBC effectively resolves the issue of surface heat flux overprediction inherent in traditional no-slip and standard Maxwell boundary conditions. 
By incorporating a distinct accommodation coefficient for the vibrational mode ($\alpha_v = 0.001$), the method correctly models the slow relaxation of vibrational 
energy at the wall. Validations against the sharp double cone (Run 35) and hollow cylinder-flare (Run 11) benchmarks demonstrate excellent agreement with experimental 
measurements.

A parametric sensitivity analysis further reveals that the vibrational thermal accommodation ($\alpha_v$) is the dominant mechanism governing the peak heat flux in 
thermal non-equilibrium flows. While the translational-rotational accommodation ($\alpha_{tr}$) affects the wall energy exchange to a lesser extent, the surface thermal 
loads are largely insensitive to variations in the momentum accommodation coefficient ($\sigma$), indicating that velocity slip effects 
are secondary in near-continuum regimes. These results confirm the reliability of the proposed GKBC and highlight the necessity of decoupling accommodation mechanisms 
for accurate aerothermal prediction.

Parametric studies with varying freestream density show that the present method accurately captures the formation and suppression of secondary separation. 
As the freestream density decreases, the primary separation bubble shrinks and the secondary separation near the corner progressively weakens and eventually vanishes. 
This behavior is shown to be governed by the evolution of the local surface pressure gradient associated with the pressure dip induced by the primary vortex core. 
The method consistently resolves the coupled changes in flow topology and surface aerothermal loads across different density conditions.

In the three-dimensional case with a $2^\circ$ angle of attack, the solver successfully captures the breakdown of axisymmetric flow topology. Driven by strong 
crossflow, the secondary separation structure in the leeward region disintegrates from continuous lines into discrete singularities (foci and saddles), forming 
tornado-like streamwise vortices. Despite the intrinsic physical unsteadiness and mesh limitations, the method yields robust predictions of the dominant flow 
features and the dual-peak heating pattern, demonstrating its capability in handling complex three-dimensional shock-boundary layer interactions.

Overall, the proposed 3D 2T-GKS framework, augmented by the GKBC, proves to be a robust and accurate tool for predicting aerothermal loads and analyzing flow physics in 
the hypersonic non-equilibrium regime.

\section*{Acknowledgments}
This work was supported by the National Natural Science Foundation of China (Nos.~12302378 and 92371201), 
and the Funding of National Key Laboratory of Computational Physics, and the Natural Science Basic Research 
Plan in Shaanxi Province of China (No.~2025SYS-SYSZD-070).

\section*{Data Availability statement}
The data that support the findings of this study are available within the article and its supplementary 
material.

% \appendix
% \renewcommand{\theequation}{A\arabic{equation}}
% \section*{Appendix A: Moments of the Maxwellian distribution function}\label{appendixA}

% reference
\bibliographystyle{elsarticle-num}
\bibliography{3D-2T-GKS}

@book{Anderson1989,
   author = {Anderson, John David},
   title = {Hypersonic and High Temperature Gas Dynamics},
   publisher = {AIAA},
   ISBN = {156347459X},
   year = {1989},
   type = {Book}
}

@article{Bose2013,
   author = {Bose, Deepak and Brown, James L and Prabhu, Dinesh K and Gnoffo, Peter and Johnston, Christopher O and Hollis, Brian},
   title = {Uncertainty assessment of hypersonic aerothermodynamics prediction capability},
   journal = {Journal of Spacecraft and Rockets},
   volume = {50},
   number = {1},
   pages = {12--18},
   ISSN = {0022-4650},
   year = {2013},
   type = {Journal Article}
}

@article{Olynick1999,
   author = {Olynick, David and Chen, Y.-K. and Tauber, Michael E.},
   title = {Aerothermodynamics of the Stardust Sample Return Capsule},
   journal = {Journal of Spacecraft and Rockets},
   volume = {36},
   number = {3},
   pages = {442--462},
   DOI = {10.2514/2.3466},
   year = {1999},
   type = {Journal Article}
}

@article{Benettin1997,
   author = {Benettin, G. and Carati, A. and Gallavotti, G.},
   title = {A rigorous implementation of the {Jeans - Landau - Teller} approximation for adiabatic invariants},
   journal = {Nonlinearity},
   volume = {10},
   number = {2},
   pages = {479},
   ISSN = {0951-7715},
   DOI = {10.1088/0951-7715/10/2/011},
   year = {1997},
   type = {Journal Article}
}

@article{Bhatnagar1954,
  author  = {Bhatnagar, P. L. and Gross, E. P. and Krook, M.},
  title   = {A Model for Collision Processes in Gases. {I}. Small Amplitude Processes in Charged and Neutral One-Component Systems},
  journal = {Physical Review},
  volume  = {94},
  number  = {3},
  pages   = {511--525},
  year    = {1954},
  doi     = {10.1103/PhysRev.94.511}
}

@book{Bird1994,
   author = {Bird, G A},
   title = {Molecular Gas Dynamics and the Direct Simulation of Gas Flows},
   publisher = {Oxford University Press},
   ISBN = {9780198561958},
   DOI = {10.1093/oso/9780198561958.001.0001},
   year = {1994},
   type = {Book}
}

@article{Cao2022-3T,
   author = {Cao, Guiyu and Shi, Yipeng and Xu, Kun and Chen, Shiyi},
   title = {Modeling and simulation in supersonic three-temperature carbon dioxide turbulent channel flow},
   journal = {Physics of Fluids},
   volume = {34},
   number = {12},
   ISSN = {1070-6631},
   DOI = {10.1063/5.0129353},
   year = {2022},
   type = {Journal Article}
}

@book{Chapman1970,
  author    = {Chapman, Sydney and Cowling, Thomas G.},
  title     = {The Mathematical Theory of Non-Uniform Gases},
  publisher = {Cambridge University Press},
  year      = {1970},
  address   = {Cambridge},
  edition   = {3rd},
}

@techreport{Edney1968,
   author = {Edney, Barry},
   title = {Anomalous heat transfer and pressure distributions on blunt bodies at hypersonic speeds in the presence of an impinging shock},
   institution = {Flygtekniska Forsoksanstalten, Stockholm (Sweden)},
   year = {1968},
   type = {Report}
}

@article{Ji2021,
   author = {Ji, Xing and Shyy, Wei and Xu, Kun},
   title = {A gradient compression-based compact high-order gas-kinetic scheme on {3D} hybrid unstructured meshes},
   journal = {International Journal of Computational Fluid Dynamics},
   volume = {35},
   number = {7},
   pages = {485--509},
   ISSN = {1061-8562},
   year = {2021},
   type = {Journal Article}
}

@article{Li2005,
   author = {Li, Qibing and Fu, Song and Xu, Kun},
   title = {Application of gas-kinetic scheme with kinetic boundary conditions in hypersonic flow},
   journal = {AIAA Journal},
   volume = {43},
   number = {10},
   pages = {2170--2176},
   ISSN = {0001-1452},
   year = {2005},
   type = {Journal Article}
}

@article{Liu2021-3T,
   author = {Liu, Hualin and Cao, Guiyu and Chen, Weifang},
   title = {Multiple-temperature gas-kinetic scheme for type {IV} shock/shock interaction},
   journal = {Communication in Computational Physics},
   pages = {853},
   year = {2021},
   type = {Journal Article}
}

@phdthesis{Lofthouse2008,
   author = {Lofthouse, Andrew J.},
   title = {Nonequilibrium Hypersonic Aerothermodynamics Using the Direct Simulation {Monte Carlo} and {Navier--Stokes} Models},
   university = {University of Michigan},
   url = {https://deepblue.lib.umich.edu/handle/2027.42/58370},
   year = {2008},
}

@article{Park1989-1,
   author = {Park, Chul},
   title = {Assessment of two-temperature kinetic model for ionizing air},
   journal = {Journal of thermophysics and heat transfer},
   volume = {3},
   number = {3},
   pages = {233--244},
   ISSN = {0887-8722},
   year = {1989},
   type = {Journal Article}
}

@article{Park1989-2,
   author = {Park, Chul and Griffith, Wayland},
   title = {Nonequilibrium Hypersonic Aerothermodynamics},
   journal = {Physics Today},
   volume = {44},
   number = {2},
   pages = {98--98},
   ISSN = {0031-9228},
   DOI = {10.1063/1.2809999},
   year = {1991},
   type = {Journal Article}
}

@article{Sutherland1893,
   author = {Sutherland, William},
   title = {{LII}. The viscosity of gases and molecular force},
   journal = {The London, Edinburgh, and Dublin Philosophical Magazine and Journal of Science},
   volume = {36},
   number = {223},
   pages = {507--531},
   ISSN = {1941-5982},
   year = {1893},
   type = {Journal Article}
}

@article{Xu2001,
   author = {Xu, Kun},
   title = {A Gas-Kinetic {BGK} Scheme for the {Navier-Stokes} Equations and Its Connection with Artificial Dissipation and {Godunov} Method},
   journal = {Journal of Computational Physics},
   volume = {171},
   number = {1},
   pages = {289--335},
   ISSN = {00219991},
   DOI = {10.1006/jcph.2001.6790},
   year = {2001},
   type = {Journal Article}
}

@article{Xu2004-2T,
   author = {Xu, Kun and Tang, Lei},
   title = {Nonequilibrium {Bhatnagar-Gross-Krook} model for nitrogen shock structure},
   journal = {Physics of Fluids},
   volume = {16},
   number = {10},
   pages = {3824--3827},
   ISSN = {1070-6631
1089-7666},
   DOI = {10.1063/1.1783372},
   year = {2004},
   type = {Journal Article}
}

@article{Lumpkin1991,
   author = {Lumpkin III, Forrest E and Haas, Brian L and Boyd, Iain D},
   title = {Resolution of differences between collision number definitions in particle and continuum simulations},
   journal = {Physics of Fluids A: Fluid Dynamics},
   volume = {3},
   number = {9},
   pages = {2282-2284},
   ISSN = {0899-8213},
   year = {1991},
   type = {Journal Article}
}

@article{Millikan1963,
   author = {Millikan, Roger C and White, Donald R},
   title = {Systematics of vibrational relaxation},
   journal = {The Journal of chemical physics},
   volume = {39},
   number = {12},
   pages = {3209-3213},
   ISSN = {0021-9606},
   year = {1963},
   type = {Journal Article}
}

@article{Gupta1990,
   author = {Gupta, Roop N and Yos, Jerrold M and Thompson, Richard A and Lee, Kam-Pui},
   title = {A review of reaction rates and thermodynamic and transport properties for an 11-species air model for chemical and thermal nonequilibrium calculations to 30000 K},
   year = {1990},
   type = {Journal Article}
}

@article{Eucken1913,
   author = {Eucken, Arnold},
   title = {Über das Wärmeleitvermögen, die spezifische Wärme und die innere Reibung der Gase},
   journal = {Phys. Z},
   volume = {14},
   number = {8},
   pages = {324-332},
   year = {1913},
   type = {Journal Article}
}

@article{Mason1962,
   author = {Mason, Eo A and Monchick, L},
   title = {Heat conductivity of polyatomic and polar gases},
   journal = {The Journal of Chemical Physics},
   volume = {36},
   number = {6},
   pages = {1622-1639},
   ISSN = {0021-9606},
   year = {1962},
   type = {Journal Article}
}

@article{Candler2003,
   author = {Nompelis, Ioannis and Candler, Graham V and Holden, Michael S},
   title = {Effect of vibrational nonequilibrium on hypersonic double-cone experiments},
   journal = {AIAA journal},
   volume = {41},
   number = {11},
   pages = {2162-2169},
   ISSN = {0001-1452},
   year = {2003},
   type = {Journal Article}
}

@article{Hao2022,
   author = {Hao, Jiaao and Fan, Jianhui and Cao, Shibin and Wen, Chih-Yung},
   title = {Three-dimensionality of hypersonic laminar flow over a double cone},
   journal = {Journal of Fluid Mechanics},
   volume = {935},
   pages = {A8},
   ISSN = {0022-1120},
   year = {2022},
   type = {Journal Article}
}

@inproceedings{Holden2003,
   author = {Holden, Michael and Wadhams, Timothy and Harvey, John and Candler, Graham},
   title = {Measurements in Regions of Low Density Laminar Shock Wave/Boundary Layer Interaction in Hypervelocity Flows and Comparison with Navier-Stokes Predictions},
   booktitle = {41st Aerospace Sciences Meeting and Exhibit},
   pages = {1131},
   type = {Conference Proceedings}
}

@inproceedings{Holden2004,
   author = {Wadhams, Timothy P and Holden, Michael S},
   title = {Summary of experimental studies for code validation in the lens facility and comparisons with recent Navier-Stokes and DSMC solutions for two- and three-dimensional separated regions in hypervelocity flows},
   booktitle = {42 nd AIAA Aerospace Sciences Meeting and Exhibit},
   ISBN = {0146-3705},
   type = {Conference Proceedings}
}

@article{Gao2025,
   author = {Gao, Xingjian and Ji, Xing and Liu, Hualin and Chen, Gang},
   title = {A two-temperature gas-kinetic scheme for hypersonic non-equilibrium flow computations},
   journal = {Physics of Fluids},
   volume = {37},
   number = {10},
   ISSN = {1070-6631},
   year = {2025},
   type = {Journal Article}
}

@book{Babinsky2011,
   author = {Babinsky, Holger and Harvey, John K},
   title = {Shock wave-boundary-layer interactions},
   publisher = {Cambridge University Press},
   volume = {32},
   ISBN = {1139498649},
   year = {2011},
   type = {Book}
}

@article{Candler2019,
   author = {Candler, Graham V},
   title = {Rate effects in hypersonic flows},
   journal = {Annual Review of Fluid Mechanics},
   volume = {51},
   number = {1},
   pages = {379-402},
   ISSN = {0066-4189},
   year = {2019},
   type = {Journal Article}
}

@inproceedings{Candler2001,
   author = {Candler, Graham and Nompelis, Ioannis and Druguet, Marie-Claude},
   title = {Navier-Stokes predictions of hypersonic double-cone and cylinder-flare flow fields},
   booktitle = {39th Aerospace Sciences Meeting and Exhibit},
   pages = {1024},
   type = {Conference Proceedings}
}

@inproceedings{Druguet2003,
   author = {Druguet, Marie-Claude and Candler, Graham and Nompelis, Ioannis},
   title = {Simulations of viscous hypersonic double-cone flows: Influence of numerics},
   booktitle = {16th AIAA Computational Fluid Dynamics Conference},
   pages = {3548},
   type = {Conference Proceedings}
}

@inproceedings{Harvey2001,
   author = {Harvey, John and Holden, Michael and Wadhams, Timothy},
   title = {Code validation study of laminar shockiboundary layer and shock/shock interactions in hypersonic flow part b: comparison with navier-stokes and dsmc solutions},
   booktitle = {39th aerospace sciences meeting and exhibit},
   pages = {1031},
   type = {Conference Proceedings}
}

@article{Holden2006,
   author = {Holden, Michael S and Wadhams, Timothy P and Harvey, John K and Candler, Graham V},
   title = {Comparisons between measurements in regions of laminar shock wave boundary layer interaction in hypersonic flows with navier-stokes and DSMC solutions},
   journal = {Report No. ADA455695 (DTIC, 2006)},
   year = {2006},
   type = {Journal Article}
}

@inproceedings{MacLean2004,
   author = {MacLean, Matthew and Holden, Michael},
   title = {Validation and comparison of WIND and DPLR results for hypersonic, laminar problems},
   booktitle = {42nd AIAA Aerospace Sciences Meeting and Exhibit},
   pages = {529},
   type = {Conference Proceedings}
}

@inproceedings{Candler2002,
   author = {Candler, G and Nompelis, Ioannis and Druguet, M-C and Holden, M and Wadhams, T and Boyd, I and Wang, W-L},
   title = {CFD validation for hypersonic flight-Hypersonic double-cone flow simulations},
   booktitle = {40th AIAA aerospace sciences meeting \& exhibit},
   pages = {581},
   type = {Conference Proceedings}
}

@article{Knight2018,
   author = {Knight, Doyle and Mortazavi, Mahsa},
   title = {CCD Report 2018-1 Simulation of Non-equilibrium Hypersonic Shock Wave Boundary Layer Interactions Using the Navier-Stokes Equations},
   journal = {Rutgers University},
   year = {2018},
   type = {Journal Article}
}

@article{Black1974,
   author = {Black, Graham and Wise, Henry and Schechter, Samuel and Sharpless, Robert L},
   title = {Measurements of vibrationally excited molecules by Raman scattering. II. Surface deactivation of vibrationally excited N2},
   journal = {The Journal of Chemical Physics},
   volume = {60},
   number = {9},
   pages = {3526-3536},
   ISSN = {0021-9606},
   year = {1974},
   type = {Journal Article}
}

@article{Xu2005,
   author = {Xu, Kun and Mao, Meiliang and Tang, Lei},
   title = {A multidimensional gas-kinetic BGK scheme for hypersonic viscous flow},
   journal = {Journal of Computational Physics},
   volume = {203},
   number = {2},
   pages = {405-421},
   ISSN = {0021-9991},
   year = {2005},
   type = {Journal Article}
}

@article{Li2025,
   author = {Cheng, LI and Jiaao, HAO},
   title = {Three-dimensional effects on establishment of hypersonic double-cone flow},
   journal = {Chinese Journal of Aeronautics},
   pages = {103583},
   ISSN = {1000-9361},
   year = {2025},
   type = {Journal Article}
}

@article{Dai2022,
   author = {Dai, Chunliang and Sun, Bo and Zhuo, Changfei and Zhou, Shengbing and Zhou, Changsheng and Yue, Lianjie},
   title = {Numerical study of high temperature non-equilibrium effects of double-wedge in hypervelocity flow},
   journal = {Aerospace Science and Technology},
   volume = {124},
   pages = {107526},
   ISSN = {1270-9638},
   year = {2022},
   type = {Journal Article}
}

@article{Grover2025,
   author = {Grover, Maninder S and Valentini, Paolo and Bisek, Nicholas},
   title = {Ab initio investigation of a hypersonic double cone experiment},
   journal = {Science Advances},
   volume = {11},
   number = {6},
   pages = {eads2147},
   ISSN = {2375-2548},
   year = {2025},
   type = {Journal Article}
}

@article{Hao2017,
   author = {Hao, Jiaao and Wang, Jingying and Lee, Chunhian},
   title = {Numerical simulation of high-enthalpy double-cone flows},
   journal = {AIAA Journal},
   volume = {55},
   number = {7},
   pages = {2471-2475},
   ISSN = {0001-1452},
   year = {2017},
   type = {Journal Article}
}

@inproceedings{Harvey2003,
   author = {Harvey, John K},
   title = {A review of a validation exercise on the use of the DSMC method to compute viscous/inviscid interactions in hypersonic flow},
   booktitle = {36 th AIAA Thermophysics Conference},
   ISBN = {0146-3705},
   type = {Conference Proceedings}
}

@inproceedings{Holden2015,
   author = {Holden, Michael S},
   title = {Experimental research and analysis in supersonic and hypervelocity flows in the LENS shock tunnels and expansion tunnel},
   booktitle = {20th AIAA international space planes and hypersonic systems and technologies conference},
   pages = {3660},
   type = {Conference Proceedings}
}

@inproceedings{Holden2013,
   author = {Holden, Michael S and Wadhams, Tim P and MacLean, Matthew G and Dufrene, Aaron T},
   title = {Measurements of real gas effects on regions of laminar shock wave/boundary layer interaction in hypervelocity flows for “blind’code validation studies},
   booktitle = {21st AIAA Computational Fluid Dynamics Conference},
   pages = {2837},
   type = {Conference Proceedings}
}

@article{Hong2022,
   author = {Hong, Qizhen and Hao, Jiaao and Uy, Ken Chun Kit and Wen, Chih-Yung and Sun, Quanhua},
   title = {Thermochemical nonequilibrium effects on high-enthalpy double-wedge flows},
   journal = {Physics of Fluids},
   volume = {34},
   number = {6},
   ISSN = {1070-6631},
   year = {2022},
   type = {Journal Article}
}

@inproceedings{Kieweg2019,
   author = {Kieweg, Sarah L and Ray, Jaideep and Weirs, V Gregory and Carnes, Brian and Dinzl, Derek and Freno, Brian and Howard, Micah and Phipps, Eric and Rider, William and Smith, Thomas},
   title = {Validation assessment of hypersonic double-cone flow simulations using uncertainty quantification, sensitivity analysis, and validation metrics},
   booktitle = {AIAA SciTech 2019 Forum},
   pages = {2278},
   type = {Conference Proceedings}
}

@inproceedings{Nompelis2010,
   author = {Nompelis, Ioannis and Candler, Graham},
   title = {Numerical investigation of double-cone flow experiments with high-enthalpy effects},
   booktitle = {48th AIAA Aerospace Sciences Meeting Including the New Horizons Forum and Aerospace Exposition},
   pages = {1283},
   type = {Conference Proceedings}
}

@article{Lighthill1963,
   author = {Lighthill, MJ},
   title = {Attachment and separation in three-dimensional flow},
   journal = {Laminar boundary layers},
   pages = {72-82},
   year = {1963},
   type = {Journal Article}
}

@techreport{Tobak1981,
   author = {Tobak, Murray and Peake, David J},
   title = {Topology of three-dimensional separated flows},
   year = {1981},
   type = {Report}
}

@article{Wang1972,
   author = {Wang, KC},
   title = {Separation patterns of boundary layer over an inclined body of revolution},
   journal = {Aiaa Journal},
   volume = {10},
   number = {8},
   pages = {1044-1050},
   ISSN = {0001-1452},
   year = {1972},
   type = {Journal Article}
}

\end{document}